\documentclass[12pt,twoside]{article}
\usepackage{geometry}                		
\usepackage{graphicx}				
\usepackage{amssymb}
\usepackage{amsmath}
\usepackage{authblk}
\usepackage{subcaption}
\usepackage{wrapfig}
\usepackage{t1enc}
\usepackage[english]{babel}
\usepackage[latin2]{inputenc}
\usepackage{graphicx}
\usepackage{babel,a4wide,amsfonts,bm}
\usepackage{slashed}
\usepackage{cite}
\usepackage{dsfont}
\usepackage{color}
\usepackage{hyperref}
\usepackage{authblk}
\newcommand{\pom}{{I\!\!P}}

\graphicspath{{figs/}}

\usepackage{color}
\usepackage[normalem]{ulem}
\definecolor{mscolor}{rgb}{0,0.1,0.7}
\definecolor{ascolor}{rgb}{1,0,1}
\definecolor{mdcolor}{rgb}{1,0,0}

\DeclareRobustCommand\msout{\bgroup\markoverwith{\color{mscolor}{\rule[0.4ex]{2pt}{0.8pt}}}\ULon}
\DeclareRobustCommand\asout{\bgroup\markoverwith{\color{ascolor}{\rule[0.4ex]{2pt}{0.8pt}}}\ULon}
\DeclareRobustCommand\mdout{\bgroup\markoverwith{\color{mdcolor}{\rule[0.4ex]{2pt}{0.8pt}}}\ULon}

\title{ High $|t|$ diffractive vector meson production at the EIC }

\author{Michal De\'ak}
\author{Anna M. Sta\'sto}
\author{ Mark Strikman}
\affil{\small \it Department of Physics, Penn State University, University Park, PA 16802, USA}

\begin{document}

\maketitle

\abstract{
	
We investigate the prospects of the diffractive production of $J/\psi$ mesons at large momentum transfer $|t|$
at the future Electron Ion Collider in electron-proton collisions.  In particular, we  focus on the measurements of the rapidity gap size.
The model used for the calculations is based on the diffractive exchange of the Balitsky-Fadin-Kuraev-Lipatov perturbative Pomeron. 
  Calculations for the cross section and  the estimates for the rates assuming  integrated luminosity of $10 \, \rm fb^{-1}$ are provided.
Two experimental strategies were considered. First, measuring the rapidity gap size directly,
by observing the activity in the forward part of the central detector, and second by putting a lower limit on the
rapidity gap size in the case when the  detector cannot measure forward activity.
We find that,  it is possible to measure at the EIC the dependence of the cross section on rapidity gap interval up to four units in rapidity.
This should allow to measure the change of the cross section by a factor 1.6 expected due to the BFKL exchange.
This is possible with the present setup of the detector which projects the coverage
up to 3.5 units of rapidity. We conclude however, that the extension of the detector up to higher
rapidity, for example to 4.5 would be desirable and provide even better lever arm for testing rapidity gap physics at the EIC.

}
\section{Introduction}

Deep Inelastic Scattering of leptons  off protons is the cleanest process to investigate the structure of the proton and it provides ample possibilities for testing Quantum Chromodynamics with great precision. The HERA machine  was the only electron - proton collider up to date, capable of colliding electrons and positrons with protons up to the center of mass energy of $\sqrt{s}=318\; {\rm GeV}$.    A particularly interesting phenomena observed at HERA were diffractive events, \cite{Ahmed:1995ns, Breitweg:1998gc}, where the proton was observed to stay intact, or dissociated into a state with proton quantum numbers and was separated from the rest of the particles by a `rapidity gap' - a region of detector devoid of any activity.  Precise experimental study  of diffraction phenomena is challenging and crucial for a complete understanding  of strong interaction dynamics.  
	
	Of particular interest is the diffractive production of the heavy vector mesons. 
The heavy vector mesons like $J/\psi$  and $\Upsilon$ have very clear detector signatures and allow for very precise access of kinematic variables associated with their detection which in turn allows to access dependence of dynamical quantities dependent on the same kinematic variables. Their diffractive production is usually described in terms of a colorless exchange with vacuum quantum numbers, which at the lowest order is given by an exchange of two gluons. At higher orders and more generally, this process can be described by an exchange, in a t-channel, of an object called a  Pomeron, which is dominated by the   gluonic  degrees of freedom. One of the most interesting questions  is the energy dependence of the Pomeron on the size of the rapidity gap, and the momentum transfer, $t$ dependence.

Most of the theoretical and experimental  measurements were focused on the region of $-t \le 1 \, {\rm GeV}^2$ where the exclusive channel  constitutes the  dominant  part of the cross section \cite{Breitweg:1997rg,Chekanov:2002xi,Aktas:2003zi,Alexa:2013xxa}. The selection of the heavy meson production allows to test process of nucleon scattering off a small quark-antiquark dipole. For  large $-t$, a different process becomes dominant - elastic scattering of  a small color dipole off a quark or a  gluon. These processes are identified by the presence of a rapidity gap between   heavy meson and the system produced in the fragmentation of parton knocked out of the target. This  process  in the limit of high energy can be described  in terms of the Pomeron exchange.  The perturbative Pomeron can be obtained as a solution to the BFKL equation in the non-forward case \cite{Balitsky:1978ic,Kuraev:1977fs,Lipatov:1985uk}. In the following we shall refer to this perturbative BFKL Pomeron simply as a Pomeron.  The dissociated target  system typically has mass much greater than the proton mass.  An advantage of this class of processes is that the Pomeron ladder is `squeezed' in this case on both ends. In addition squeezing on the $J/\psi$ end leads to the suppression of the multi-Pomeron exchanges which may fill the rapidity gap.   By that we mean that two large comparable scales are present at both ends of the Pomeron, thus largely suppressing the diffusion of the transverse momenta within the Pomeron into the infrared regime. 
This is the best kinematics to study energy dependence of the vacuum exchange amplitude without having to separate  the effects originating from two sets of large logarithms, $\ln 1/x$ and $\ln(Q^2)$. In fact  in this case the rapidity gap dependence of the cross section is directly converted into intercept of the Pomeron exchange at a given $t$. 
Roughly speaking the dependence on the rapidity gap of the cross section should scale as $2(\alpha_{\pom}(t)-1)$ which in our case is about 0.4-0.5 for the BFKL Pomeron. 

The  process of  the diffractive production of heavy vector mesons at large values of  $|t|$ was measured at H1 \cite{Aktas:2003zi} and ZEUS  \cite{Chekanov:2002rm} experiments at HERA.  The theoretical description of this process was first discussed in \cite{Abramowicz:1995hb, Forshaw:1995ax} and detailed studies using the 
 exchange of the BFKL Pomeron have been performed in series of works, see for example  
\cite{Ginzburg:1996vq,Forshaw:2001pf,Enberg:2002zy,Enberg:2003jw,Poludniowski:2003yk}.
These calculations were  applied to these  data  \cite{Chekanov:2002rm,Aktas:2003zi}
and shown that they can successfully describe the experimental data. More recently, the formalism with the BFKL Pomeron exchange was utilized to evaluate the vector meson diffraction in DIS and related to the contribution to the  $J/\psi$ hadroproduction due to the Pomeron loops \cite{Kotko:2019kma}.

  One limitation of the experimental study at HERA was the fact that the  detectors had a rather limited acceptance  in rapidity and could not measure directly the dependence on  the rapidity gap in this process. As a result, the determination of the energy dependence of the Pomeron amplitude was sensitive to details of the $t$-dependence of the amplitude. In particular the  analysis of the data performed within DGLAP approximation found $\alpha_{\pom}(t)-1$ close to zero at large $-t$ \cite{Blok:2010ds}.

There are several planned DIS machines which have a potential to explore the diffraction with much higher precision than at HERA, on a variety of targets (protons and nuclei) and possibly at higher center-of-mass energy. The US based Electron Ion Collider machine \cite{Accardi:2012qut, NAP25171, Aschenauer:2017jsk}, planned in Brookhaven National Laboratory, will be  a  high luminosity machine,  with the center of mass energy up to about  $140 \;  {\rm GeV}$. It will be also capable of colliding electrons with a wide range of nuclei, thus offering access to a completely novel kinematic regime in $eA$ scattering. On the higher energy end is the Large Hadron-electron Collider  \cite{Dainton:2006wd, AbelleiraFernandez:2012cc, Klein:2018rhq,Agostini:2020fmq}, a CERN based proposal, with a projected center-of-mass energy up to about $\sqrt{s} = 812 \; {\rm GeV}$ and its future extension, the Future Circular Collider in electron - proton option, with the energy reach potentially up to $3.5 \; {\rm TeV}$ \cite{FCC_CDRv1,FCC_CDRv3}.
The prospects of the inclusive diffraction at the EIC and LHeC and FCC-eh machines were studied recently in \cite{Armesto:2019gxy}.  Pseudodata were simulated as well as extraction of the diffractive parton densities, and the potential for their constraining was evaluated.

In this paper we shall  analyze in detail the prospects of the dissociative diffractive photoproduction of $J/\psi$ at the possible future $ep$ collider EIC (Electron Ion Collider) planned in Brookhaven National Laboratory.
The main goal of this paper is to map out the details of the kinematics of this process at energies relevant to the EIC, and to find out the specific requirements on the acceptances of the detectors, which would allow for the tests of the energy dependence of the Pomeron in a large rapidity gap range for given energy. 
The high integrated luminosity of  EIC, of the order of $10$ fb$^{-1}$, allows for more precise analysis of this process.

The structure of the paper is as follows:  in the next section we recall the kinematics of the process, in particular the photoproduction limit and the expression for the rapidity gap, in Sec. 3 we discuss the cross section. in Sec. 4 we discuss various experimental scenarios and the numerical results and finally in Sec. 5 we state the conclusions.

\section{Kinematics of dissociative diffractive vector meson production}

\begin{figure}[h]
\begin{center}
\includegraphics[width=8cm, height=6cm]{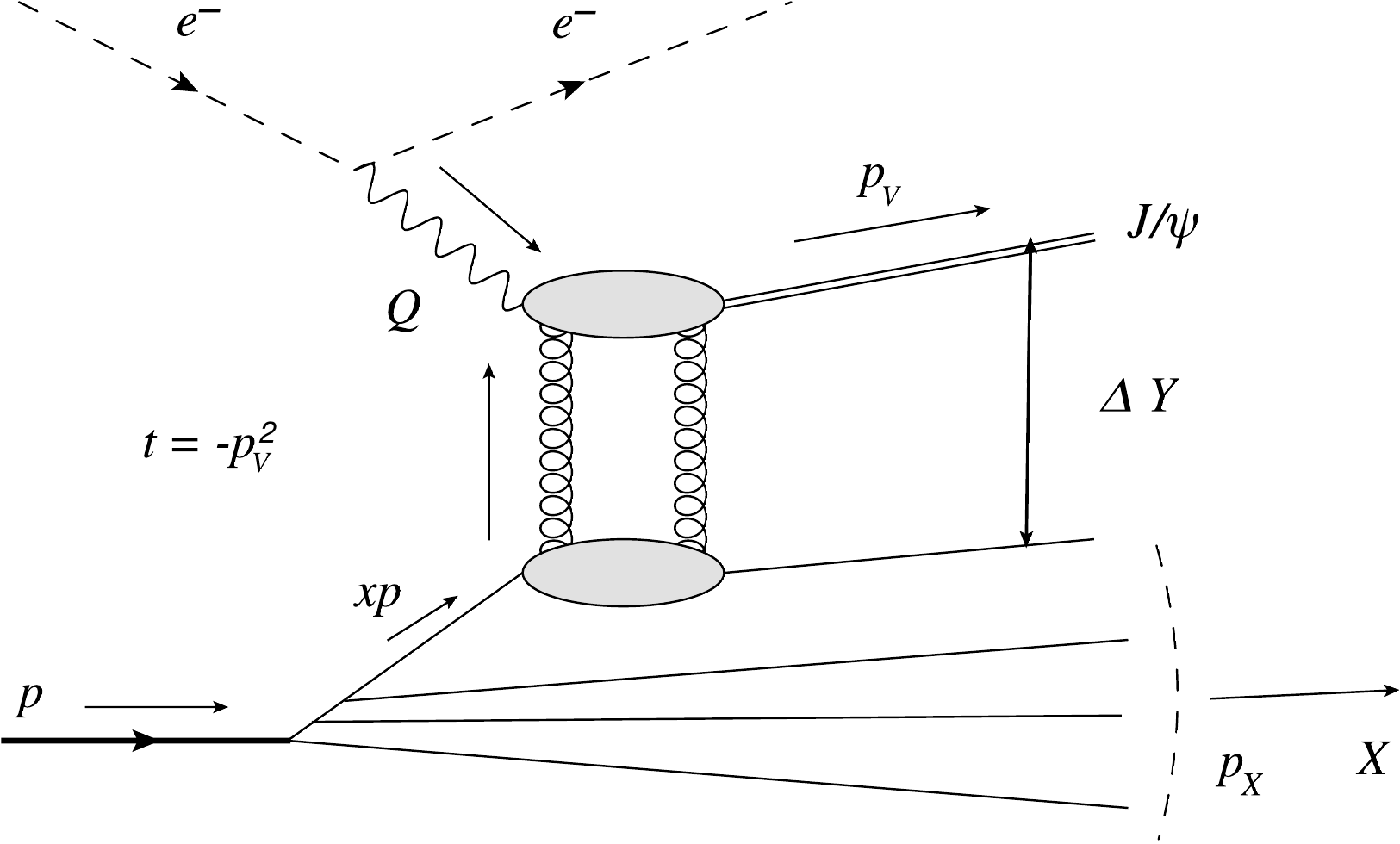}
\end{center}
\caption{Diagram of the amplitude of the process for the diffractive dissociative vector meson (in this case $J/\psi$) production. The exchanged photon carries four-momentum $q$ and incoming proton carries four momentum $p$. The proton then dissociates into the state $X$, which is separated by a rapidity gap $\Delta Y$ from the vector meson. }
\label{fig:diagram}
\end{figure}

The diagram for the amplitude for the process in question is illustrated in Fig.~\ref{fig:diagram}. The electron scatters off the proton, via an exchange of the photon and the Pomeron, producing $J/\psi$. The vector meson is separated from the dissociated proton via rapidity gap $\Delta Y$. In the process studied the proton dissociates into the final  state $X$. In the approach considered in this paper we shall model the $t$-channel exchange via the non-forward BFKL Pomeron \cite{Balitsky:1978ic, Kuraev:1977fs, Lipatov:1985uk}. The Pomeron interacts with the parton from the proton that carries fraction $x$ of the longitudinal momentum of the incoming proton.

 It is crucial to understand the detailed  kinematics of the studied diffractive process $e+p\rightarrow e+J/\psi+X$, where $J/\psi$ and rest of the produced particles - $X$ -  are separated by a rapidity gap of size $\Delta Y$. For this purpose it is important to outline the implications of the photoproduction limit $Q^2\simeq 0$ and recall how the size of the rapidity gap $\Delta Y$ depends on the scattering energy and invariant energies of subsystems occurring in the process, as well as on the momentum transfer.

\subsection{Photoproduction limit}

The photoproduction limit means that the photon four-momentum squared is very small, practically  $Q^2\simeq 0$. We can then write the four-momentum of the $J/\psi$ meson in following way
\begin{equation}
p_{V}=x_{V}p+z_{V}q+{p_T}_{V} \; ,
\end{equation}
where $p$ is the proton four-momentum, $q$ is photon four-momentum and ${p_T}_{V}$ is the transverse component of the $J/\psi$ four-momentum $p_{V}$. We use here collider frame.
The coefficients $x_{V}$ and $z_{V}$ are not independent. The coefficient $x_{V}$ can be derived using the on-shell condition for the four-momentum $p^2_{V}=M^2_{V}$ and is given by
\begin{equation}
x_{V}=\frac{{\bf p}^2_{V}+M^2_{V}}{z_{V}W^2}\;,
\end{equation}
where ${\bf p}_{V}$ (${\bf p}_{V}^2=-{p_T}^2_{V}$) is the transverse momentum (two-momentum) of the $J/\psi$ meson, $M_{V}$ is its mass and $W$ is the energy of the photon-proton collision.
The energy $W$ can be written in the following way:
\begin{equation}
W^2=\left(p+q\right)^2=ys-Q^2+m_p^2\simeq ys\; ,
\end{equation}
where $y$ is the inelasticity and $q$ is the four-momentum of the photon.

Practically the whole photon momentum is transferred to the  $J/\psi$-meson, $1-z_{V}\ll 1$ and ${\bf p}^2_{V}$, $M^2_{V}\ll W^2$. We also have $x_{V}\ll 1$ thus the $p$ component of $p_{V}$ can be neglected. As a result the transverse momentum flowing in the $t$-channel of the process is ${\bf p}_{V}$ and $t=-{\bf p}^2_{V}$, see Fig.~\ref{fig:diagram}. Given the approximation, the four-momentum $p_X=p-{p_T}_{V}$ and the particle in the system $X$ with the smallest rapidity  interacting with the photon via an exchange of the BFKL ladder has the final momentum $p_j=xp-{p_T}_{V}$ (where $x$ is the  proton's longitudinal momentum fraction carried by this  particle). This variable   can be used to calculate the size of the rapidity gap and its value will determine if the  additional activity accompanying the rapidity gap  be observable  in the forward part of the main detector.
\subsection{The definition of the  rapidity gap}

The size of the rapidity gap $\Delta Y$ is an important variable essential for the comparison of  the data to a model containing the BFKL dynamics. In the approach adopted in this paper,  the absence of activity in the rapidity gap region is generated by an exchange of the non-forward BFKL Pomeron between the photon-$J/\psi$ vertex and the proton. Hence, this process can be viewed as a sensitive probe of the BFKL dynamics.

To simplify the discussion let us consider the limit $-t / (xW^2) \ll 1 $.
Expressions in the case when  $-t / (xW^2) $ are  comparable to unity are more complicated and anyway this limit does not correspond to our other approximations.
Let us work in the c.m. frame of initial $\gamma$ and parton of the nucleon 
 to  which the two gluon  ladder is attached. In this frame photon and parton four
 momenta are $\left({\tilde p}_V,{\tilde p}_V\right)$, $\left({\tilde p}_V,-{\tilde p}_V\right)$. Accordingly $xW^2 = 4{\tilde p}_V^2$.
 
The four momenta of $J/\psi$ and the recoil jet   are 
\begin{equation} 
\left(\sqrt{{\tilde p}_V^{\prime 2} + {p_T}_{V}^2 +M_{V}^2}, {\tilde p}_V^{\prime}, {\bf p_T}_{V}\right)\quad \text{and}
\quad\left(\sqrt{{\tilde p}_V^{\prime 2} + {p_T}_{V}^2 }, -{\tilde p}_V^{\prime}, -{\bf p_T}_{V}\right) \;.
\end{equation}

 In the discussed limit  we can 
use ${\tilde p}_V^{\prime}={\tilde p}_V$. Similarly we can  use approximation $t=-{p_T}_{V}^2$.
We can write rapidity for jet and for $J/\psi $  as

\begin{equation}
 y_j=  \ln \left({E_j+{\tilde p}_V\over \sqrt{-t}}\right),
\end{equation}
and
\begin{equation}
 y_{V} = \ln \left(\sqrt{-t +M_V^2 }\over E_V+{\tilde p}_V\right).
\end{equation}
Hence the rapidity interval can be written as
\begin{equation}\label{eq:deltaY}
\Delta Y= y_j - y_{V}= \ln {xW^2 \over \sqrt{ -t (-t +M_{V}^2})} \; ,
\end{equation}
where we have used approximation that $-t/(xW^2) \ll 1 $.
The expression in this form  was previously used in \cite{Frankfurt:2008er}.
 It differs slightly from the one commonly used in the literature in which  the denominator  is equal to $-t +M_{V}^2$.

\section{The partonic cross section}

\begin{figure}[h]
\hspace{-0.5cm}
\centering
\begin{subfigure}{6.5cm}
\includegraphics[width=7cm, height=4.7cm]{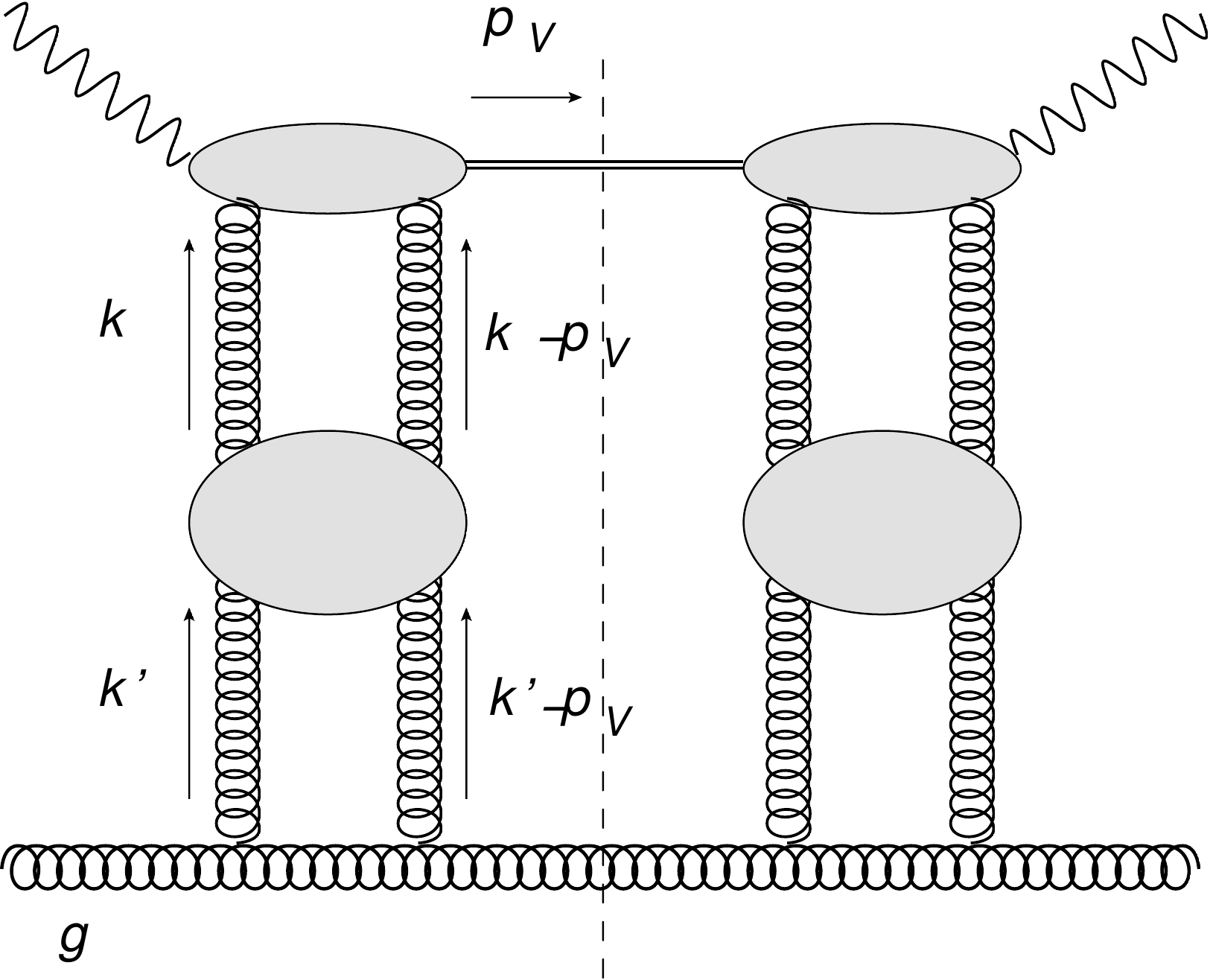}
\caption{Diagram for the diffractive scattering off the gluon from the target.}
\end{subfigure}
\hspace{1.0cm}
\centering
\begin{subfigure}{6.5cm}
\includegraphics[width=7cm, height=4.7cm]{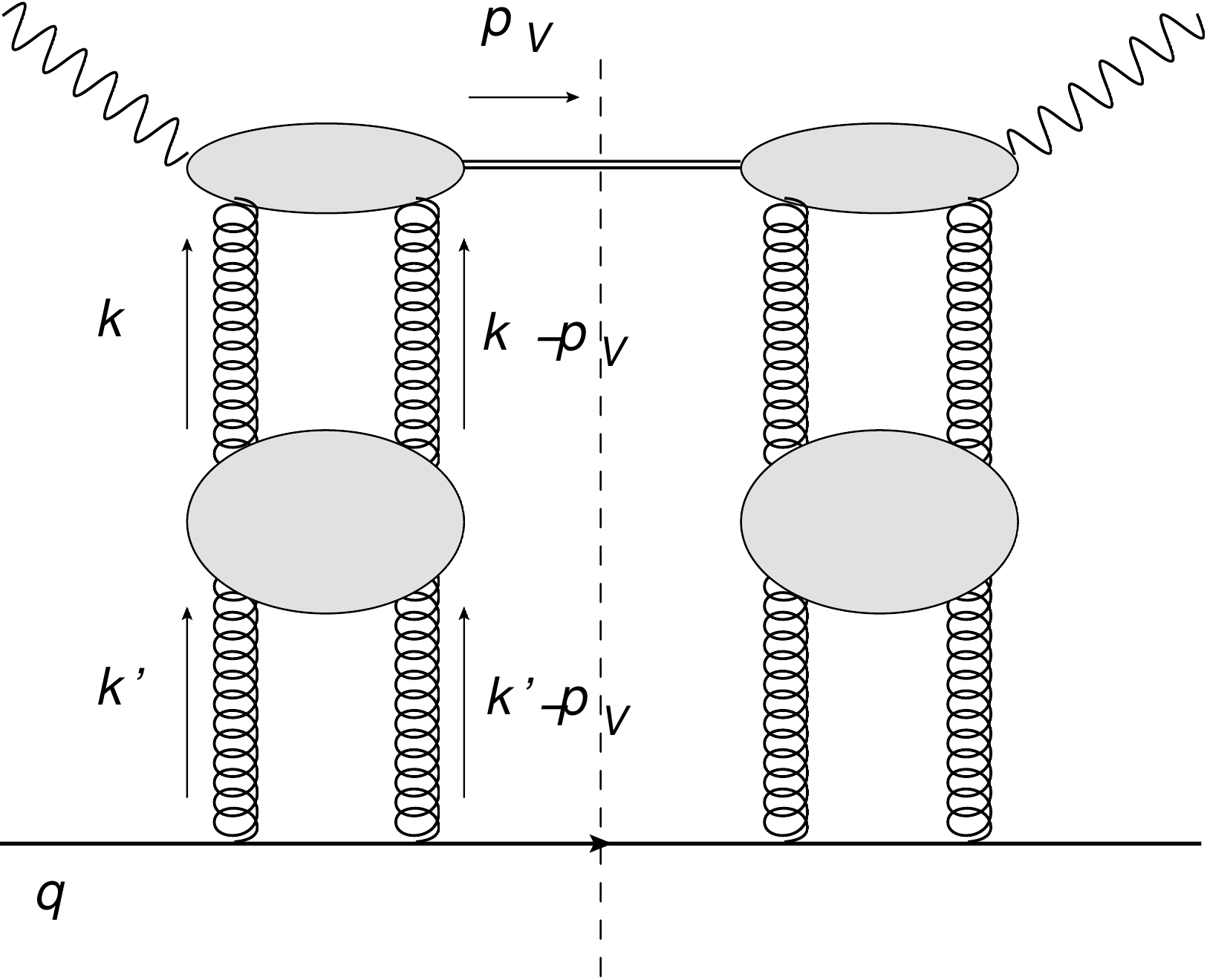}
\centering\caption{Diagram for the diffractive scattering off the quark from the target.}
\end{subfigure}
\centering\caption{Diagrams for the partonic sub-process of the  diffractive $J/\psi$ production in scattering of the photon off the parton. Vertical dashed line indicates the diffractive cut, the upper blob denotes the photon-meson impact factor and the lower blob indicates the gluon Green's function of the BFKL hard Pomeron.}
\label{fig:pomeronex}
\end{figure}

The cross section of the diffractive process $e+p\rightarrow e+J/\psi+{\rm  \;gap}+X$ in the high energy limit can be seen as composed of three main objects:
\begin{itemize}
\item the partonic cross section ${\hat\sigma}_{\gamma^{\ast}i}$ - scattering of virtual photon $\gamma^{\ast}$ and the parton of species $i$. In our approach ${\hat\sigma}_{\gamma^{\ast}i}$ is calculated in the BFKL framework as a convolution of the impact factor of the photon-gluon into meson transition, the non-forward  BFKL ladder - corresponding to Pomeron exchange and the parton impact factor.

\item the photon flux -  describing the coupling of the $\gamma^{\ast}p$ process to the electron.

\item the collinear parton density functions - PDFs - containing the non-perturbative information about the proton structure. PDFs are convoluted with ${\hat \sigma}_{\gamma^{\ast}i}$ to obtain ${\sigma}_{\gamma^{\ast}p}$.
\end{itemize}

\subsection{The $\gamma^{\ast}p$ cross section}

The formalism that we summarize below has been developed and used to compare with the experimental data in series of works  \cite{Ginzburg:1996vq,Motyka:2001zh,Enberg:2002zy,Enberg:2003jw,Forshaw:2001pf}.

In this work we shall use the results and the notation as well as  conventions from \cite{Kotko:2019kma}.
The $\gamma^{\ast}$-proton scattering cross section, in the limit of high energy, can be written in terms of of $\gamma^{\ast}$-parton cross section convoluted  with the corresponding PDFs and  summed over all relevant parton species:
\begin{equation}
d\sigma=\sum\limits_{i=q,{\bar q},g}\int dx\;f_i\left(x,\mu\right) \, d{\hat\sigma}_{\gamma^{\ast}i}\left({ \hat{s}},t\right) \; ,
\end{equation}
where $\hat{s}=xW^2$ is the photon-parton invariant mass squared.
The diffractive $\gamma^{\ast}p$ cross section is given, in the high energy limit,  by the convolution of the photon-gluon to meson impact factor, the  BFKL Pomeron and the parton impact factor.

The evolved parton impact factor is defined as
\begin{equation}
\Phi_q\left(\Delta Y,{\bf k},{\bf p}\right)=\int d^3{\bf k}^\prime\,\Phi_{q,0}\left({\bf k}^\prime,{\bf p}\right)\,{\mathcal G}_{\Delta Y}\left({\bf k},{\bf k}^\prime,{\bf p}\right),
\end{equation}
where ${\mathcal G}_{\Delta Y}$ is the non-forward gluon Green's function and it is the solution of the non-forward BFKL equation with Dirac $\delta$-function $\delta\left({\bf k}-{\bf k}^\prime\right)$ as the initial condition. The function $\Phi_q$ is then a solution of non-forward BFKL equation with initial condition $\Phi_{q,0}=\alpha_s$, the leading order quark impact factor.

The latter one is taken in the leading order approximation.
The diffractive gluon impact factor $\Phi_g^{ab}\left(\Delta Y,{\bf k},{\bf p}\right)$ is given by
\begin{equation}
\Phi_g^{ab}\left(\Delta Y,{\bf k},{\bf p}\right)=\Phi_q\left(\Delta Y,{\bf k},{\bf p}\right)\frac{N_c}{N_c^2-1}\delta^{ab}\; .
\end{equation}

The quark impact factor differs from the gluon impact factor just by the color factor:
\begin{equation}
\Phi_q^{ab}\left(\Delta Y,{\bf k},{\bf p}\right)=\Phi_q\left(\Delta Y,{\bf k},{\bf p}\right)\frac{\delta^{ab}}{N_c}\;,
\end{equation}

The differential photon-parton elastic cross section  can be written in the following way:
\begin{equation}\label{eq:dsgpAmp}
d{\hat\sigma}_{\gamma^{\ast}i}\left({ \hat{s}},t\right) =\frac{C_{\gamma i}}{16\pi { \hat{s}}^2}\left|{\mathcal A}_i\left({ \hat{s}},t\right)\right|^2\frac{d^2{\bf p}}{\pi}
\end{equation}
where the amplitude to produce the vector meson through a single Pomeron exchange ${\mathcal A}$ (Fig.~\ref{fig:pomeronex}) and $C_{\gamma i}$ is a color factor.  The amplitude is dominated by the its imaginary part. The real part enters the calculation at higher orders of the logarithmic expansion. One can calculate the imaginary part of ${\mathcal A}$ in the following way 
\begin{equation}
\text{Im}{\mathcal A}\left({ \hat{s}},t\right)={ \hat{s}}\int\frac{d^2{\bf k}}{2\pi}\frac{\Phi_{V}\left({\bf k},{\bf p}\right)\Phi_q\left(\Delta Y,{\bf k},{\bf p}\right)}{\left({{\bf k}}^2+s_0\right)\left[\left({\bf p}-{\bf k}\right)^2+s_0\right]} \; ,
\end{equation}
where $\Phi_{V}$ and $\Phi_q$ are the impact factors for the
vector meson and for the quark, respectively and $s_0$ infrared cut-off ($s_0=0.5$\;GeV$^2$, the cut-off is also applied in the BFKL evolution, see \cite{Kotko:2019kma}).

The lowest order photon to vector meson impact factor is taken within the non-relativistic approximation and  reads \cite{Ryskin:1992ui,Ginzburg:1996vq,Bzdak:2007cz}
\begin{equation}
\Phi^{ab}_{\gamma V} \left(\bm{k},\bm{p}\right)  =   \Phi_{V}\left(\bm{k},\bm{p}\right)\, {\delta^{ab} \over N_c}\,,
\label{eq:Phi_gamV}
\end{equation}
where $a$, $b$ are color indices of the exchanged gluons, and the kinematic part of the impact factor reads
\begin{equation}
\Phi_{V}\left(\bm{k},\bm{p}\right) = 
16\pi e e_q \alpha_s M_V g_{V}  \left[\frac{1}{M_V^2+\bm{p}^2}-\frac{1}{M_V^2+(\bm{p}-2\bm{k})^2}\right] \, ,
\label{eq:Phi_V}
\end{equation}
where 
\begin{equation}
g_{V} = \sqrt{\frac{3 M_V\Gamma_{V\rightarrow ll}}{16\pi\alpha^2 _{em} e_q^2}}\; ,
\label{eq:gJpsi}
\end{equation}
with $e_q$ being the charge of the quark in the meson in units of the elementary charge $e$, $M_V$ -- the mass of the vector meson, and $\Gamma_{V\rightarrow ll}$ its leptonic decay width. 
The photon to vector meson impact factor \eqref{eq:Phi_V} is valid for the transverse polarizations of the photon and of the vector meson. For the case of the quasi-real photon there exists a contribution from the amplitude of the transition between transverse photon and longitudinally polarized vector meson, however that amplitude was estimated to be small \cite{Ryskin:1992ui}. It was also  shown \cite{Frankfurt:2008er} that  the large $t$ behavior of the non-spin-flip contribution is different from the spin-flip term with strong sensitivity on the form of the  vector meson-photon coupling. However, the H1 data on $J/\psi$  photoproduction
\cite{Aktas:2003zi} indicate that the spin flip contribution remains a small correction in the
whole studied range of $t$. Hence, given the results of \cite{Ryskin:1992ui} and experimental measurements  \cite{Aktas:2003zi} we will neglect the spin-flip contribution.

\subsection{The QCD coupling and the PDFs}

The QCD coupling $\alpha_S$ was kept fixed inside of the BFKL ladder, and the non-forward BFKL equation has been taken at the LL approximation, similarly to the approach in \cite{Kotko:2019kma}.
The strong coupling inside the BFKL Pomeron has been tuned (in practice reduced) so that the calculation describe the HERA data on diffractive vector meson dissociation at high $t$ and the resulting intercept is reduced so that matches that of the resummed model. Very good description of the experimental data from HERA was obtained \cite{Kotko:2019kma}.  In principle, more refined approach could be utilized with BFKL NLL or resummation. The LL approach is however sufficient for our purposes, where we are more focused on the requirements on the detector and mapping out possible range of kinematics accessible at the EIC. The $\alpha_s$ in the coupling of the PDFs runs with the scale $\mu^2=-t+M_V^2$ and we have used the CT14nlo PDF set~\cite{Dulat:2015mca}, similarly to \cite{Kotko:2019kma}.

\subsection{The photon flux}

The electron-proton collision cross section can be written as a convolution of  the $\sigma_{\gamma^{\ast}p}$~\eqref{eq:dsgpAmp} and the photon flux $f_{\gamma/e}$:
\begin{equation}
\sigma_{ep}=\int\limits_{y_{min}}^{y_{max}}dy\int\limits_{Q_{min}^2}^{Q_{max}^2}dQ^2\;f_{\gamma/e}\left(y,Q^2\right)\sigma_{\gamma^*p}\left(y,Q^2\right) \;,
\label{eq:photon_flux_convolution}
\end{equation}
where the expression for the photon flux is \cite{Budnev:1974de,Frixione:1993yw}
\begin{equation}
f_{\gamma/e}\left(y,Q^2\right)=\frac{\alpha}{2\pi Q^2y}\left[1+(1-y)^2-\frac{2m_e^2y^2}{Q^2}\right]\; .
\end{equation}
 Here, the variable $\alpha$ denotes the fine structure constant and $m_e$ is the electron mass.
The inelasticity variable $y$ can be defined as $y=W^2/s$, with $s$ is the total $ep$ collision energy squared.

Limits for $Q^2$ integration~\cite{Aktas:2003zi} in \ref{eq:photon_flux_convolution} are
\begin{equation}
Q^2_{\rm min}=m_e^2\;\frac{y^2}{1-y};\quad\quad Q^2_{\rm max}=4\,{\rm GeV}^2\;.
\end{equation}

We can integrate $f_{\gamma/e}\left(y,Q^2\right)$ over $Q^2$ because $\sigma_{\gamma^{\ast}p}\left(y,Q^2\right)\approx \sigma_{\gamma^{\ast}p}\left(y\right)$. 

\begin{equation}
\frac{d\sigma\left(ep\rightarrow eJ/\psi X\right)}{dy}= {\tilde\Phi}_{\gamma/e}\left(y\right)\;\sigma_{\gamma^{\ast}p}\left(y\right),
\end{equation}
where
\begin{equation}
{\tilde\Phi}_{\gamma/e}\left(y\right)=\frac{\alpha}{2\pi  y} \left\{\left[1+\left(1-y\right)^2\right] \log
   \left[\frac{Q_{\rm max}^2 (1-y)}{m_{e}^2 y^2}\right]+2\left[
   \frac{m_{e}^2}{Q_{\rm max}^2}y^2-(1-y)\right]\right\}\;.
\end{equation}

\subsection{Pomeron intercept}

The motivation to use BFKL dynamics to model the diffractive photo-production of $J/\psi$ in electron-proton scattering stems from the presence of two comparable scales - the factorization scale (at the lower end of the gluon ladder Fig.~\ref{fig:pomeronex}) and the scale associated with the production of the $J/\psi$-meson $~M_V^2$. In scenarios where the evolution scales are comparable applicability of the DGLAP evolution is limited.

One of the experimental signatures of the BFKL dynamics, if present, is the following asymptotic behavior
\begin{equation}
\sigma_{\gamma^\ast p\rightarrow V+gap+X}\approx\beta\left(Q^2\right)e^{\delta\Delta Y}+...
\end{equation}
where $\beta$ is some function and $\delta$  is related to the Pomeron intercept $2(\alpha_{\pom}(t0)-1)=\delta$.

Using the relations above we can extract the Pomeron $\alpha_{\pom}(t)$ from data of the cross section $\sigma_{\gamma^{\ast}p}$ at fixed $t$ and $x$ using the logarithmic derivative:
\begin{equation}\label{eq:logder}
\alpha_{\pom}(t)=\frac{1}{2}\left(\frac{d\log{\sigma_{\gamma^\ast p\rightarrow V+gap+X}}}{d{\Delta Y}}+2\right)\;.
\end{equation}

The value of  $\alpha_{\pom}(t)$ extracted from data and its dependence on kinematical variables can be used as a discriminant between different models.

\section{Experimental scenarios}

\begin{figure}[h]
\hspace{2cm}
\begin{center}
\begin{subfigure}{6cm}
\includegraphics[width=7cm, height=4cm]{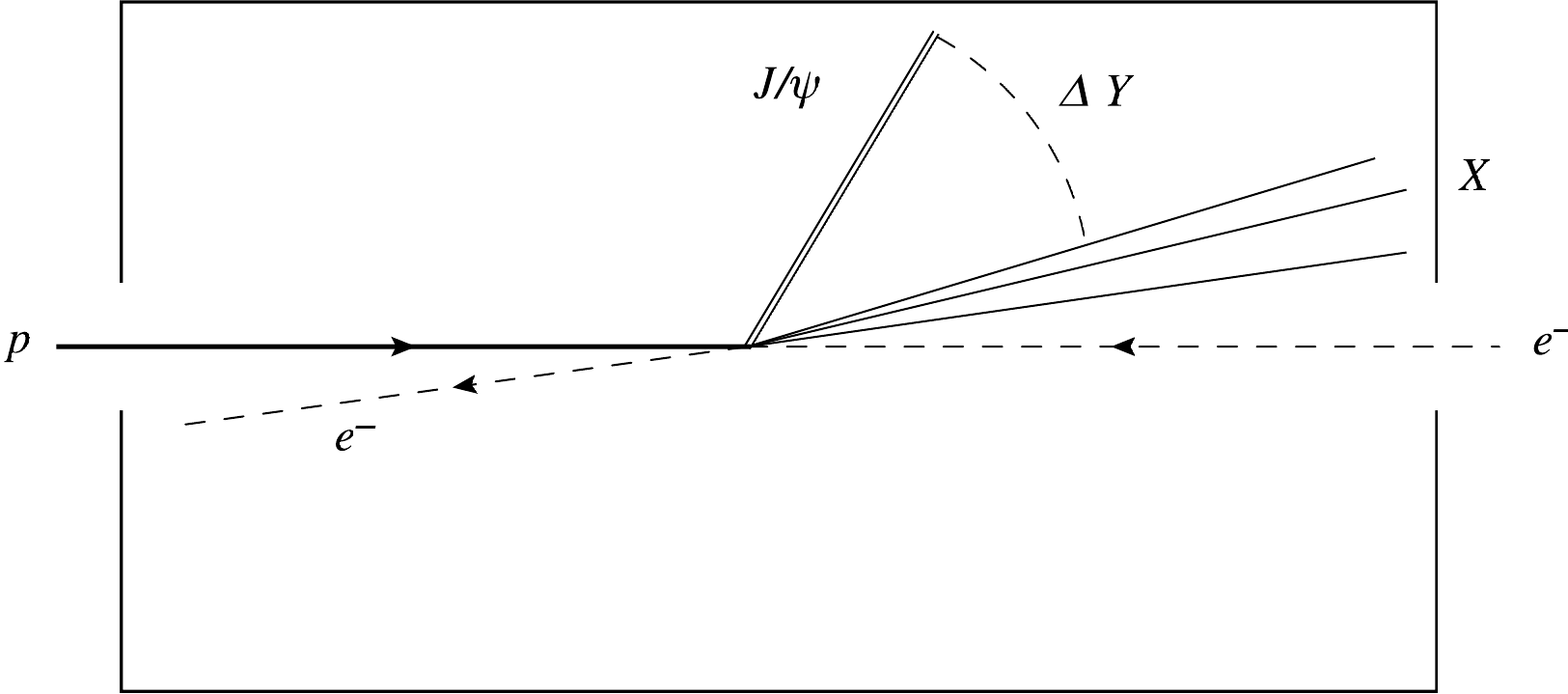}
\caption{}
\end{subfigure}
\hspace*{1.75cm}
\begin{subfigure}{6cm}
\includegraphics[width=7cm, height=4cm]{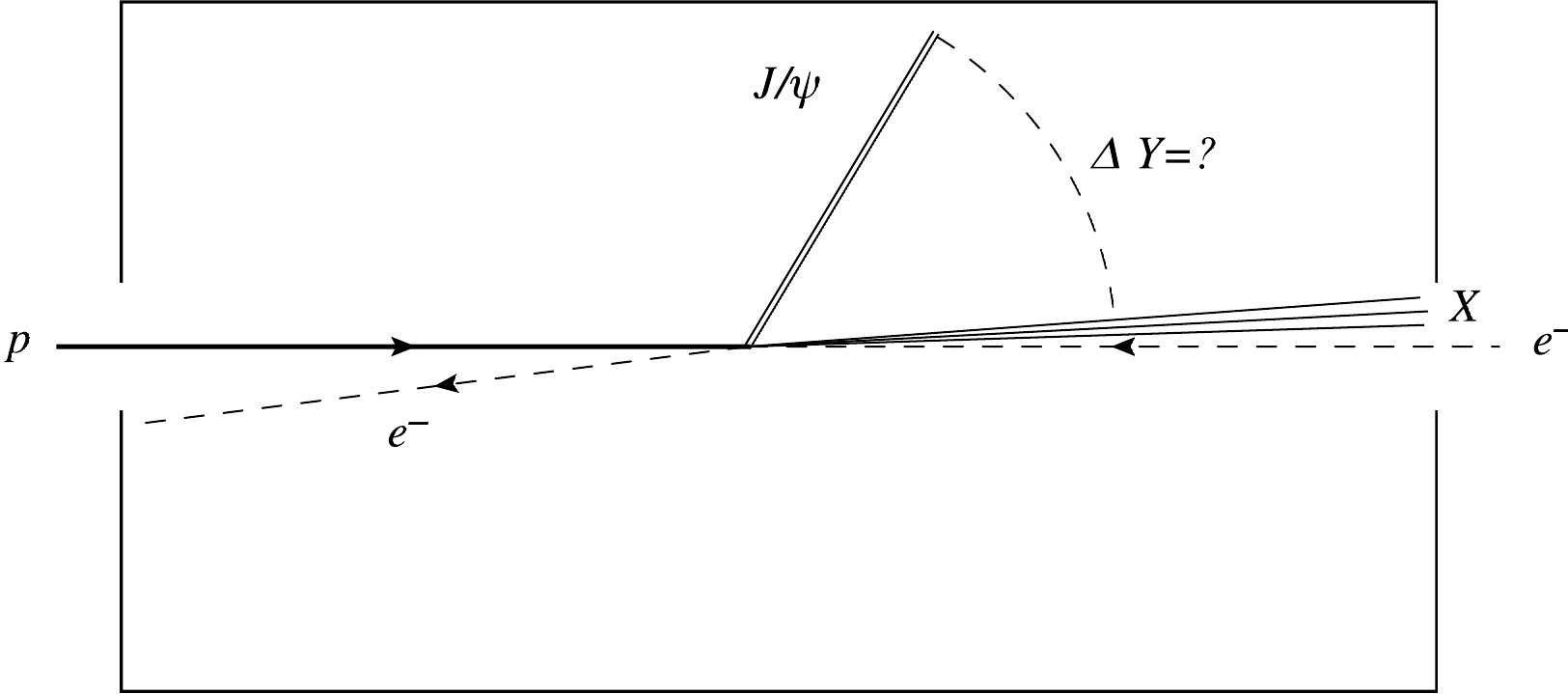}
\caption{}
\end{subfigure}
\end{center}
\centering\caption{Two event selection scenarios. Left (a): Activity in the forward part of the main detector is required, separated from vector meson by the rapidity gap. Right (b): Only rapidity gap is required. No activity in the central detector in the forward region. Exact size of the rapidity gap is unknown. }
\label{fig:detector}
\end{figure}

In the following we shall consider two different experimental  scenarios (sketched in the Fig.~\ref{fig:detector}):
\begin{enumerate}  
\item Request the  detection of the $J/\psi$ meson (its reconstruction via the decay products), rapidity gap - a  region with no activity in the detector and activity in the detector in the direction of the proton beam - cartoon (a) in Fig.~\ref{fig:detector}. The latter is separated from the vector meson by the rapidity gap.
\item Request detection of the $J/\psi$ meson (its reconstruction via decay products), rapidity gap - a \ region with no activity in the detector - cartoon (b) in Fig.~\ref{fig:detector}.
\end{enumerate}

In the second case, given the limitations in the coverage of the central part of the detector, one does not have a knowledge of the exact size of the rapidity gap $\Delta Y$. In that scenario, it is thus necessary to integrate over the longitudinal proton momentum fraction $x$ in the parton density functions in the range of rapidity not accessible by the detector.  Eq.~\eqref{eq:deltaY} relates $\Delta Y$ and $x$ for given $W$ and $t$. The minimum polar angle $\theta_{min}$ covered by the detector is related to the maximum rapidity covered. Given the proton beam energy $E_p$ we can write the polar angle of the last particle at the edge of rapidity gap (the angle between the $z$-axis  and the particle) using following formula:
\begin{equation}
\theta_j=\arctan{\frac{|{\bf p_{j}}|}{xE_p}} \; .
\end{equation}
In the numerical calculation we have applied a cut $\theta_j>4^{\circ}$ directly on the angle $\theta_j$ as given by the formula above.

\subsection{Numerical Results and Discussion}

For the purpose of plotting the phase space range and study the number of events produced in the experiment we define variable $\Delta Y_{\rm min}$ as the minimum size of the rapidity gap - for given $\Delta Y_{\rm min}$: $\Delta Y\geq \Delta Y_{\rm min}$.

\begin{figure}[h]
\hspace*{-2cm}\centering{
\begin{subfigure}{6cm}
\includegraphics[width=8cm, height=5.4cm]{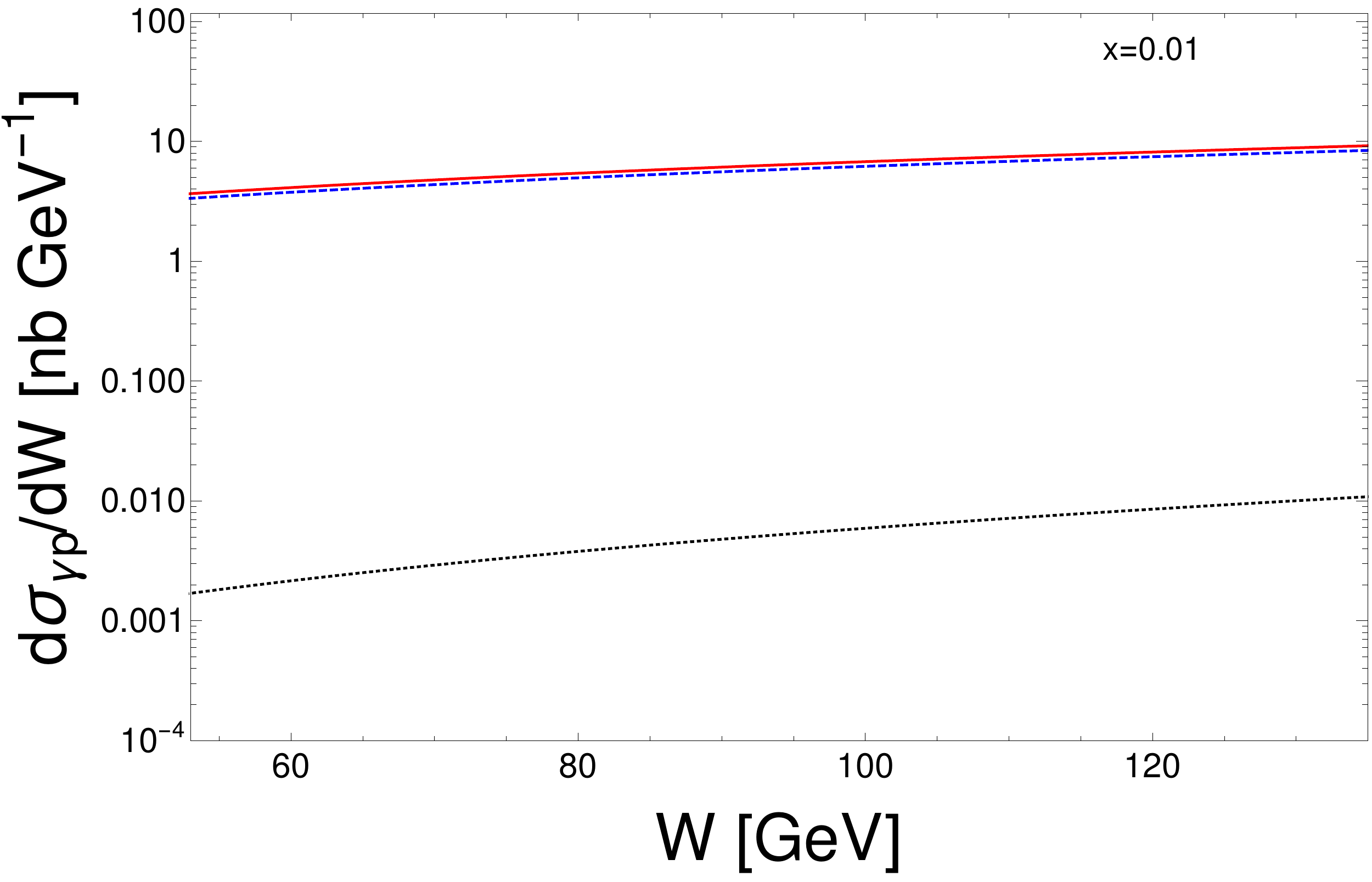}
\end{subfigure}
\hspace{2cm}
\begin{subfigure}{6cm}
\includegraphics[width=8cm, height=5.4cm]{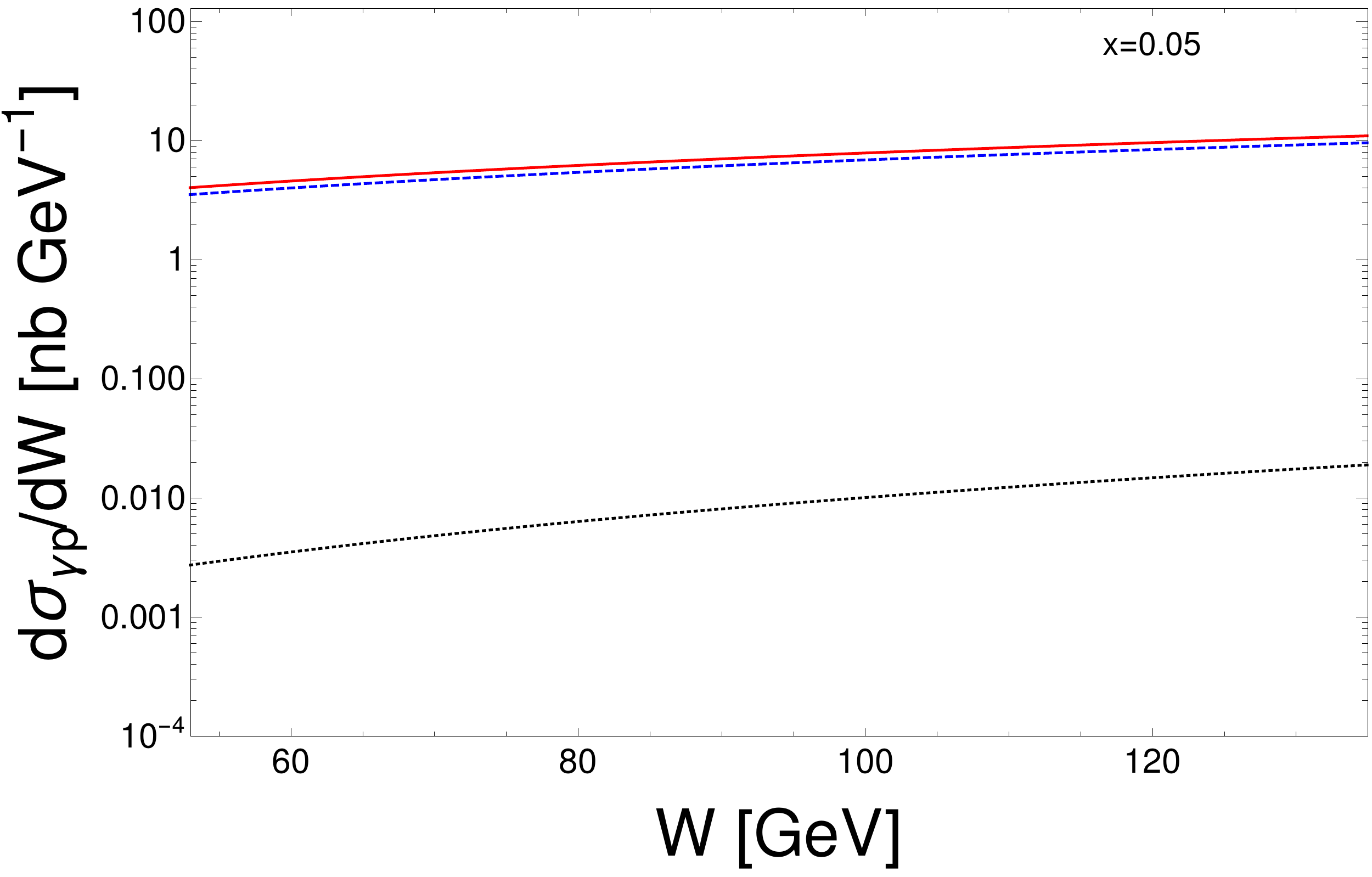}
\end{subfigure}}
\hspace*{-2cm}\centering{
\begin{subfigure}{6cm}
\includegraphics[width=8cm, height=5.4cm]{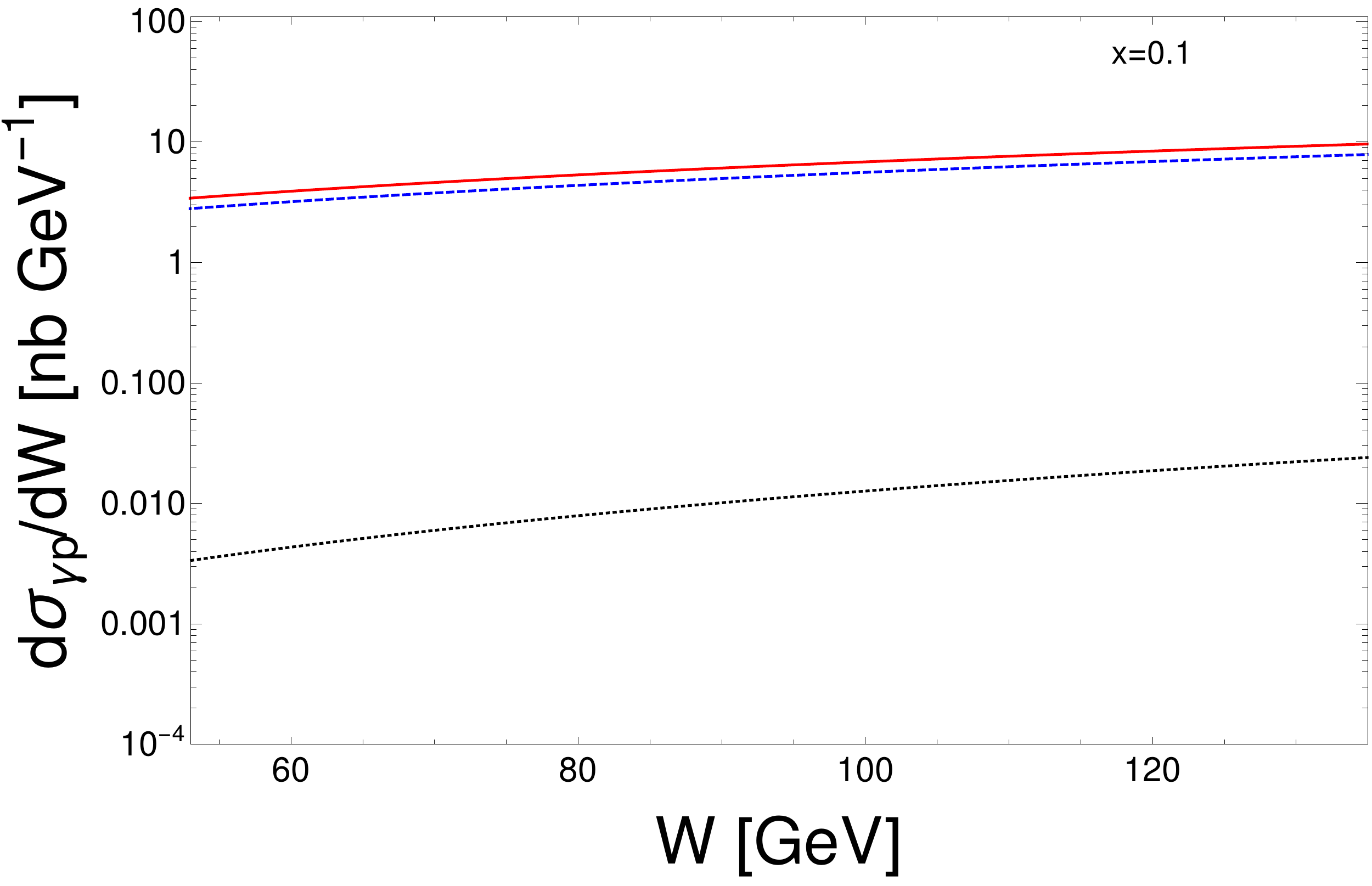}
\end{subfigure}
\hspace{2cm}
\begin{subfigure}{6cm}
\includegraphics[width=8cm, height=5.4cm]{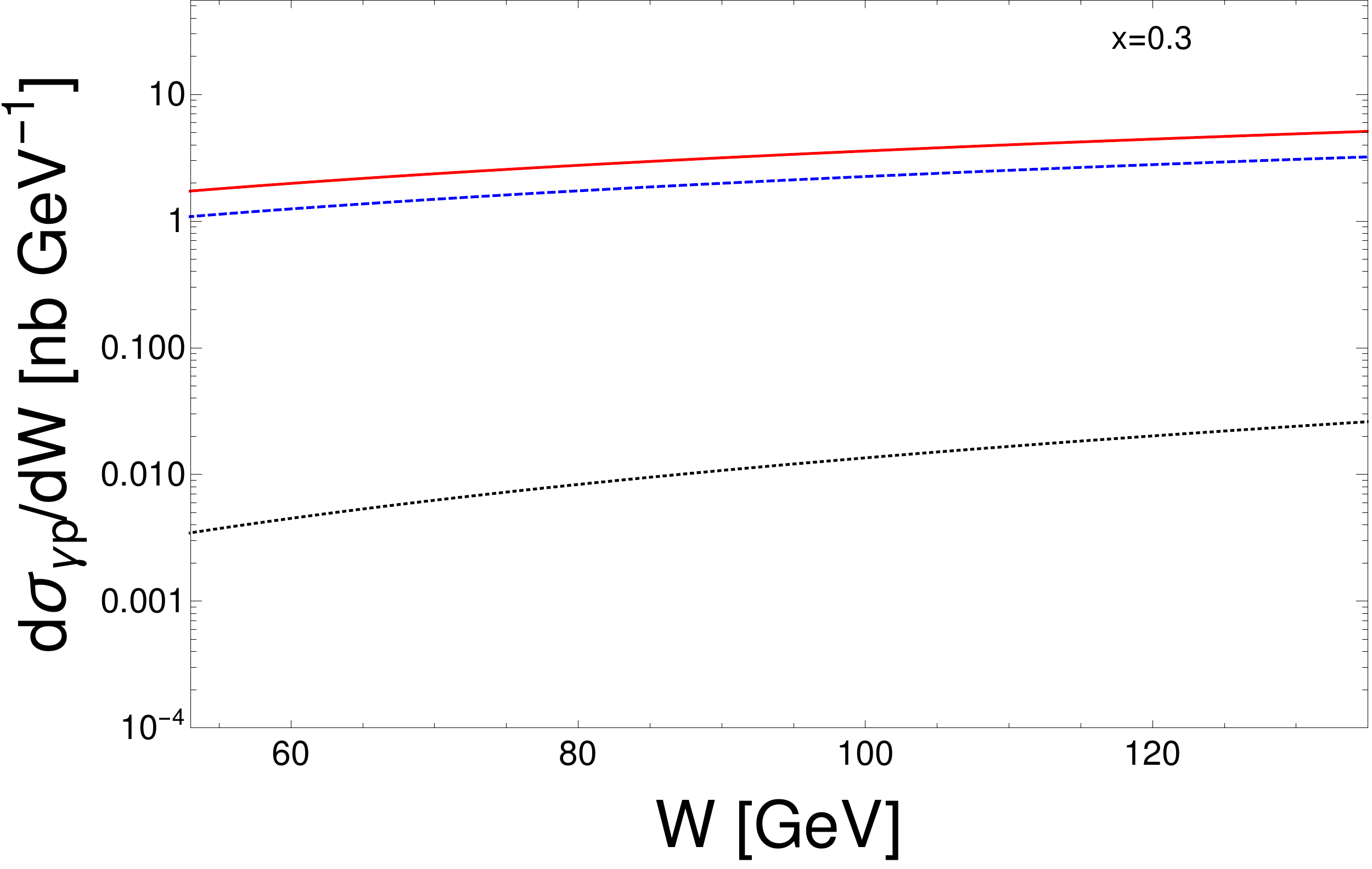}
\end{subfigure}}
\caption{The dependence of the $\gamma p \rightarrow V+gap+X$ cross section  on $W$ with $x$ and $-t=1\;$GeV$^2$ fixed and $\Delta Y_{min}=2$. Black line: quark contribution. Blue line: gluon contribution. Red line: sum of contributions.}
\label{fig:oDpW1}
\end{figure}

We start the analysis from the calculations of the $\gamma p$ cross section and investigating its behavior. Later on, when we discuss the rates, we shall refer to the $ep$ cross section, which is obtained by convolution of the $\gamma p$ cross section with the  photon flux.
Since there are three variables $W,t$ and $x$, we first present the cross section as a function of each variable, keeping  the other two fixed.
First, in Fig.~\ref{fig:oDpW1} the $\gamma p$ cross section dependence on the energy $W$ for fixed $x$ and $t$ with $\Delta Y_{\rm min}=2$ is plotted.  We see the approximately power-like growth  of the cross section with the energy $W$, which is an expected result of the BFKL Pomeron exchange.

\begin{figure}[h]
\hspace*{-2cm}\centering{
\begin{subfigure}{6cm}
\includegraphics[width=8cm, height=5.4cm]{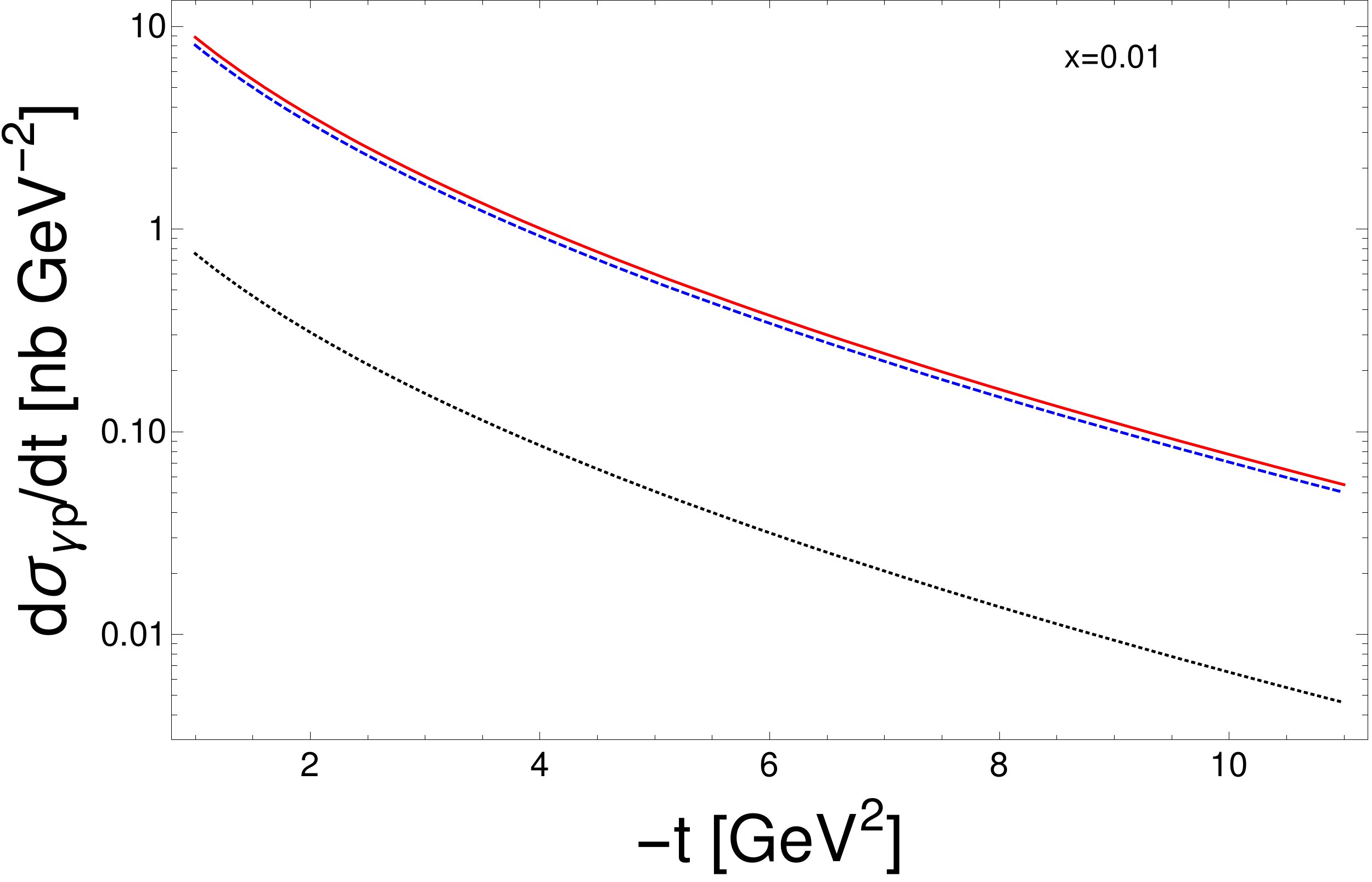}
\end{subfigure}
\hspace{2cm}
\begin{subfigure}{6cm}
\includegraphics[width=8cm, height=5.4cm]{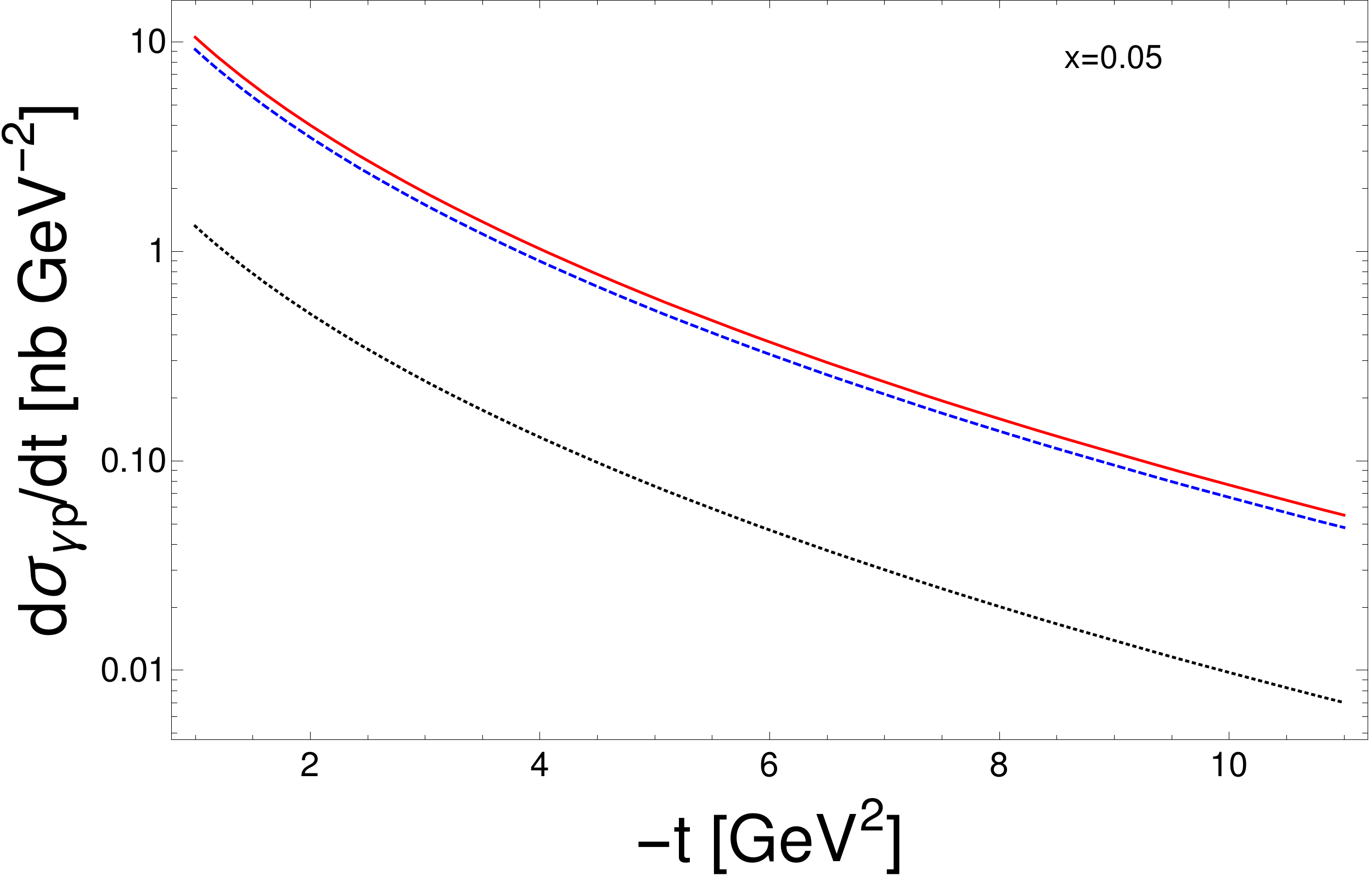}
\end{subfigure}}
\hspace*{-2cm}\centering{
\begin{subfigure}{6cm}
\includegraphics[width=8cm, height=5.4cm]{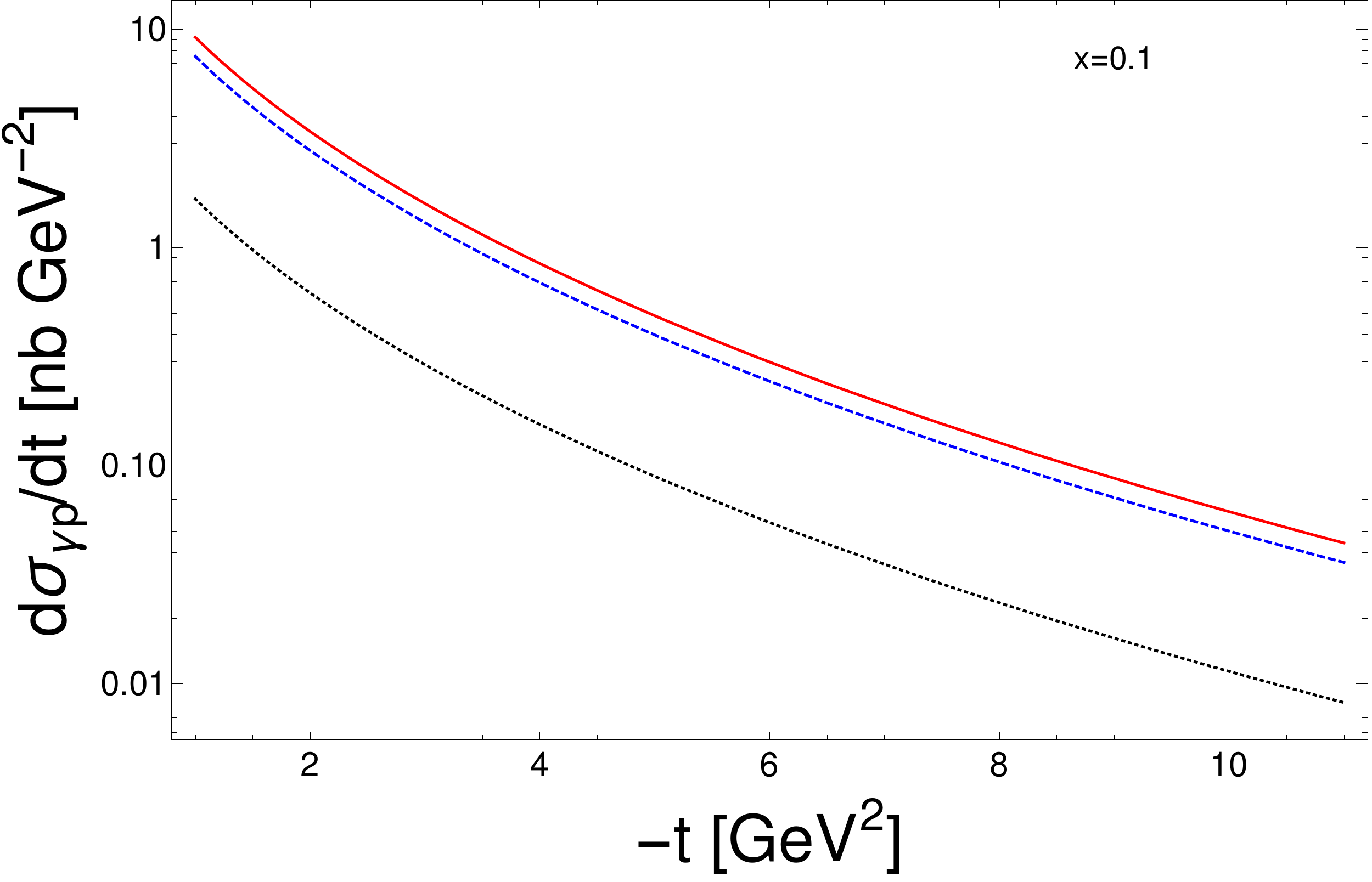}
\end{subfigure}
\hspace{2cm}
\begin{subfigure}{6cm}
\includegraphics[width=8cm, height=5.4cm]{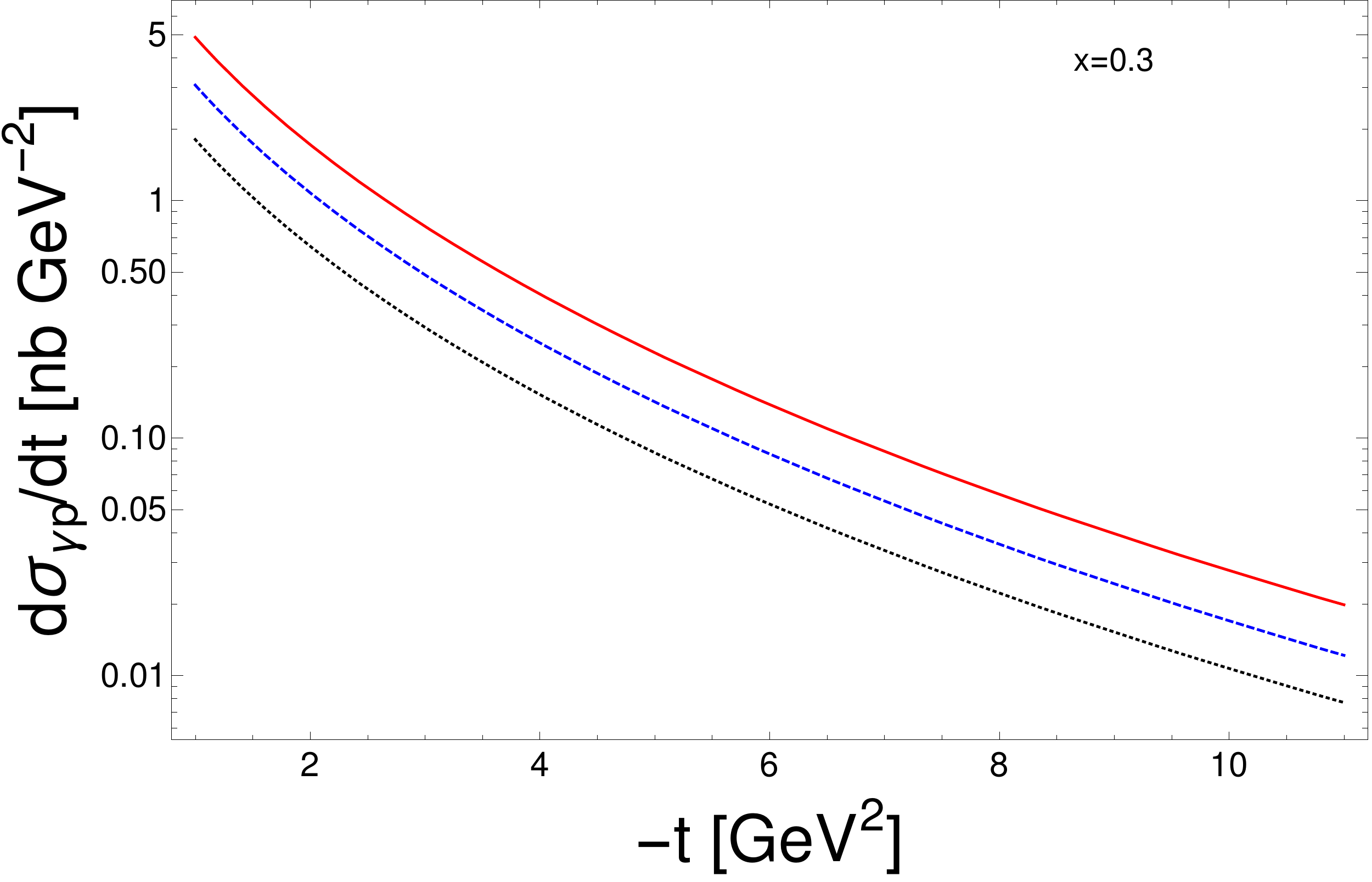}
\end{subfigure}}
\caption{The dependence of the $\gamma p \rightarrow V+gap+X$ cross section  on $t$ with $x$ and $W=130\;$GeV fixed and $\Delta Y_{min}=2$. Black line: quark contribution. Blue line: gluon contribution. Red line: sum of contributions.}
\label{fig:oDpt5}
\end{figure}

\begin{figure}[h]
\hspace*{-2cm}\centering{
\begin{subfigure}{6cm}
\includegraphics[width=8cm, height=5.4cm]{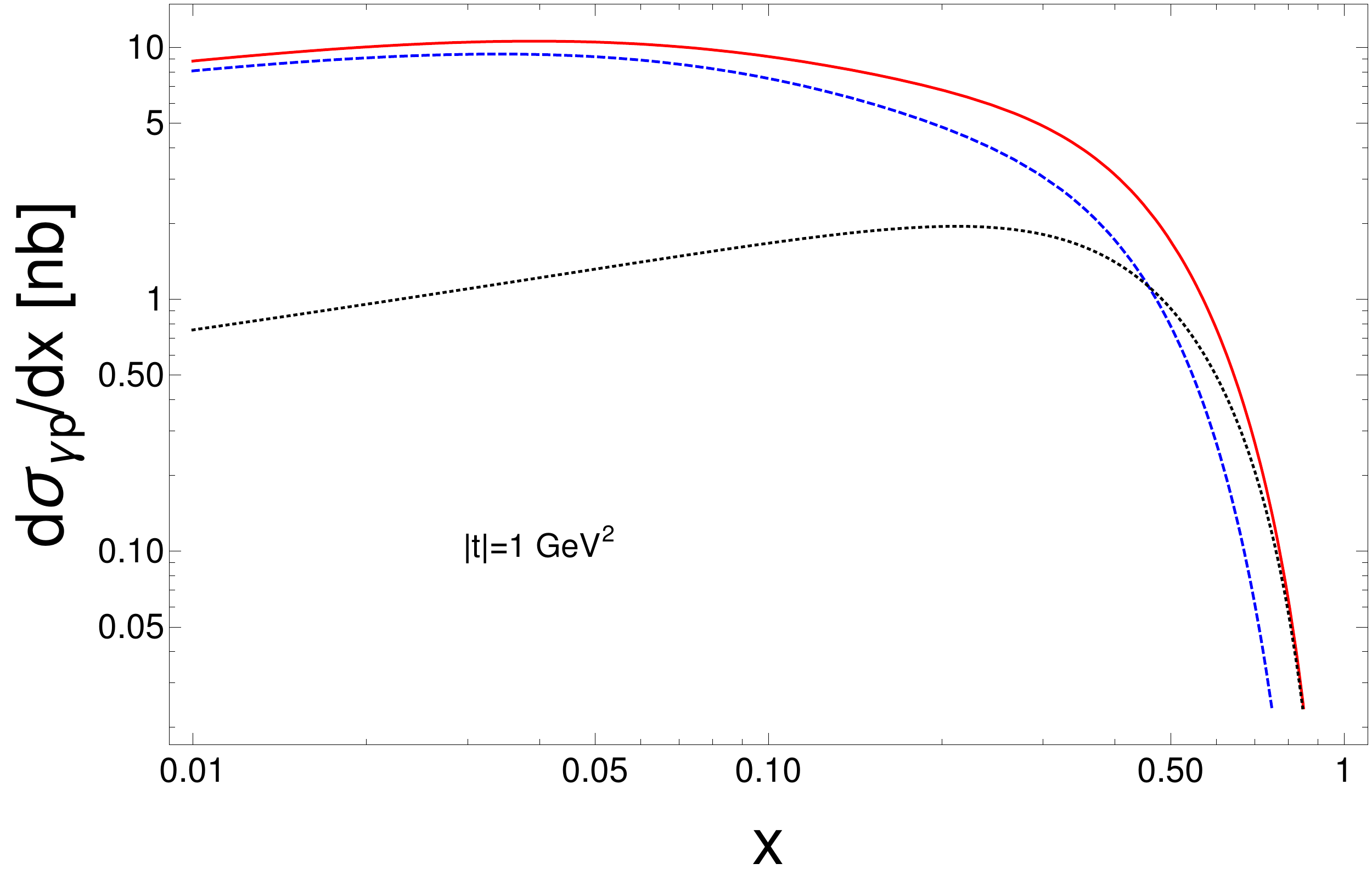}
\end{subfigure}
\hspace{2cm}
\begin{subfigure}{6cm}
\includegraphics[width=8cm, height=5.4cm]{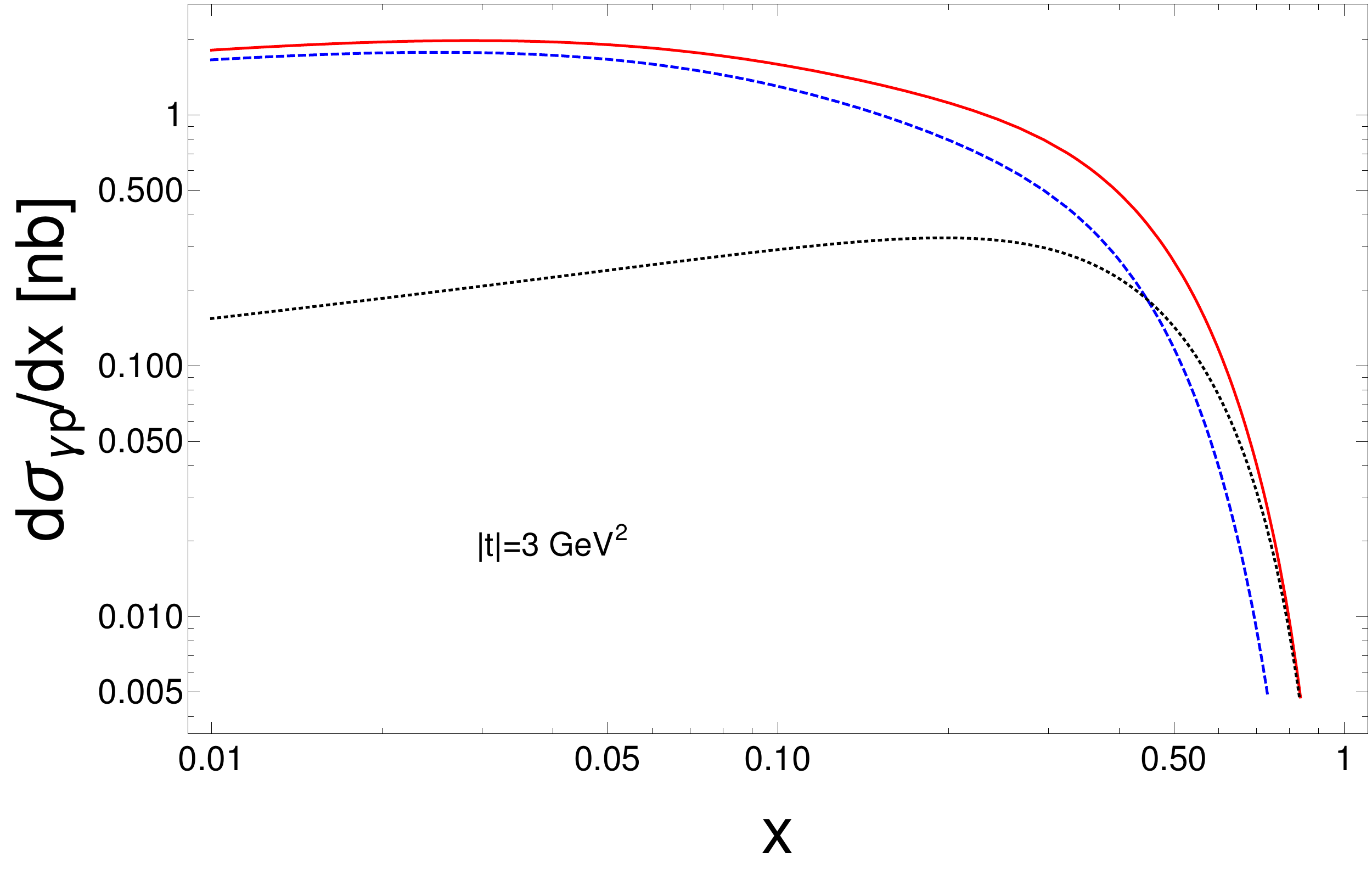}
\end{subfigure}}
\hspace*{-2cm}\centering{
\begin{subfigure}{6cm}
\includegraphics[width=8cm, height=5.4cm]{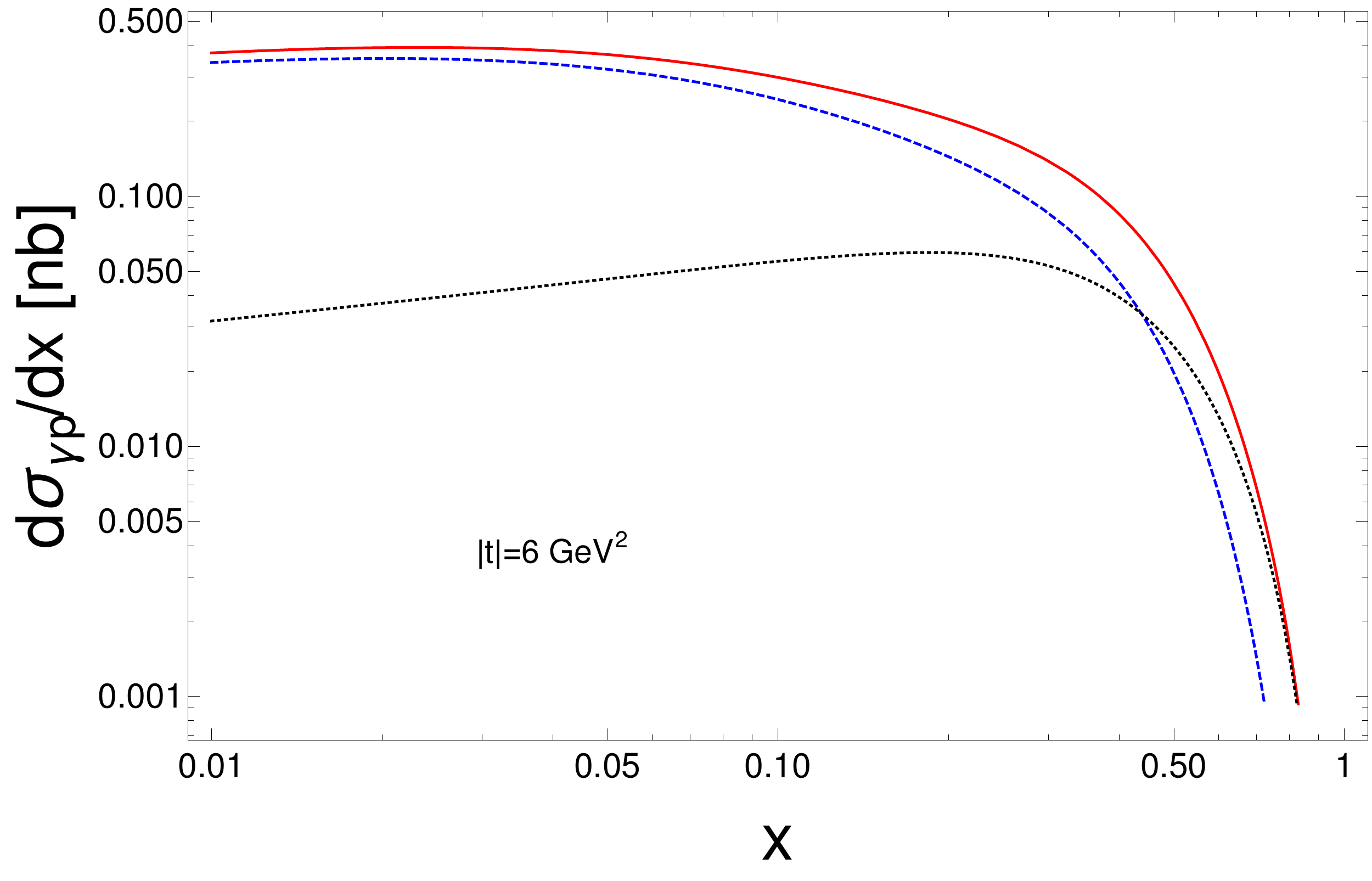}
\end{subfigure}
\hspace{2cm}
\begin{subfigure}{6cm}
\includegraphics[width=8cm, height=5.4cm]{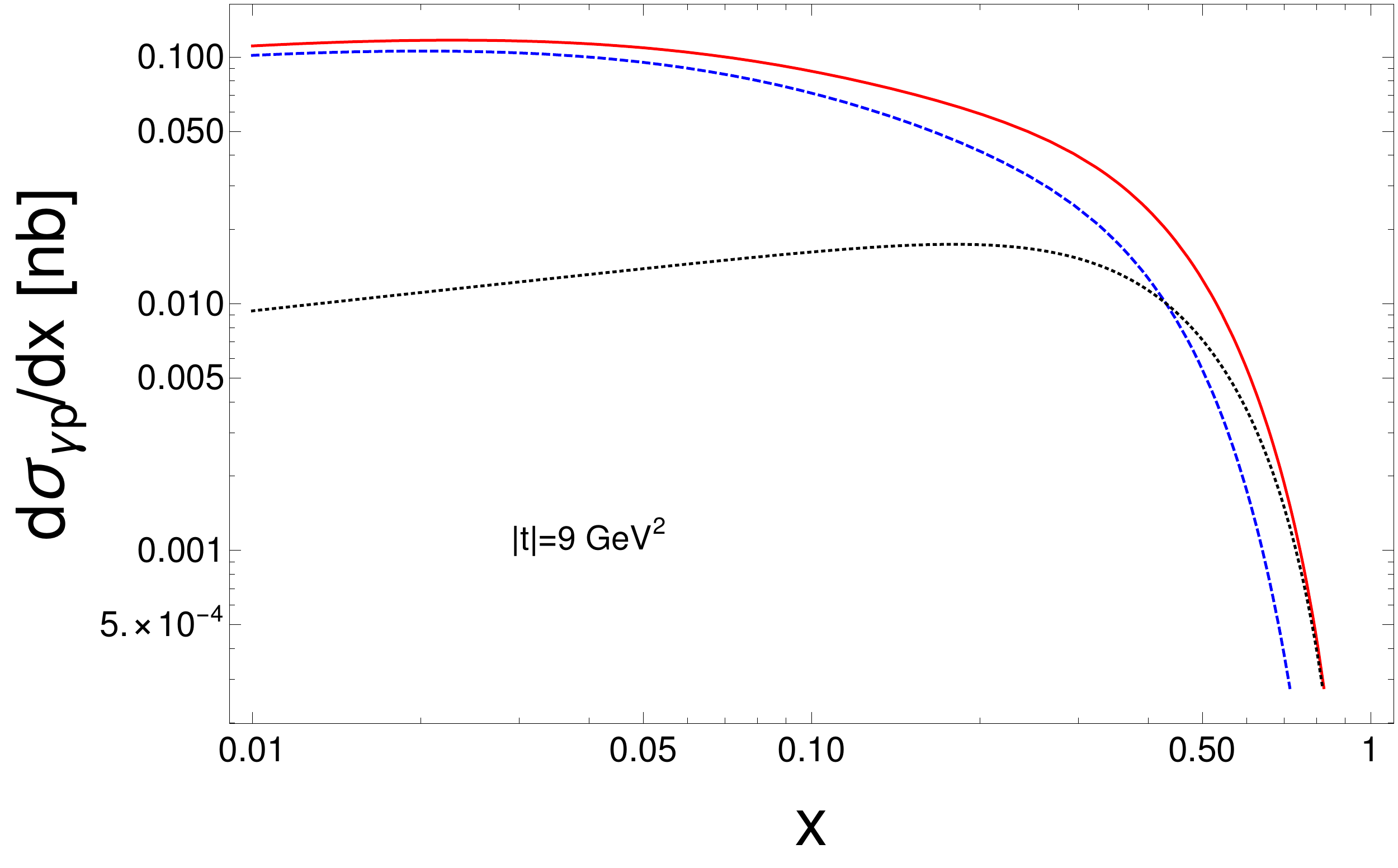}
\end{subfigure}}
\caption{The dependence of the $\gamma p \rightarrow V+gap+X$ cross section  on $x$ with $t$ and $W=130\;$GeV fixed fixed and $\Delta Y_{min}=2$. Black line: quark contribution. Blue line gluon: contribution. Red line sum of contributions.}
\label{fig:oDpx5}
\end{figure}

In Fig.~~\ref{fig:oDpt5} the cross section dependence on $t$ for fixed $x$ and $W$ with $\Delta Y_{\rm min}=2$ is plotted. We observe  drop off of the cross section with the increasing momentum transfer $|t|$.
In the next plots shown in  Fig.~\ref{fig:oDpx5} cross section dependence on $x$ for fixed $t$ and energy $W$  is shown. 
In all the plots in Figs.~\ref{fig:oDpW1}-\ref{fig:oDpx5} we show contributions of the channel where the Pomeron attaches to either the gluon (blue curve) or the  quark (black curve) from the target  as well as their  sum (red curve). We can see in  Figs.~\ref{fig:oDpW1}-\ref{fig:oDpx5}, that the gluon  contribution is the dominant one, but  its relative size to the quark  contribution depends on the value of $x$. Approaching large $x\sim 0.3$ the importance of the quark contribution grows. This behaviour is  expected and depends solely on the relative magnitude  of gluon and quark parton distribution functions. The dependence can be seen directly  in Fig.~\ref{fig:oDpx5},
where it is observed  that the quark contribution only becomes sizeable at $x>0.3$ and dominant at about  $x>0.4$. The different $x$ dependence of the cross section as a function of $x$ is of course the consequence of the different behavior for the quark and gluon distributions.

\begin{figure}[h]
	\hspace*{-2cm}\centering{
		\begin{subfigure}{6cm}
			\includegraphics[width=8cm, height=5.4cm]{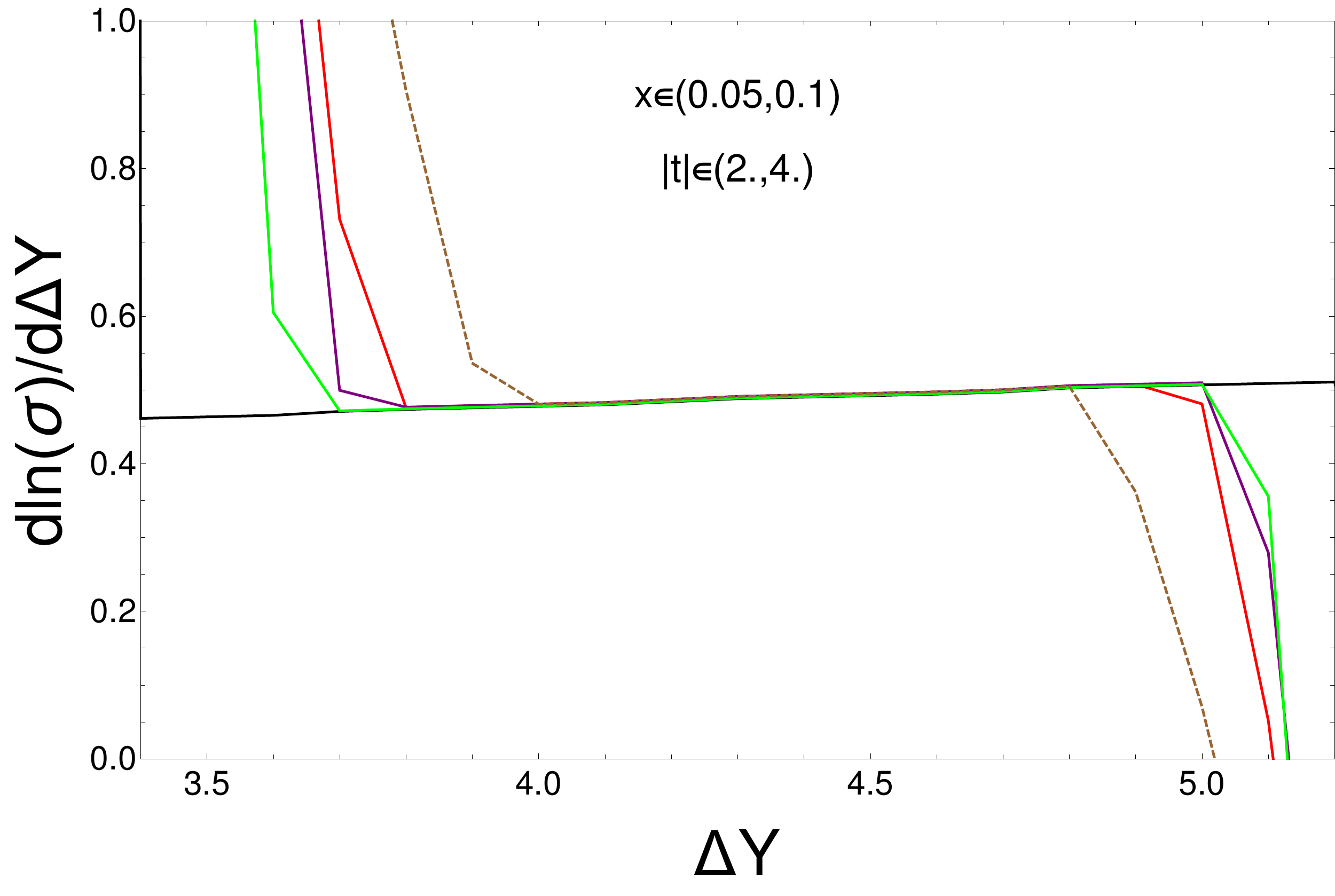}
		\end{subfigure}
		\hspace{2cm}
		\begin{subfigure}{6cm}
			\includegraphics[width=8cm, height=5.4cm]{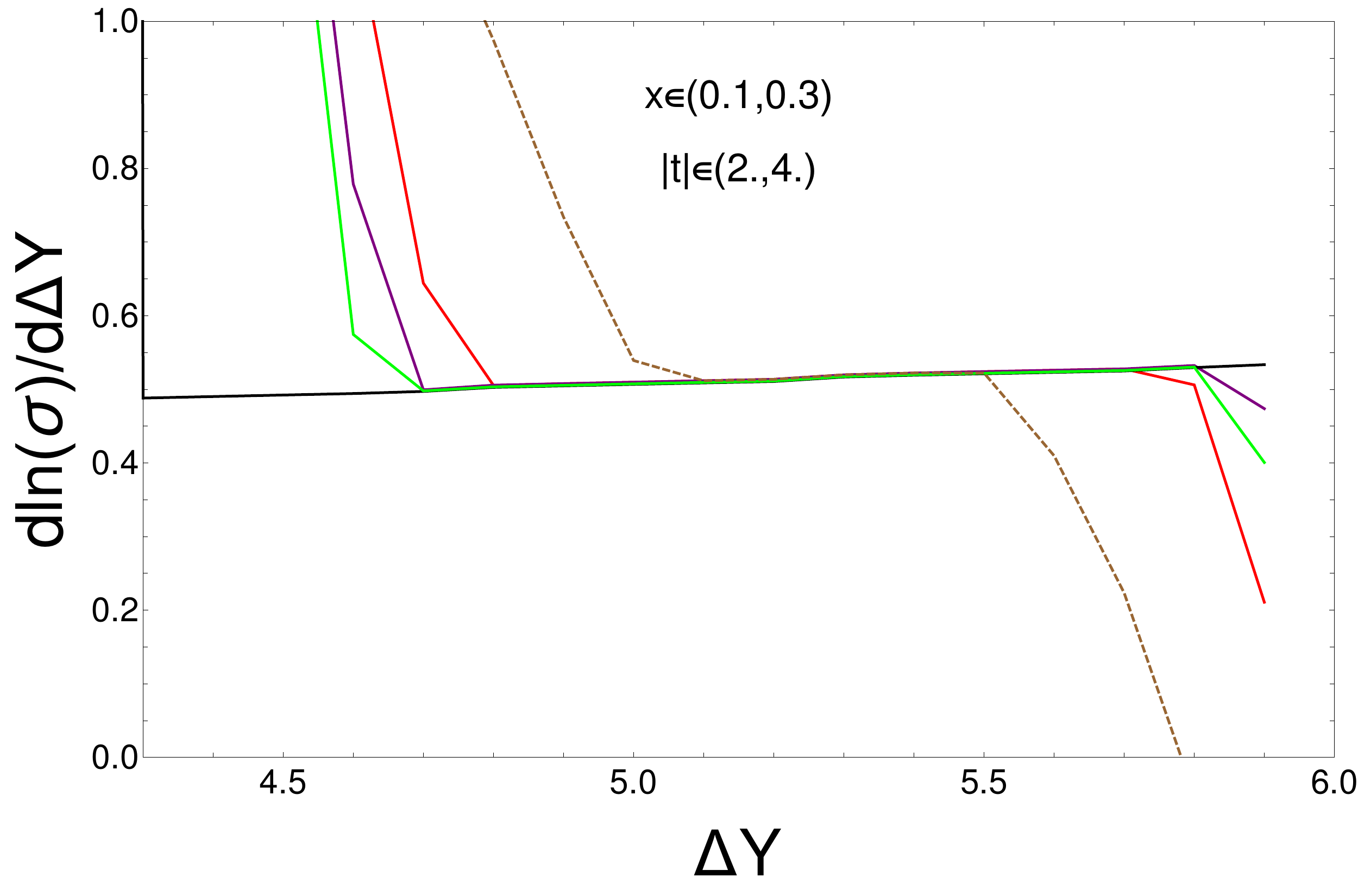}
	\end{subfigure}}
	\hspace*{-2cm}\centering{
		\begin{subfigure}{6cm}
			\includegraphics[width=8cm, height=5.4cm]{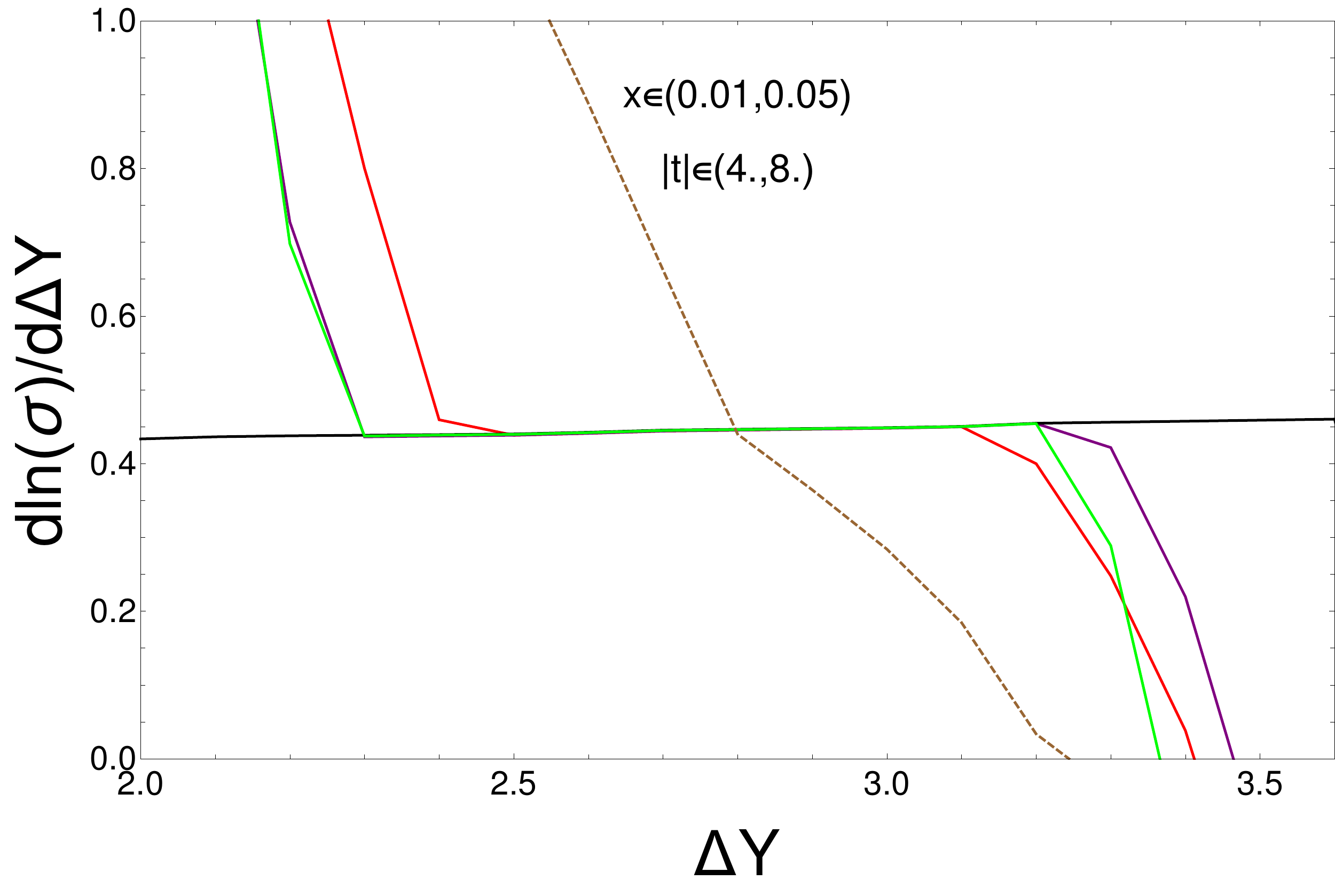}
		\end{subfigure}
		\hspace{2cm}
		\begin{subfigure}{6cm}
			\includegraphics[width=8cm, height=5.4cm]{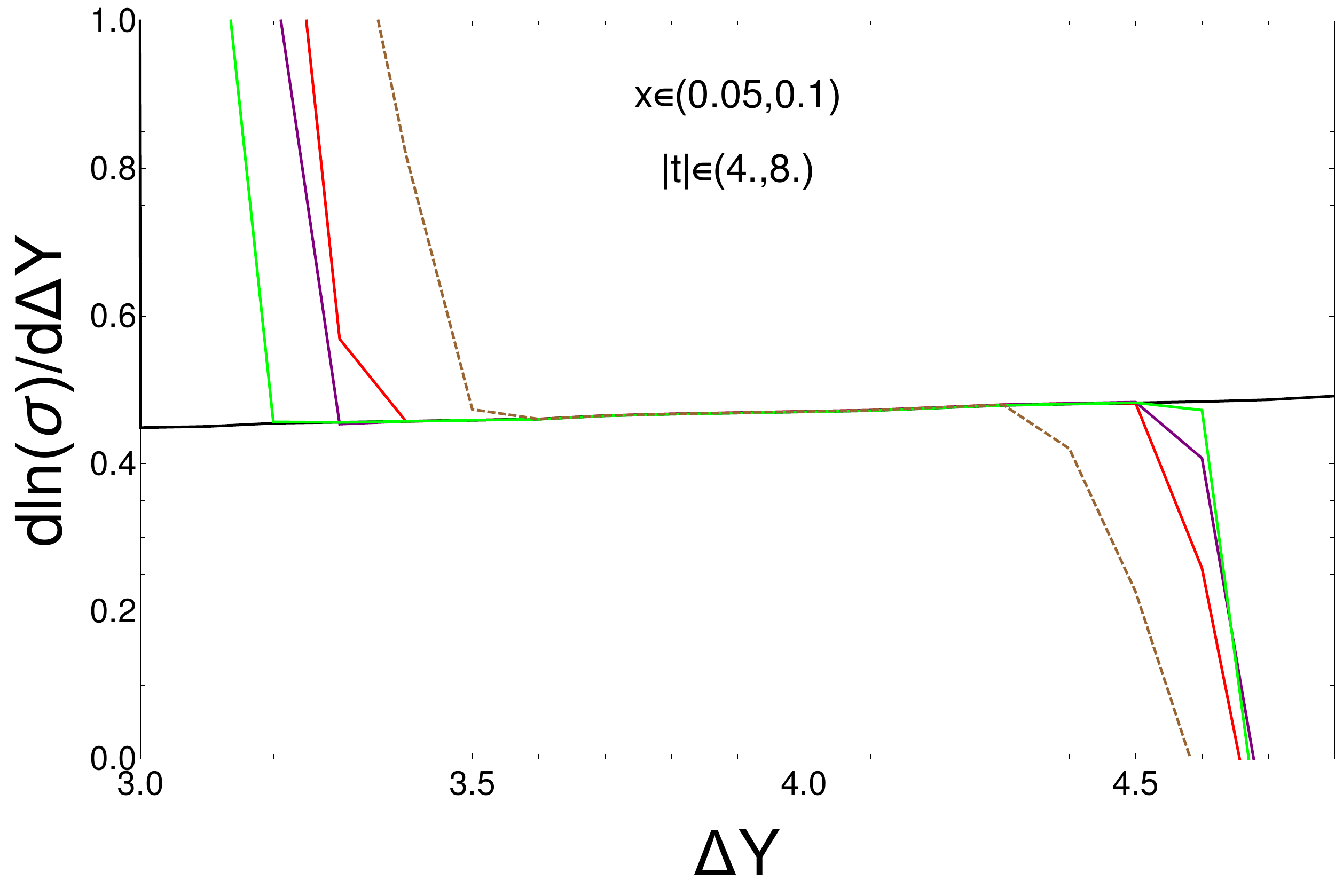}
	\end{subfigure}}
	\caption{The  logarithmic derivative  of the $\gamma p \rightarrow V+gap+X$ cross section in $\Delta Y$ evaluated in center of bins (black line) versus calculated from cross section averaged over bins. The brown line: original size of bins; the red line: bins in $x$ halved ; the green line: bins in $x$ and $t$ halved; the purple line: bin in $x$ one third of its size.}
	\label{fig:dYM2}
\end{figure}

Finally, to complete the analysis of the $\gamma p$ cross section, 
we have studied numerically also the logarithmic derivative of the cross section in $\Delta Y$ - $d\ln\sigma/d\Delta Y$~\eqref{eq:logder}. We have compared the logarithmic derivative of the cross section evaluated in the center of the bins with logarithmic derivative of the cross section averaged over the respective bins. We have found, that the range of values of $\Delta Y$ for which the Pomeron intercept is accessible is limited and depends on the size of the bins in which the variables $t$ and $x$ are measured. The larger the size of bins of $t$ and $x$ the smaller the range in $\Delta Y$ for which the Pomeron intercept is accessible as illustrated in Fig.~\ref{fig:dYM2}. The theoretical range (black line in Fig.~\ref{fig:dYM2}, experimentally inaccessible) of $\Delta Y$ in case the $t$ and $x$ would be measured in exact points. Pomeron intercept can not be accessed in lowest $x$ bins in the case of the full size of bins. Since this kind of measurement would be crucial in distinguishing between different models of vector meson production in this process a lot of experimental effort must be invested in measuring $x$ and $t$ in as small bins as possible, which would be determined by the available statistics.
The main conclusion from this study is the observation of the value of the logarithmic derivative to be around $0.4-0.5$ which is expected from the exchange of the BFKL Pomeron in the diffractive amplitude, and a rather  weak dependence on $t$ and $x$.

\begin{figure}[h]
\hspace*{-2cm}\centering{
\begin{subfigure}{6cm}
\includegraphics[width=8cm, height=5.4cm]{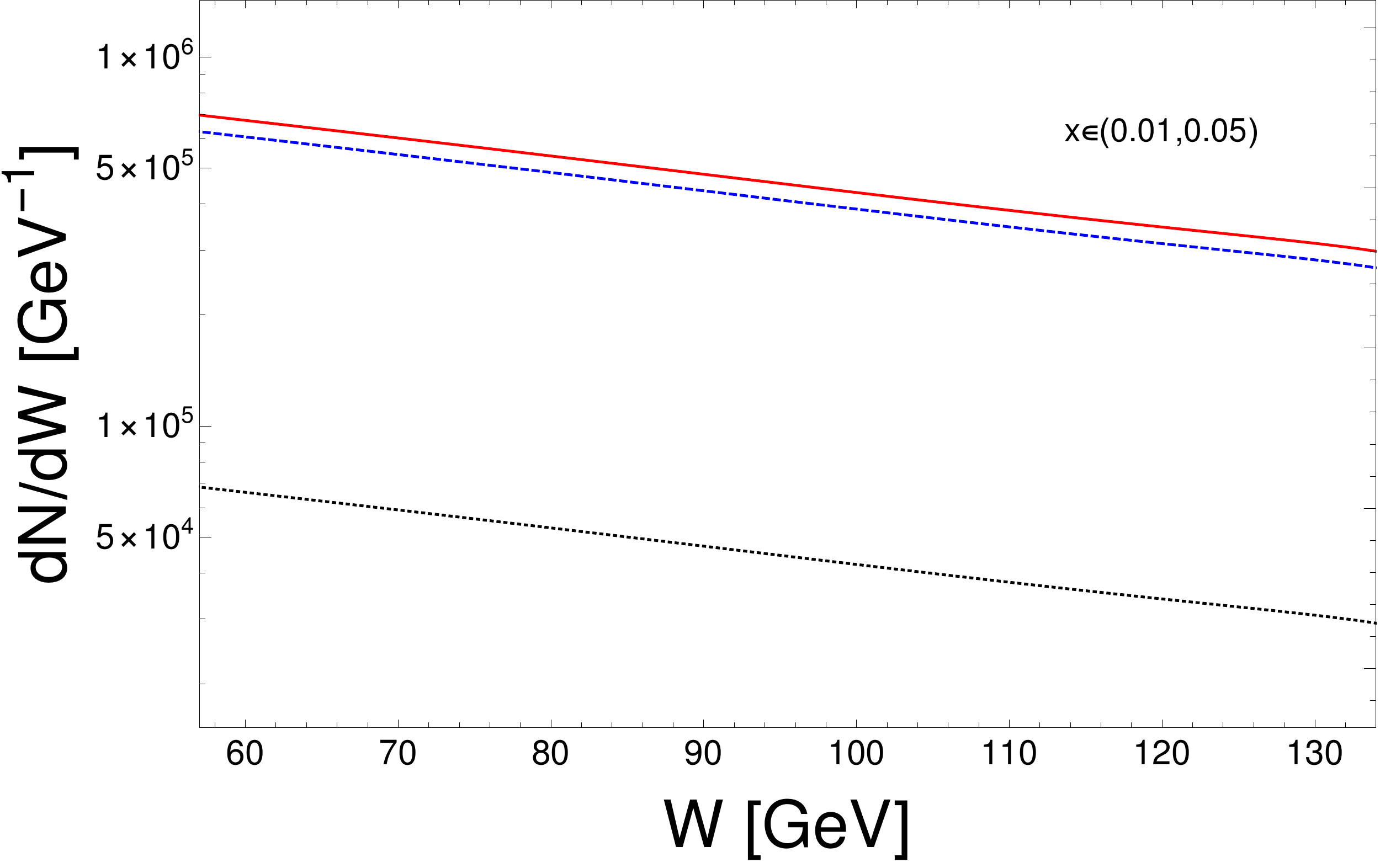}
\end{subfigure}
\hspace{2cm}
\begin{subfigure}{6cm}
\includegraphics[width=8cm, height=5.4cm]{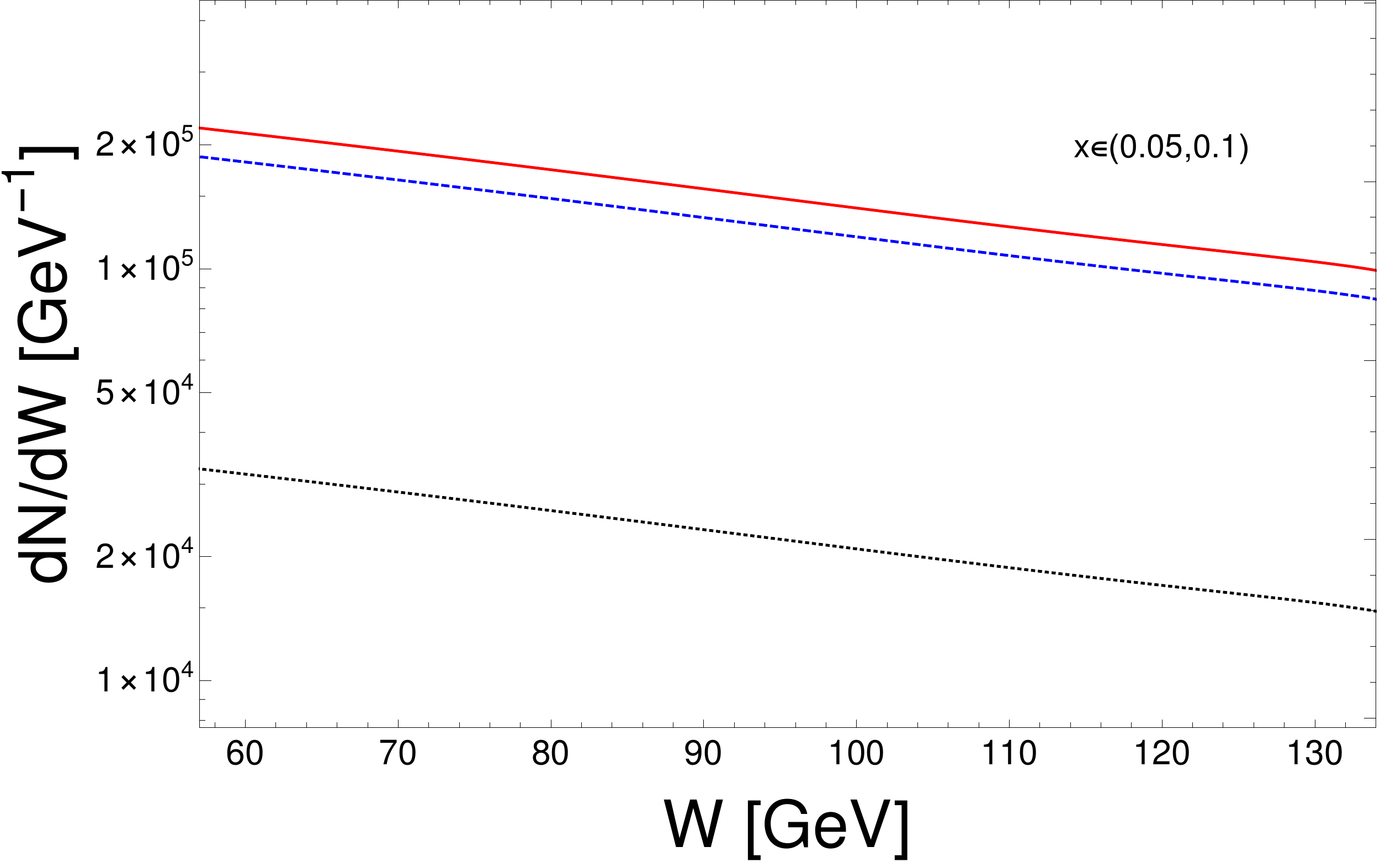}
\end{subfigure}}
\hspace*{-2cm}\centering{
\begin{subfigure}{6cm}
\includegraphics[width=8cm, height=5.4cm]{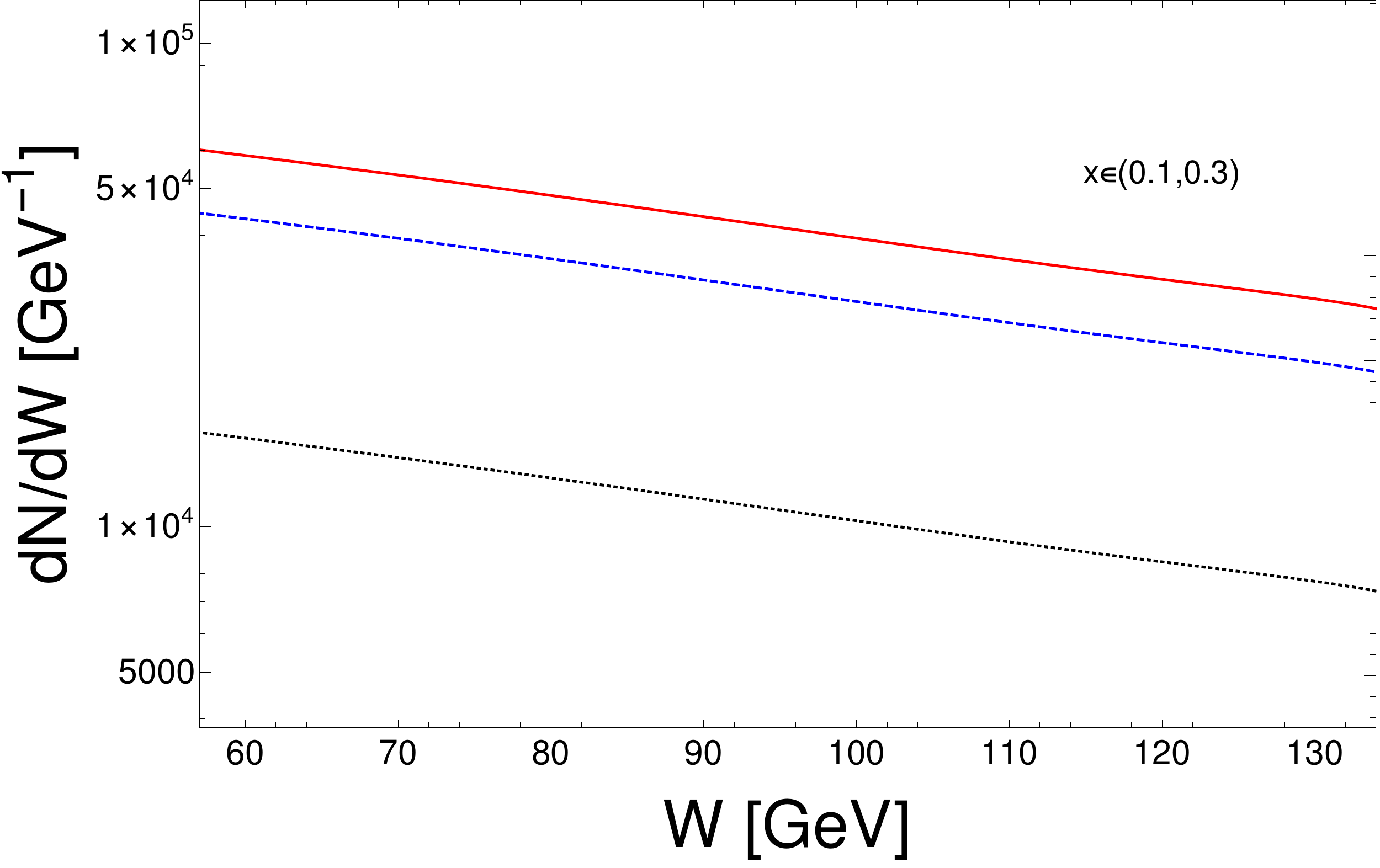}
\end{subfigure}}
\caption{Energy dependence of the number of events for various bins in $x$ and the bin in $t\in\left(1,2\right)\;$GeV$^2$. Integrated luminosity ${\cal L }=10 \;\rm fb^{-1}$. 
	}
\label{fig:oDpW1fx}
\end{figure}

We next proceed to study  the rates,  and  in particular the range of the rapidity gaps (we don't include the decay branching factor which for the dimuon channel is about $5\%$).
First, in Fig.~\ref{fig:oDpW1fx} we show the  plots of the energy dependence of the number of events for various bins of $x$ and $t$.  This is essentially $\gamma p$ cross section convoluted with the photon flux and assumed integrated luminosity of ${\cal L} = 10 \, \rm fb^{-1}$ at an EIC.
 The decrease of the number of events with the energy $W$ is the result of the $y$ (inelasticity $y$ grows with $W$) dependence of the photon flux which decreases with growing $y$.

\begin{figure}[h]
\centering{
\begin{subfigure}{6cm}
\includegraphics[width=8cm, height=5.4cm]{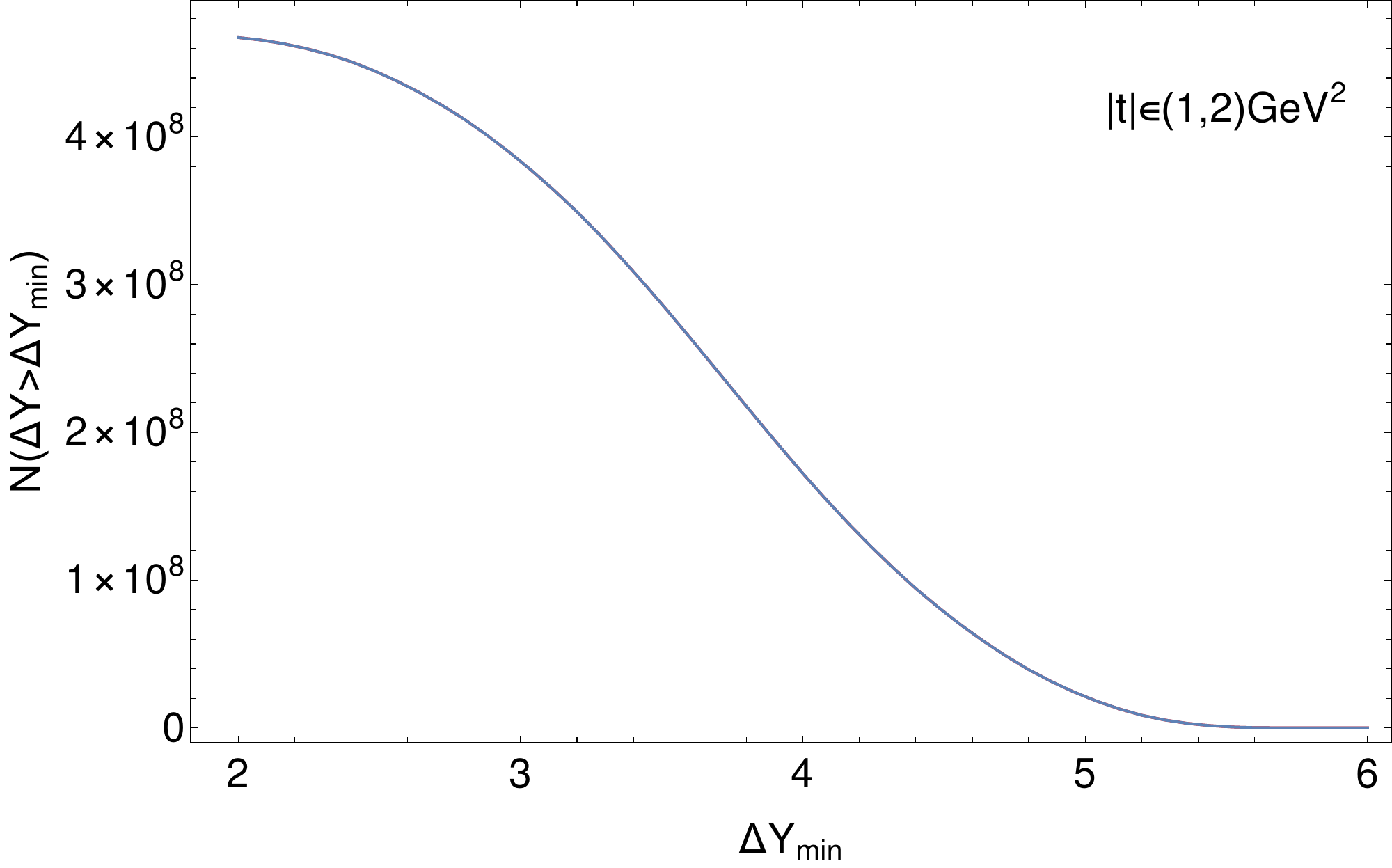}
\end{subfigure}
\hspace{2cm}
\begin{subfigure}{6cm}
\includegraphics[width=8cm, height=5.4cm]{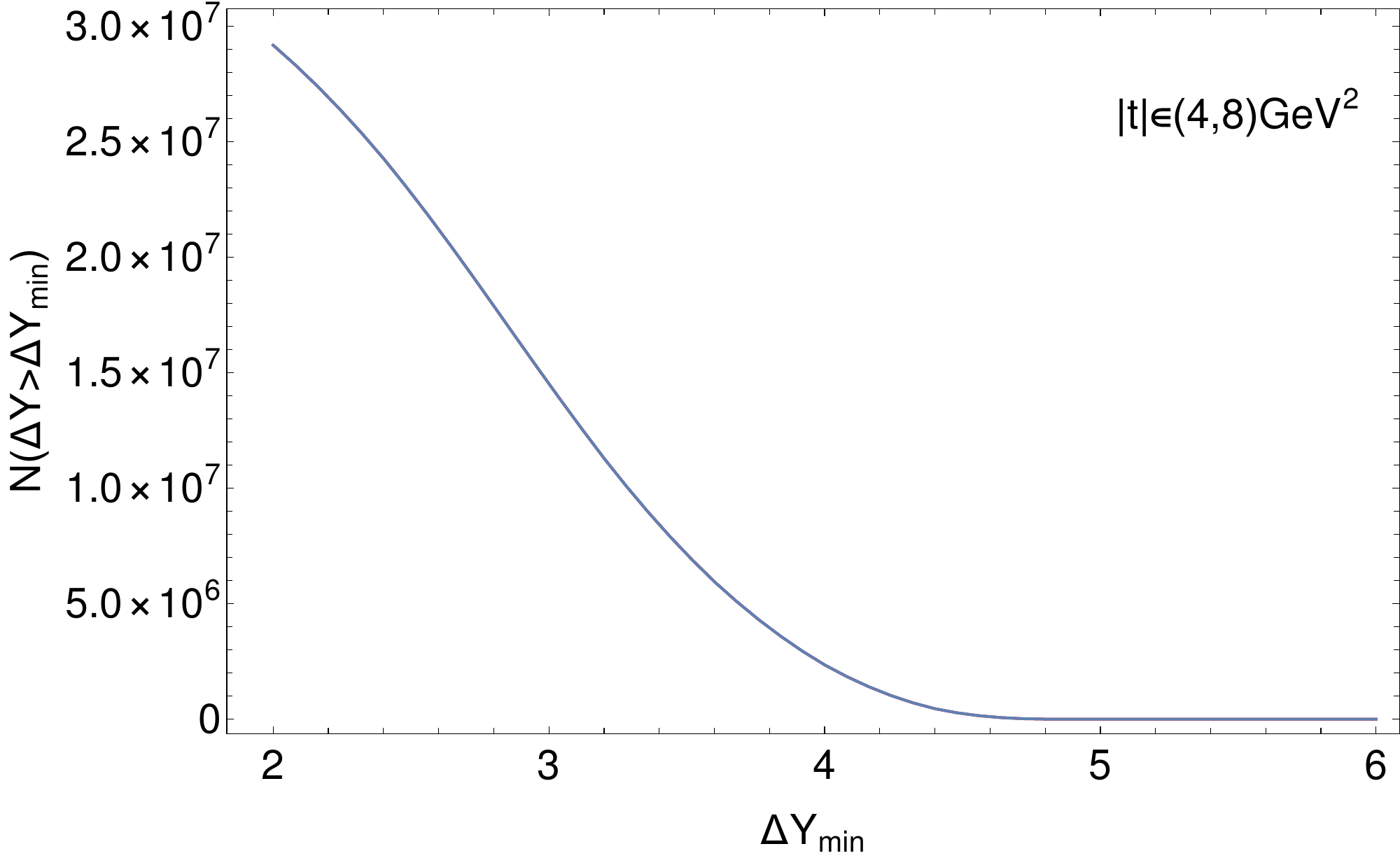}
\end{subfigure}
}
\centering{
\begin{subfigure}{6cm}
\includegraphics[width=8cm, height=5.4cm]{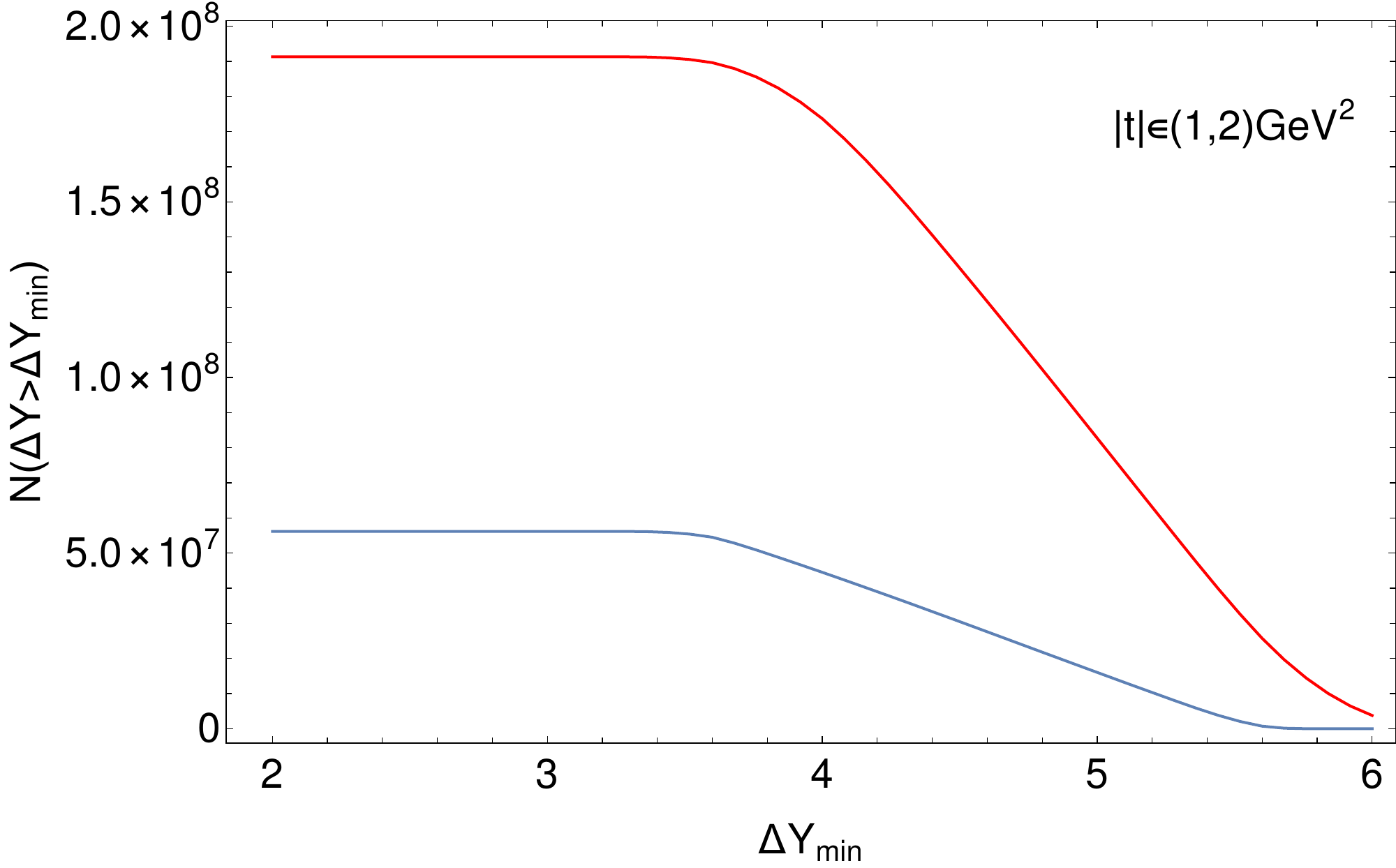}
\end{subfigure}
\hspace{2cm}
\begin{subfigure}{6cm}
\includegraphics[width=8cm, height=5.4cm]{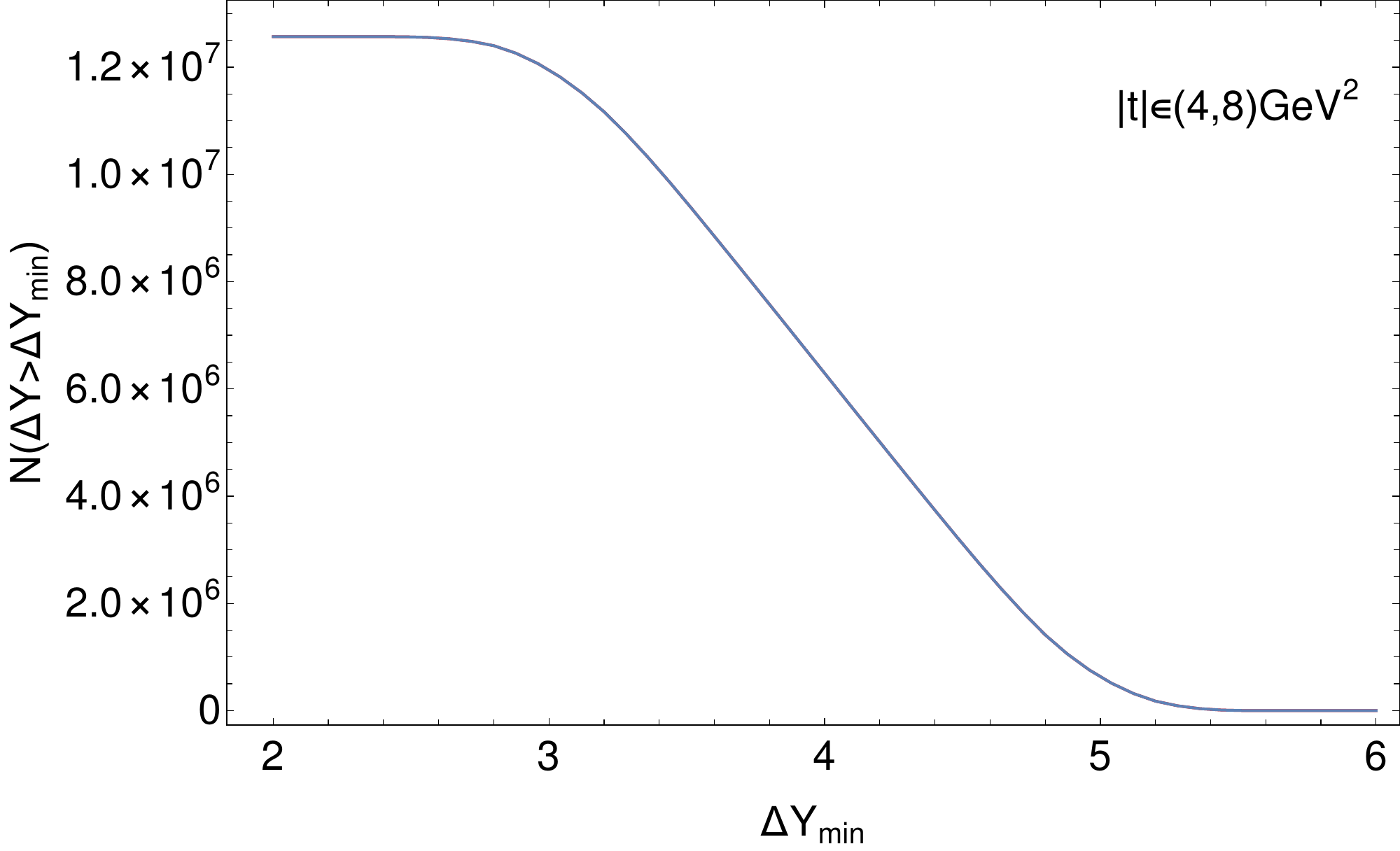}
\end{subfigure}}
\centering{
\begin{subfigure}{6cm}
\includegraphics[width=8cm, height=5.4cm]{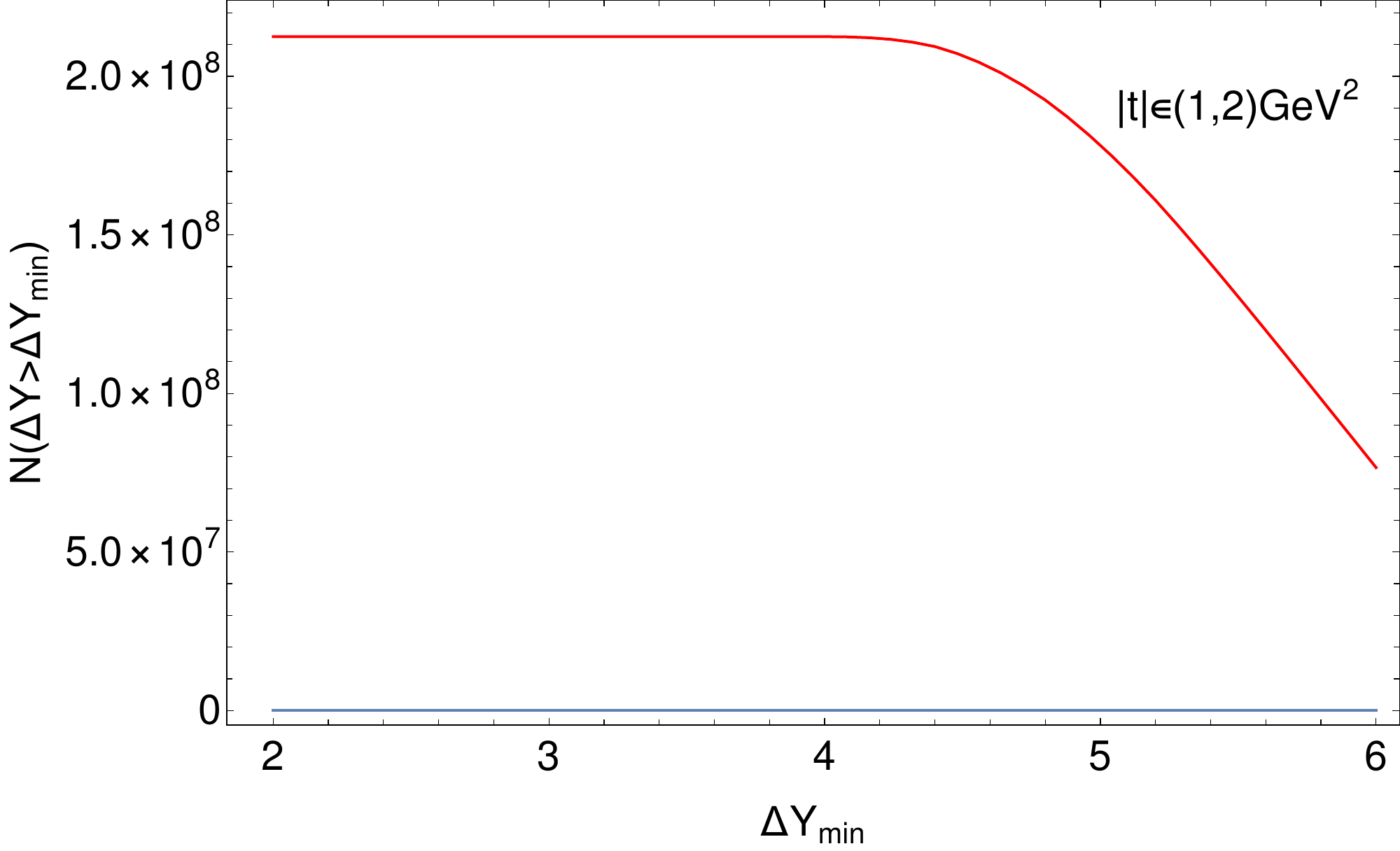}
\end{subfigure}
\hspace{2cm}
\begin{subfigure}{6cm}
\includegraphics[width=8cm, height=5.4cm]{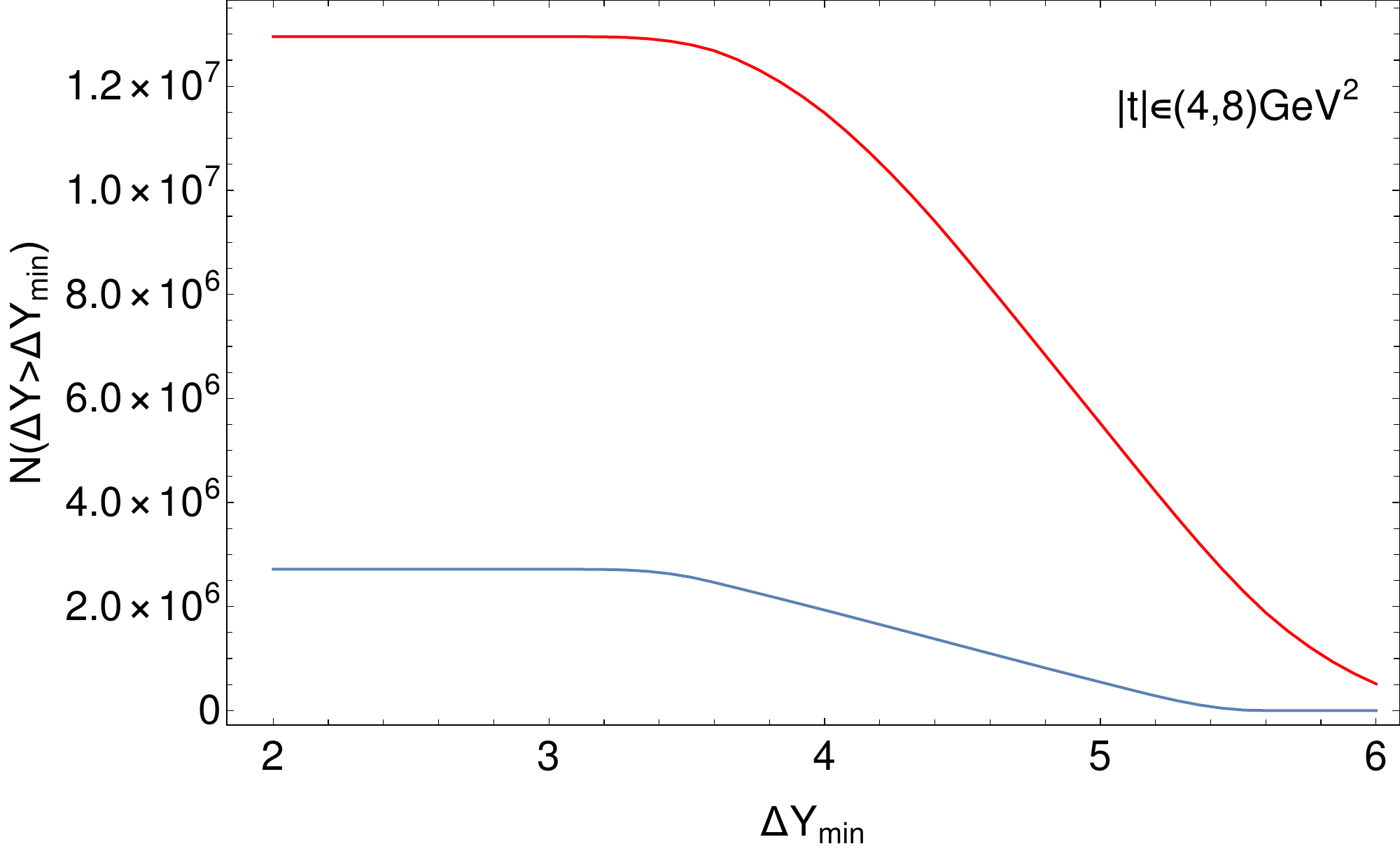}
\end{subfigure}}

\caption{Rates as defined in Eq.\eqref{eq:Wintegrated_rates}. Left column:  bin in $|t| \in \left(1,2\right)$ GeV$^2$, right column:  bin in $|t| \in \left(4,8\right)$ GeV$^2$. First row (from top to bottom):  bin in  $x\in\left(0.05,0.1\right)$, second row: $x\in\left(0.05,0.1\right)$, third row $x\in\left(0.1,0.3\right)$. No cuts on angle the - red line, restriction on angles $4^{\circ}$ - the blue line. 
 Integrated luminosity ${\cal L }=10 \;\rm fb^{-1}$. 
 }
\label{fig:intW1}
\end{figure}

\begin{figure}[h]
	\centering{
		\begin{subfigure}{6cm}
			\includegraphics[width=8cm, height=5.4cm]{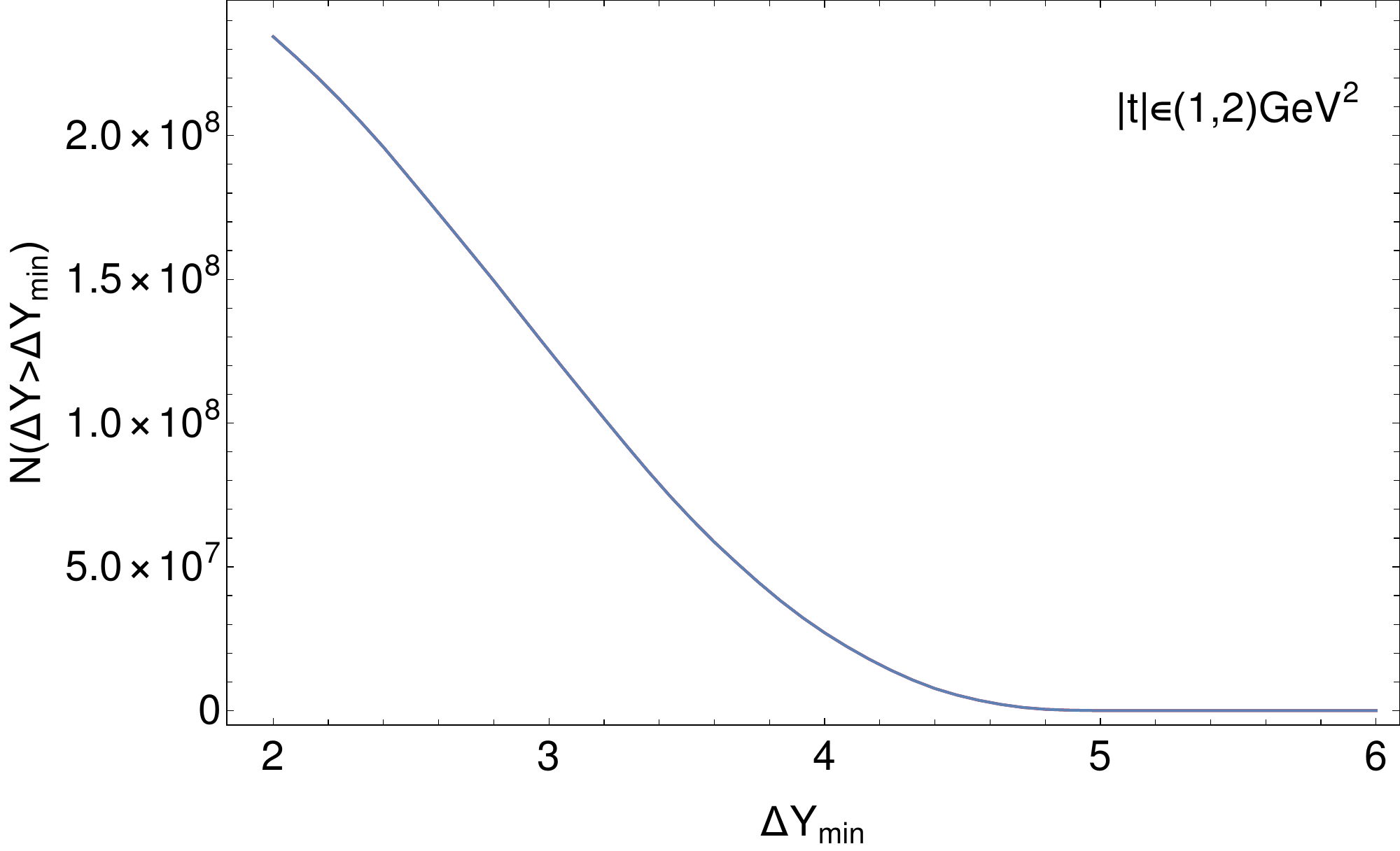}
		\end{subfigure}
		\hspace{2cm}
		\begin{subfigure}{6cm}
			\includegraphics[width=8cm, height=5.4cm]{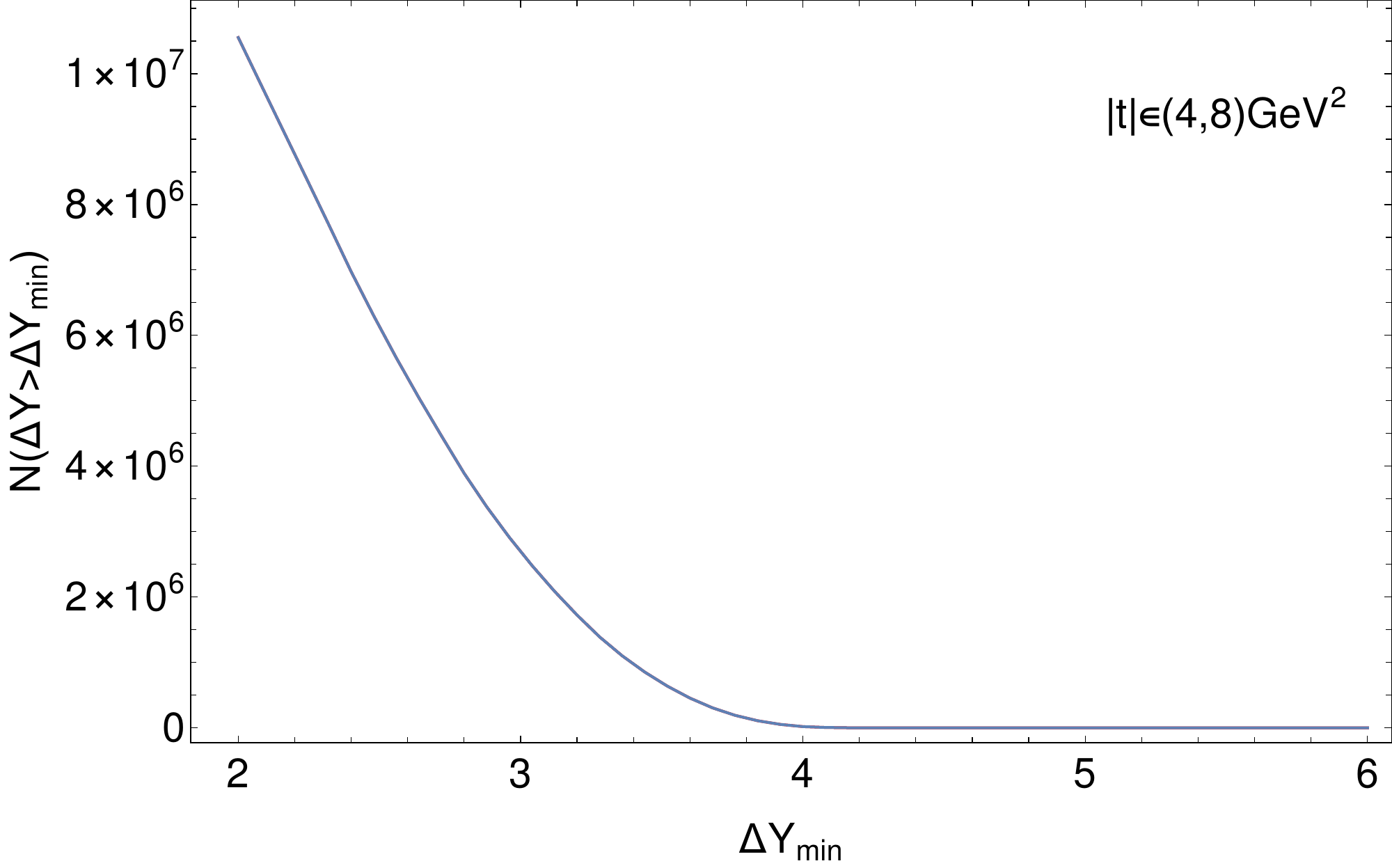}
		\end{subfigure}
	}
	\centering{
		\begin{subfigure}{6cm}
			\includegraphics[width=8cm, height=5.4cm]{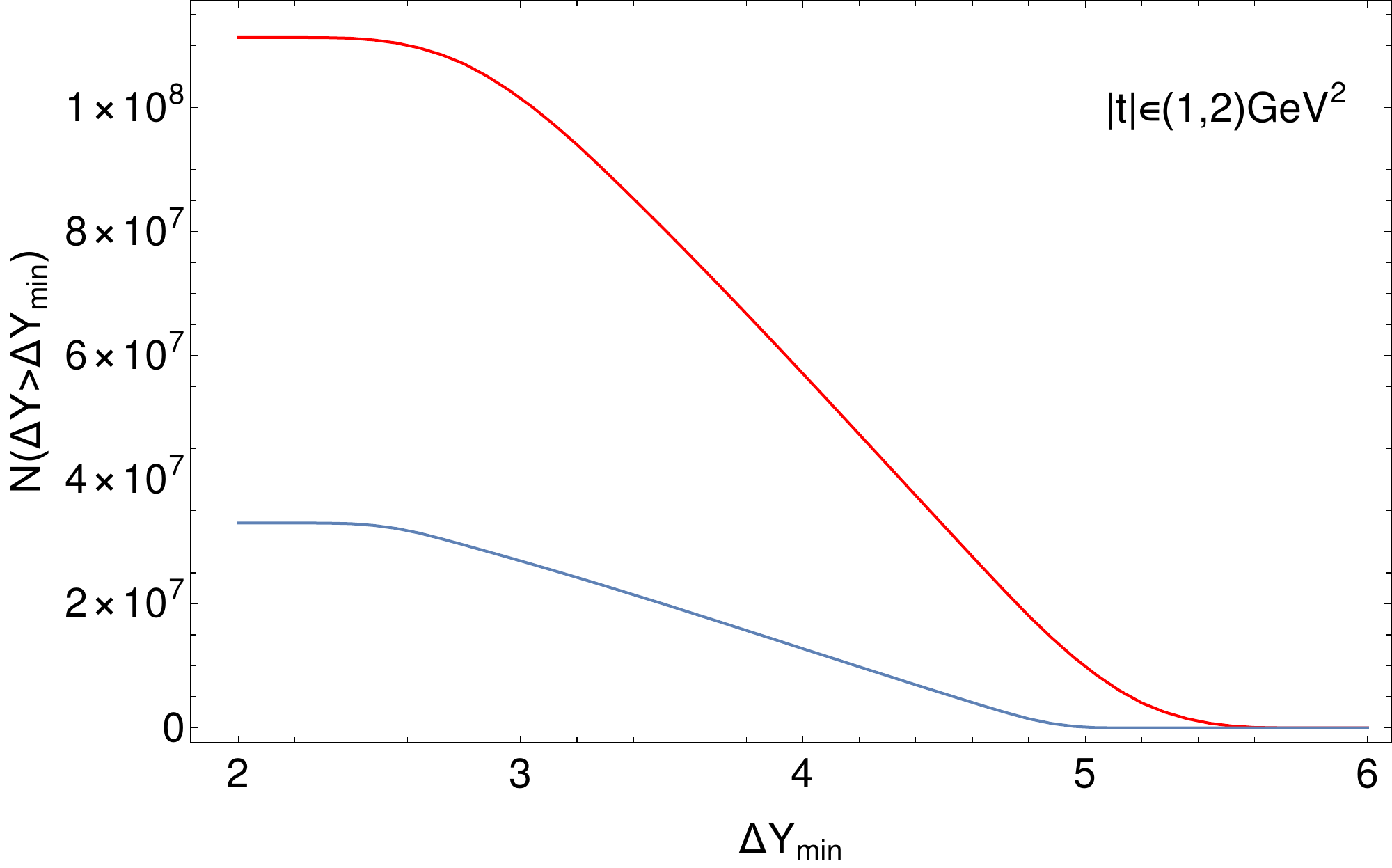}
		\end{subfigure}
		\hspace{2cm}
		\begin{subfigure}{6cm}
			\includegraphics[width=8cm, height=5.4cm]{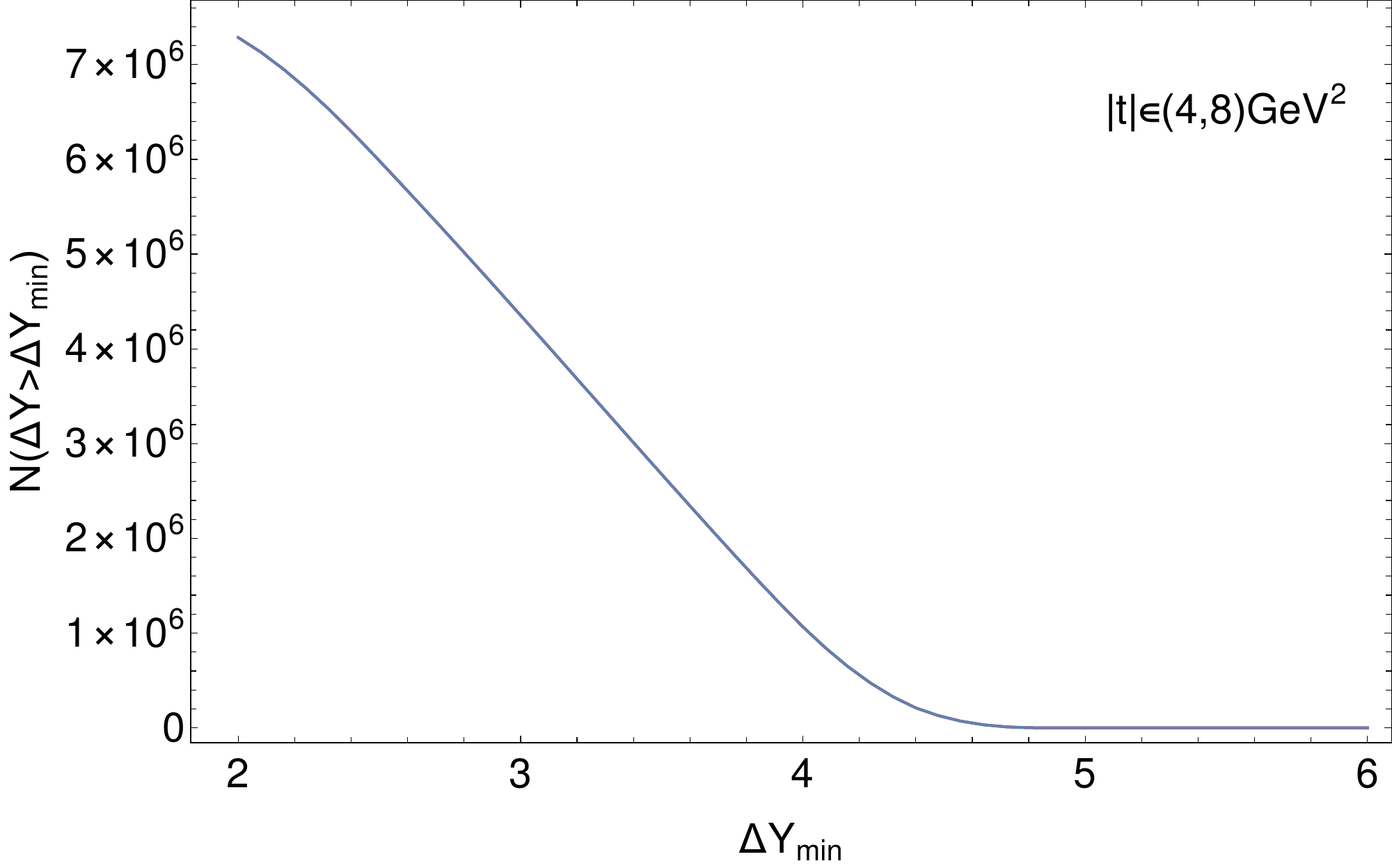}
	\end{subfigure}}
	\centering{
		\begin{subfigure}{6cm}
			\includegraphics[width=8cm, height=5.4cm]{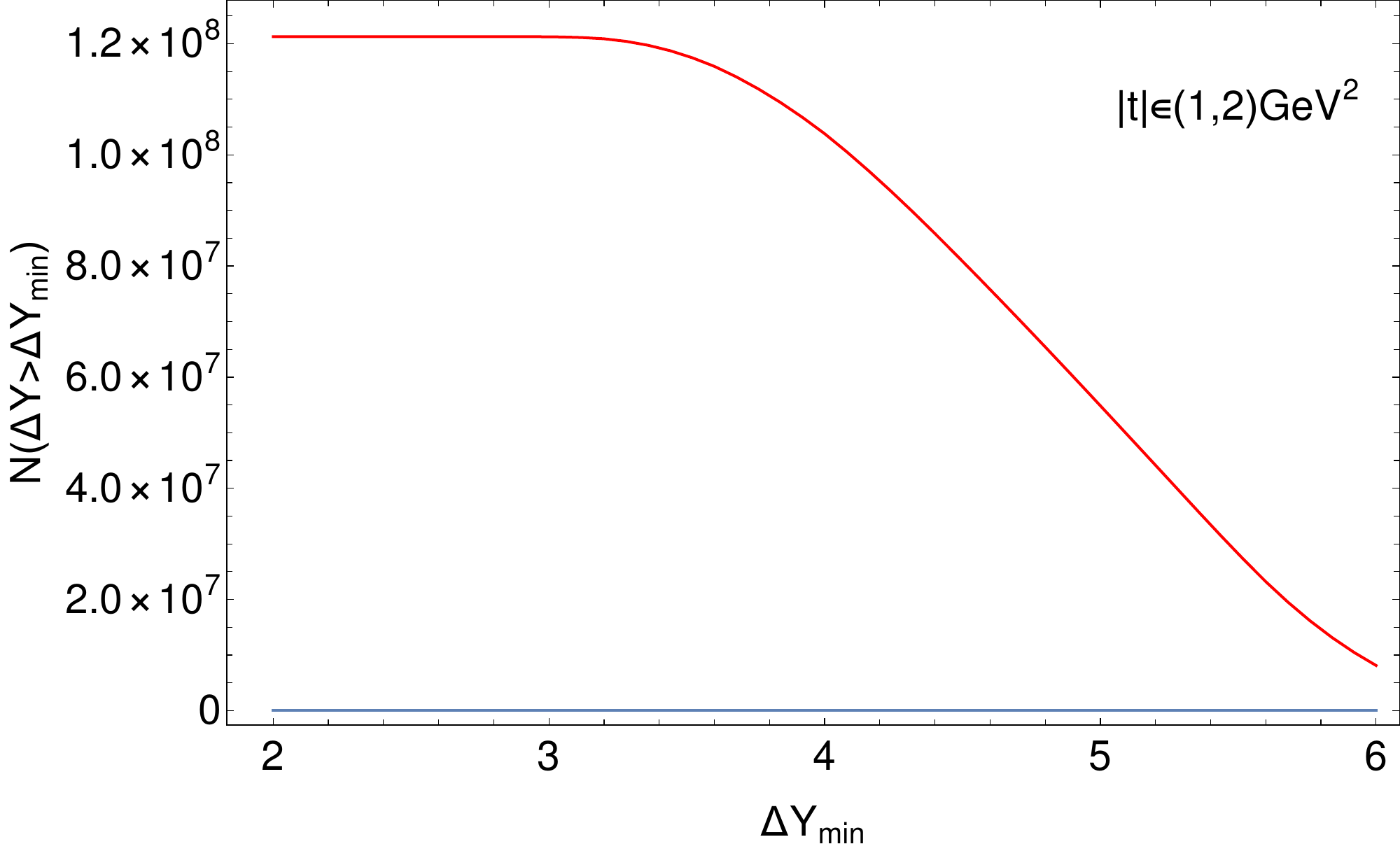}
		\end{subfigure}
		\hspace{2cm}
		\begin{subfigure}{6cm}
			\includegraphics[width=8cm, height=5.4cm]{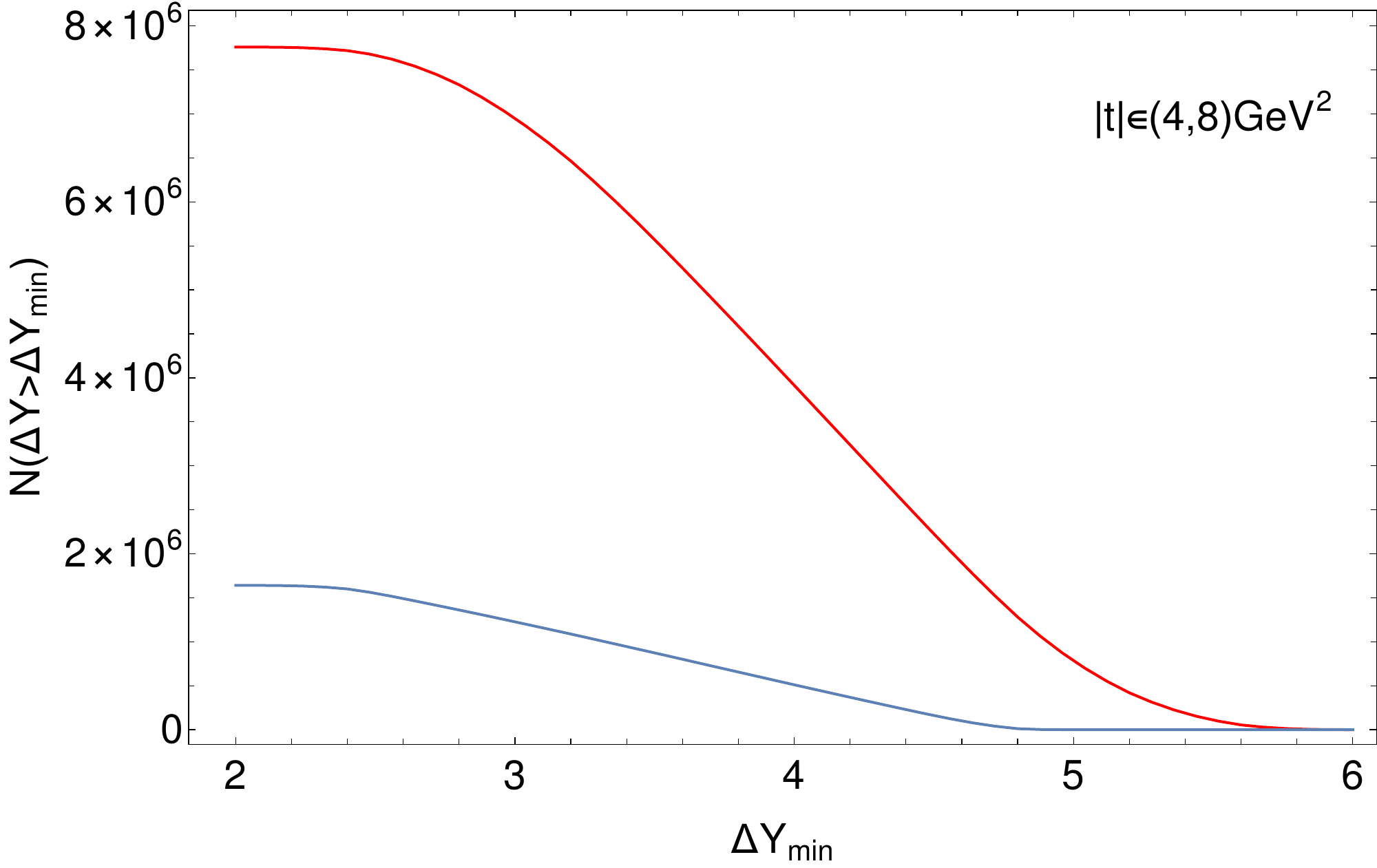}
	\end{subfigure}}
	
	\caption{Rates as defined in Eq.\eqref{eq:Wintegrated_rates}. Left column:  bin in $|t| \in \left(1,2\right)$ GeV$^2$, right column:  bin in $|t| \in \left(4,8\right)$ GeV$^2$. First row (from top to bottom):  bin in  $x\in\left(0.05,0.1\right)$, second row: $x\in\left(0.05,0.1\right)$, third row $x\in\left(0.1,0.3\right)$. No cuts on angle the - red line, restriction on angles $4^{\circ}$ - the blue line. For $W$ energy range $\left(30,100\right)\;$GeV. 
 Integrated luminosity ${\cal L }=10 \;\rm fb^{-1}$. 
}
	\label{fig:intW1b}
\end{figure}

We next  analyze the range of the rapidity gaps which can be accessible in various kinematic setups. 
To illustrate better the range in rapidity gaps  
we show in Figs.~\ref{fig:intW1}-\ref{fig:intW1b}  number of events integrated over the energy $W$ as a function of $\Delta Y_{\rm min}$. 
To be precise we show the quantity
\begin{equation}
N(\Delta Y_{\rm min}) = {\cal L} \int_{\Delta x} dx \int_{\Delta t} dt \, \int_{y_{min}}^{y_{max}} dy \, \Theta(\Delta Y - \Delta Y_{\rm min}) \,  {\tilde\Phi}_{\gamma/e}\left(y\right)\; \frac{d \sigma_{\gamma^{\ast}p}\left(y\right)}{dt dx}  \; ,
\label{eq:number_ev1}
\end{equation}
where the integrals  with subscripts $\Delta x$ and $\Delta t$ mean that one integrates over different bins in $x$ and $t$. Limits $y_{min}$ and $y_{max}$ are defined by the limits on $W$ which are taken to be equal to $50 \, \rm GeV$, the same as the lower limit at H1   \cite{Aktas:2003zi} and $140 \, \rm GeV$ the maximum energy of EIC.  Of course, in reality the cuts  on $W$ will have to be determined by the specific acceptance of the EIC detector, and it is likely that they will be lower given the lower energy at EIC than HERA. For comparison 
in Fig.~\ref{fig:intW1b} we have changed the $W$ integration range from $\left(50,140\right)\;$GeV to $\left(30,100\right)\;$GeV.

We  also analyze the case when there is a cut imposed  on the angle, below which we require activity in the detector. 
We have chosen the cut on the angle  to be equal to $4^{\circ}$, which  corresponds  to pseudorapidty of $3.3$. This is in line with current detector project at an EIC, which assumes coverage of the central detector up to $3.5$  units in rapidity in the forward direction. 
We then compare
the number of events defined in this way  without this angular cut (red line) to the number of events where the cut is applied (blue line).
We see that for small $x$ bin, $0.01<x<0.05$ the blue and red lines are on top of each other, top row in Fig.\ref{fig:intW1}, since in this case the
cut on the particle angle is not effective. This is because     for bins where $x$ is rather small, there is activity in the detector which passes the cut. On the other hand for larger values in $x$ bin, and rather small to moderate $t$, (second row in Fig.\ref{fig:intW1} and bin in lower $t$ and a third row), the cut has a substantial effect since many particles produced in the dissociative state may escape the detector. We also clearly see that the region in $\Delta Y_{\rm min}$ where there is substantial number of events, increases with increasing $x$ as expected.  For the smallest $x$ bin there is a rapidity gap of size $2$, whereas for largest values of $x$ one can reach gap sizes of the order of $4$. This is substantially changed in the case when the range of the $W$ integrated is changed to 
$\left(30,100\right)\;$GeV, see Fig\ref{fig:intW1b}. We conclude that the different range on the energy does change significantly the range on the $\Delta Y$ which is possible.

\begin{figure}[h]
\hspace*{-2cm}\centering{
\begin{subfigure}{6cm}
\includegraphics[width=8cm, height=5.4cm]{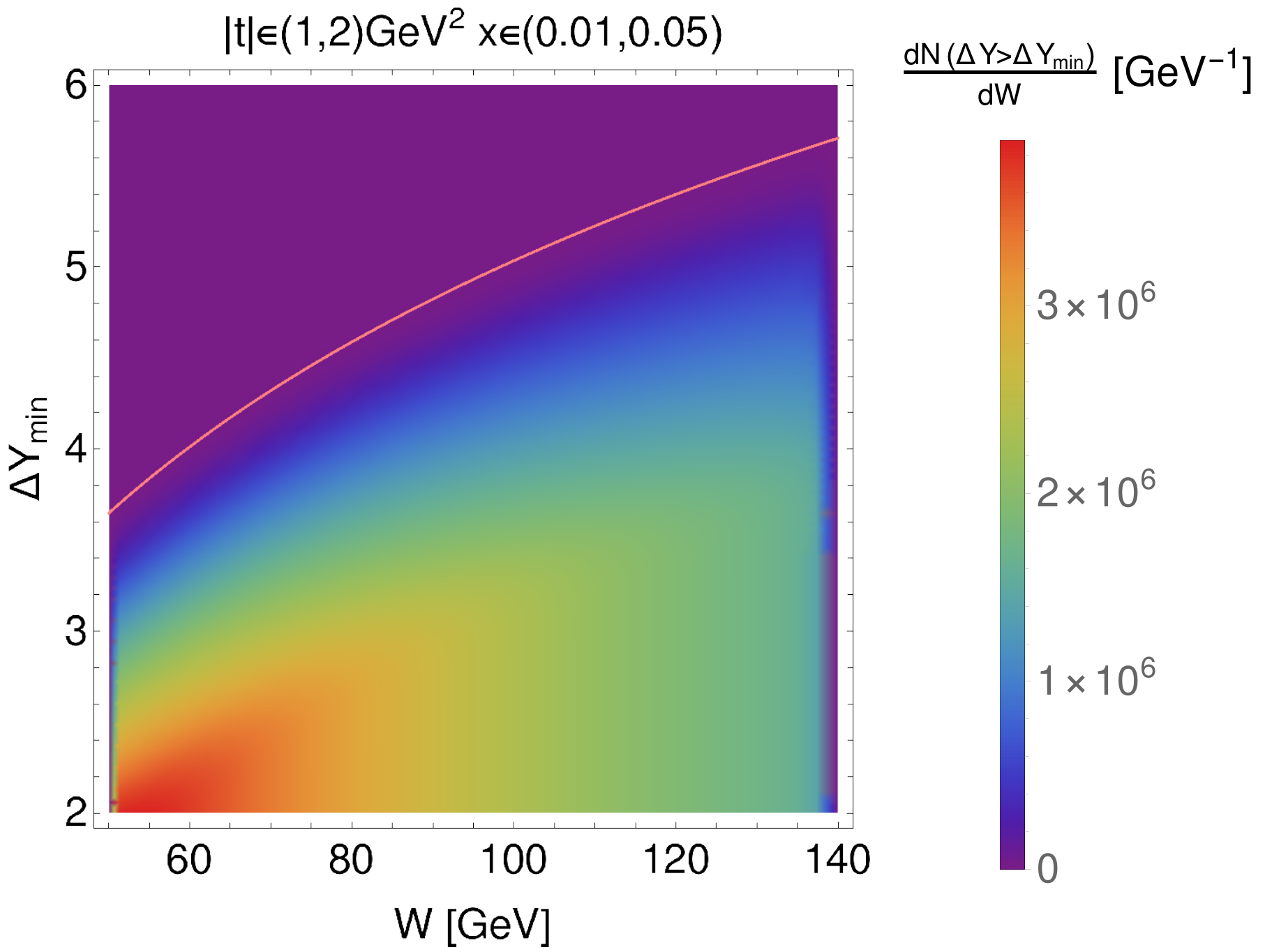}
\end{subfigure}
\hspace{2cm}
\begin{subfigure}{6cm}
\includegraphics[width=8cm, height=5.4cm]{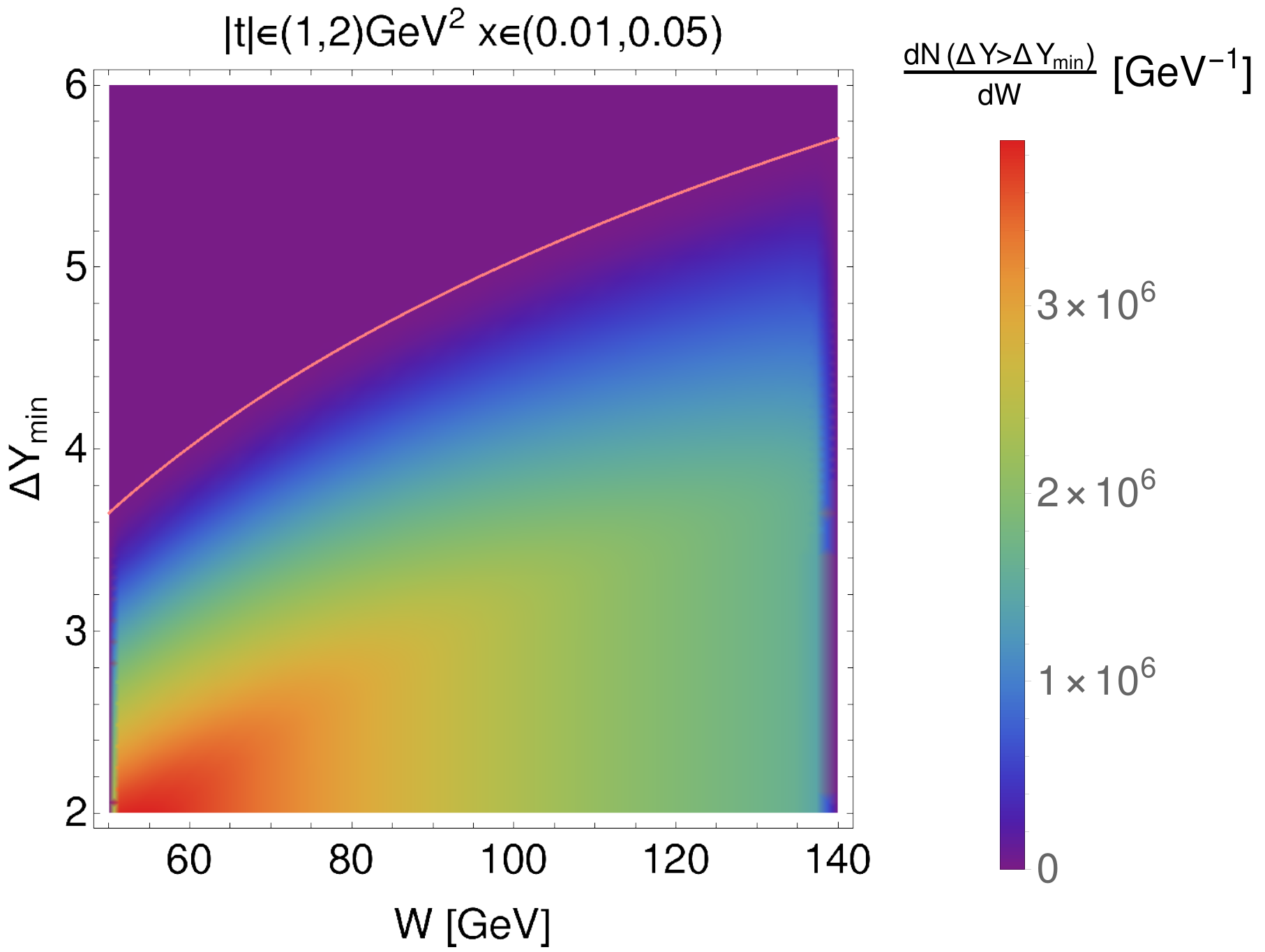}
\end{subfigure}}
\hspace*{-2cm}\centering{
\begin{subfigure}{6cm}
\includegraphics[width=8cm, height=5.4cm]{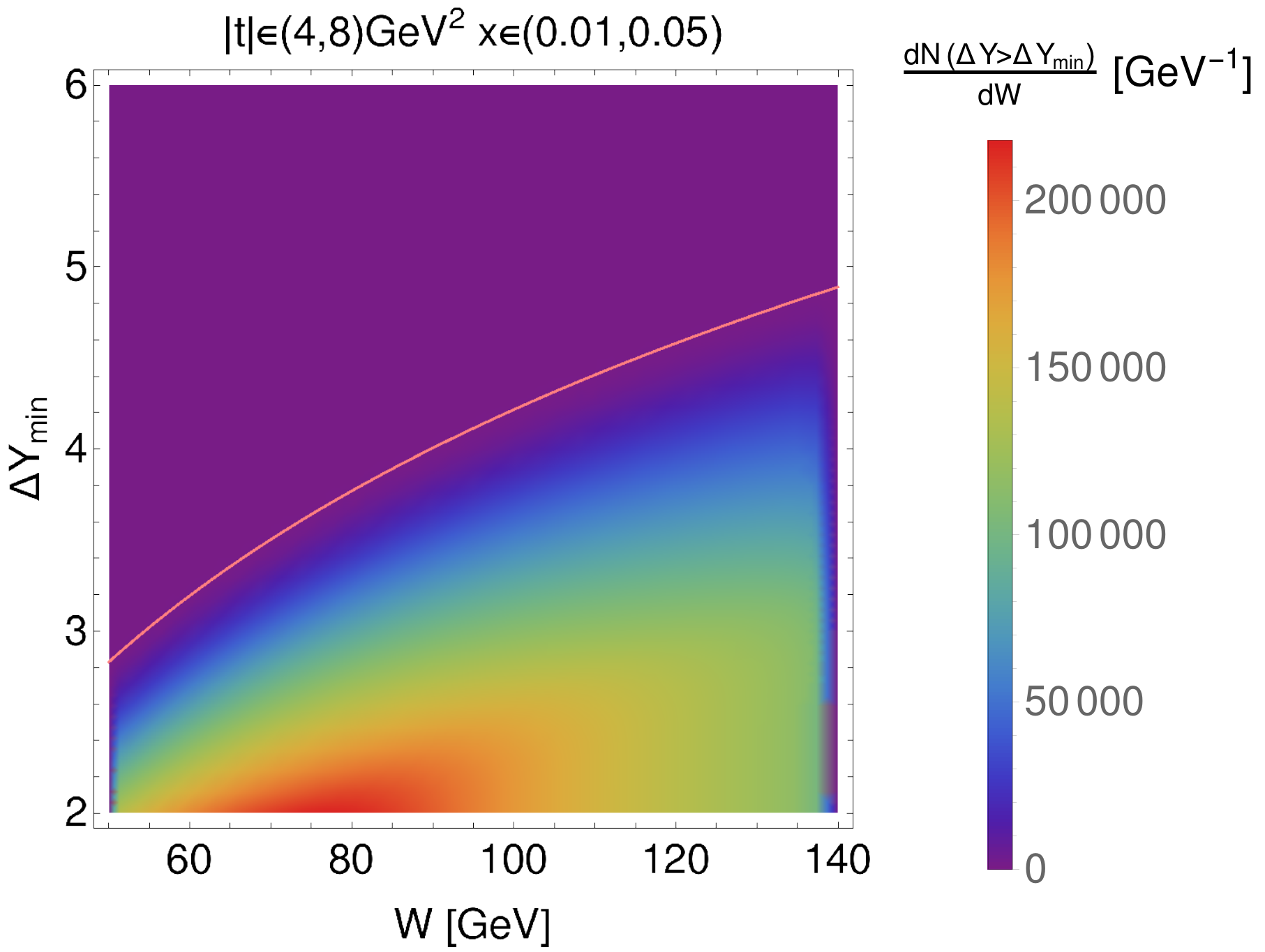}
\end{subfigure}
\hspace{2cm}
\begin{subfigure}{6cm}
\includegraphics[width=8cm, height=5.4cm]{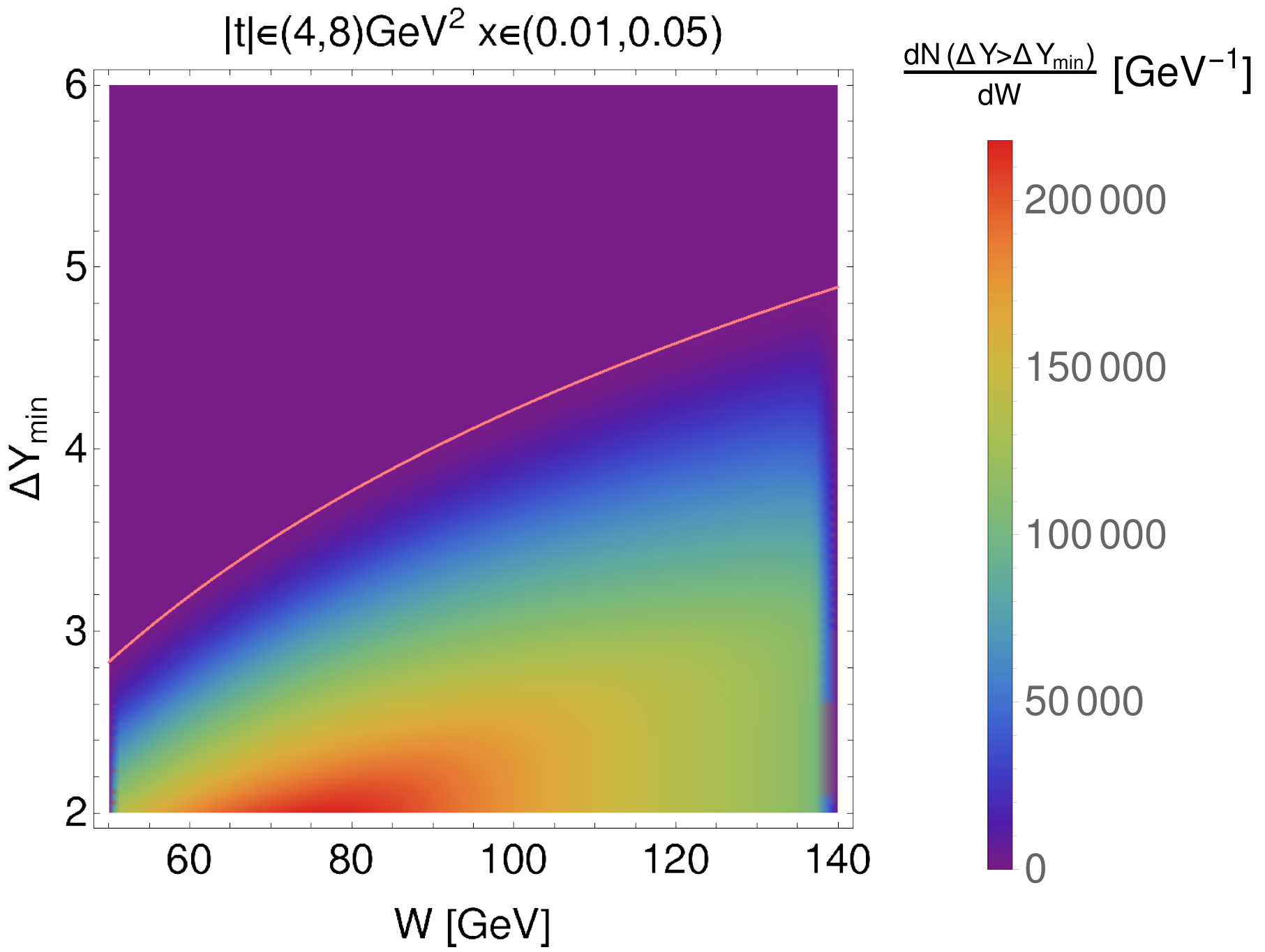}
\end{subfigure}}
\caption{Differential number of events in $W$ in bins in $t$ and $x$ as a two-dimensional function of $W$ and $\Delta Y_{\rm min}$. Left column: no cuts on angle, right column: restriction on angles $4^{\circ}$. Bin in $x\in\left(0.01,0.05\right)$. Upper row: bin in $|t|\in (1,2) { \rm GeV^2}$, lower row:   bin in $|t|\in (4,8) { \rm GeV^2}$.
 Integrated luminosity ${\cal L }=10 \;\rm fb^{-1}$. 
 } 
\label{fig:tbinnoacut1}
\end{figure}

\begin{figure}[h]
\hspace*{-2cm}\centering{
\begin{subfigure}{6cm}
\includegraphics[width=8cm, height=5.4cm]{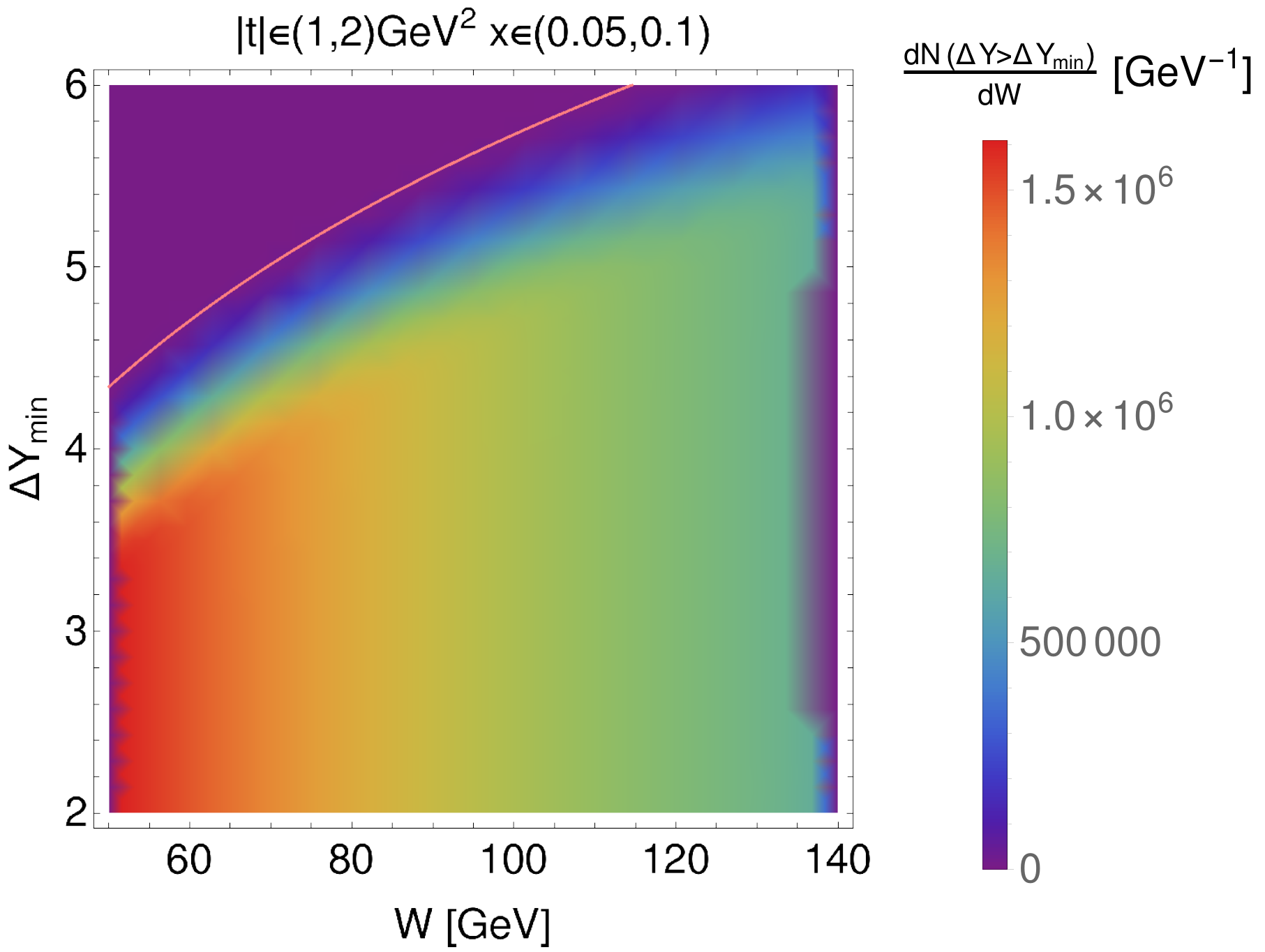}
\end{subfigure}
\hspace{2cm}
\begin{subfigure}{6cm}
\includegraphics[width=8cm, height=5.4cm]{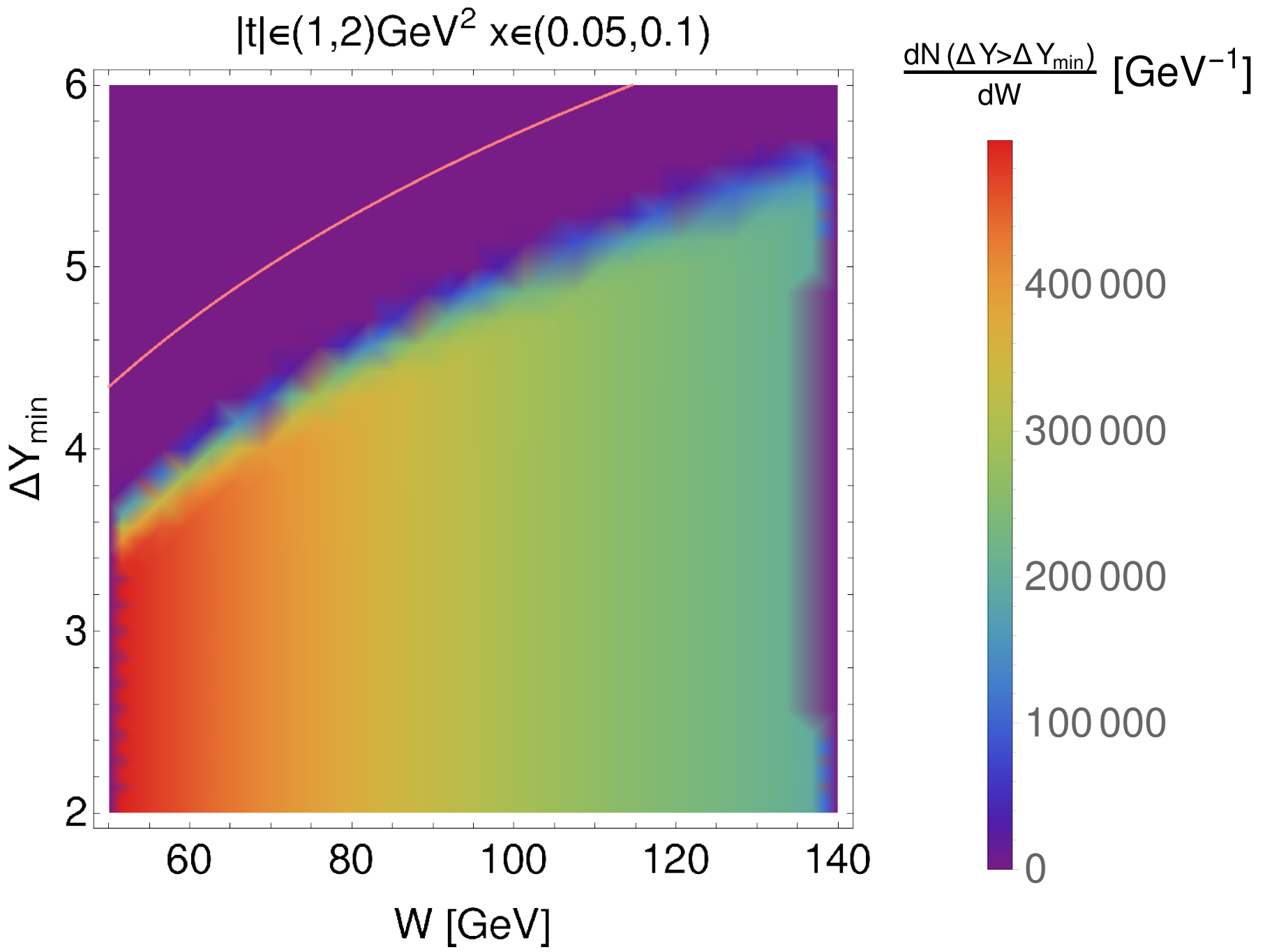}
\end{subfigure}}
\hspace*{-2cm}\centering{
\begin{subfigure}{6cm}
\includegraphics[width=8cm, height=5.4cm]{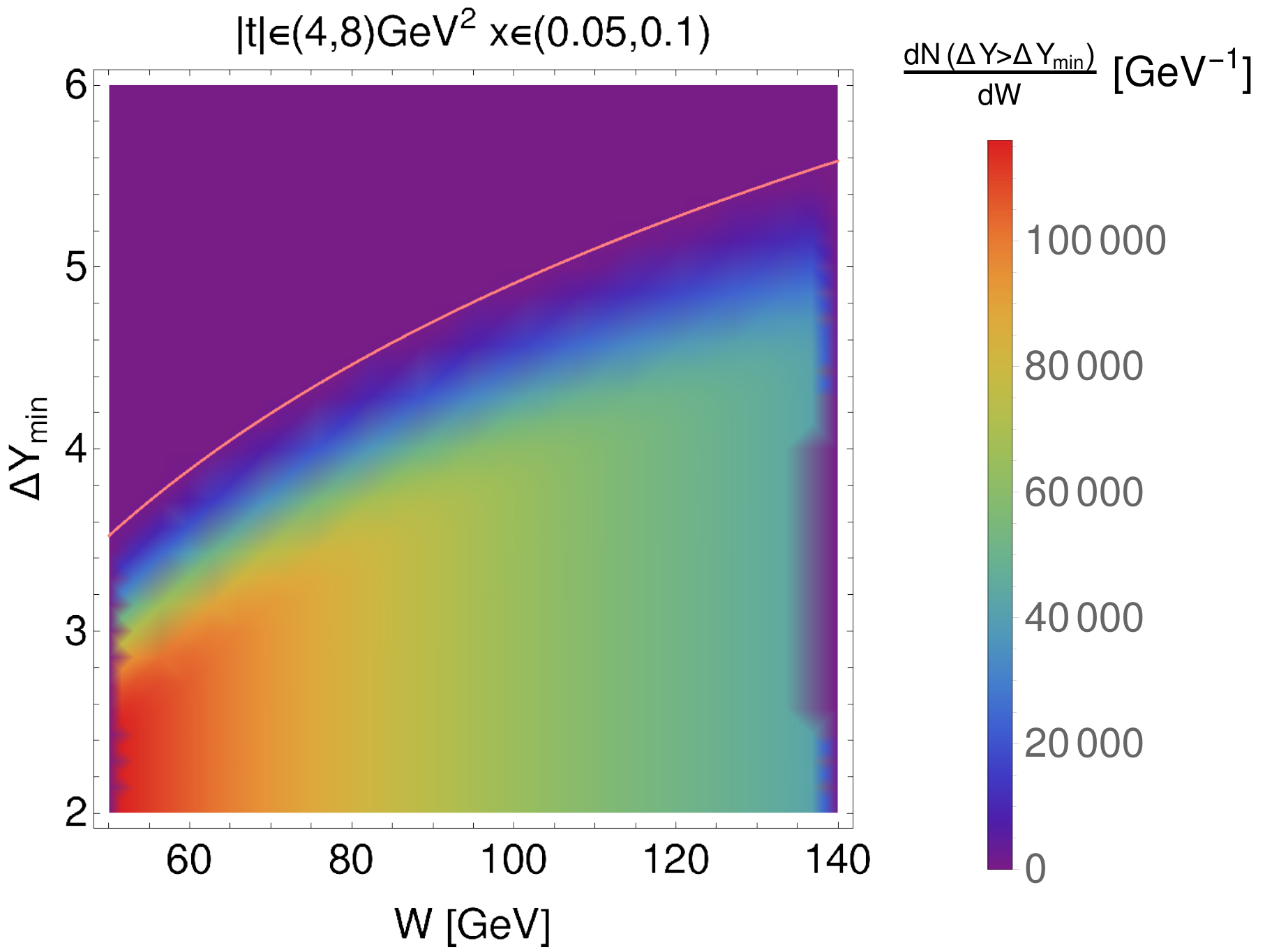}
\end{subfigure}
\hspace{2cm}
\begin{subfigure}{6cm}
\includegraphics[width=8cm, height=5.4cm]{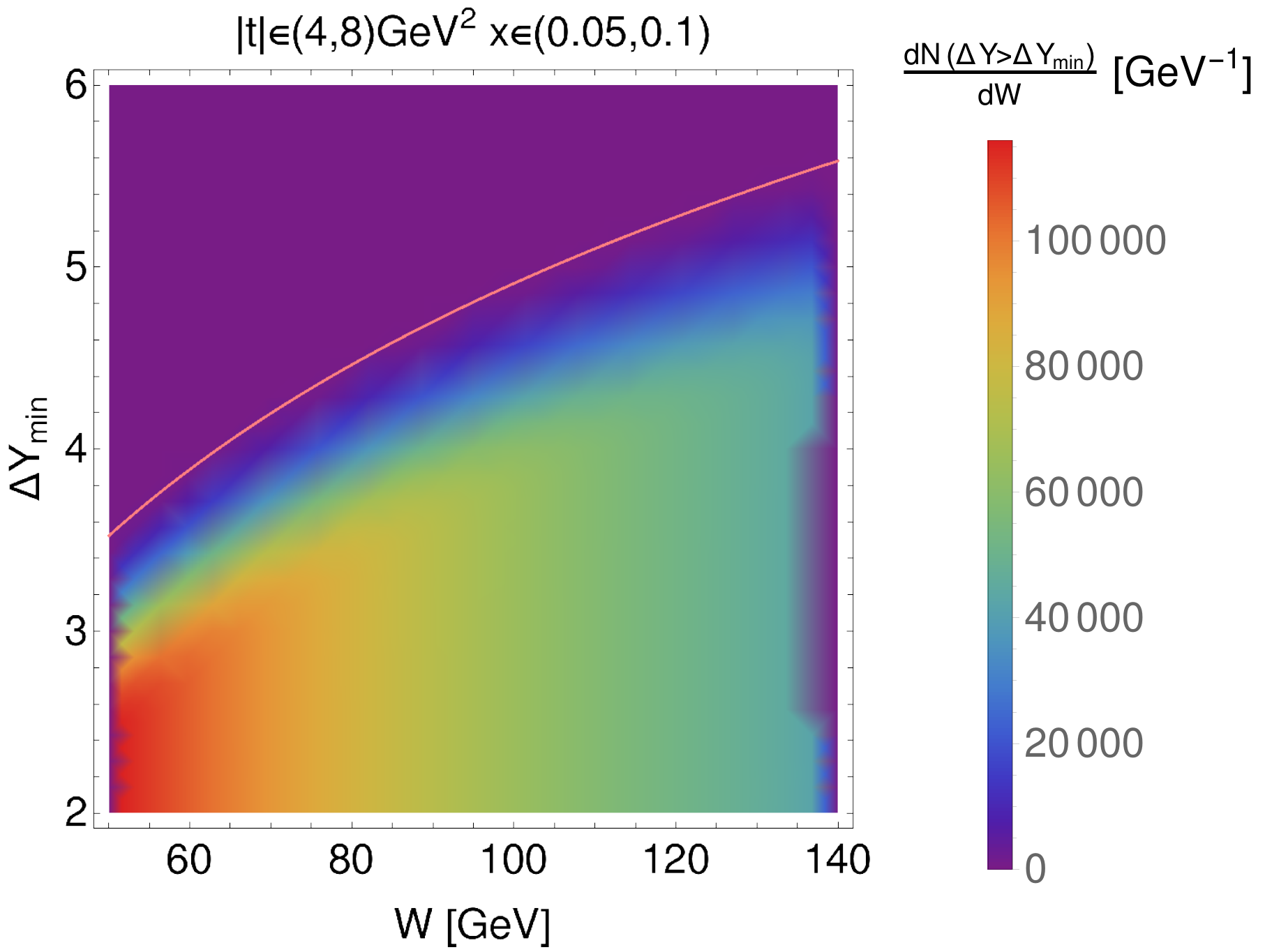}
\end{subfigure}}
\caption{Differential number of events in $W$ in bins in $t$ and $x$ as a two-dimensional function of $W$ and $\Delta Y_{\rm min}$. Left column: no cuts on angle, right column: restriction on angles $4^{\circ}$. Bin in  $x\in\left(0.05,0.1\right)$. Upper row: bin in $|t|\in (1,2) { \rm GeV^2}$, lower row:   bin in $|t|\in (4,8) { \rm GeV^2}$.
 Integrated luminosity ${\cal L }=10 \;\rm fb^{-1}$. 
 }
\label{fig:tbinnoacut2}
\end{figure}

\begin{figure}[h]
\hspace*{-2cm}\centering{
\begin{subfigure}{6cm}
\includegraphics[width=8cm, height=5.4cm]{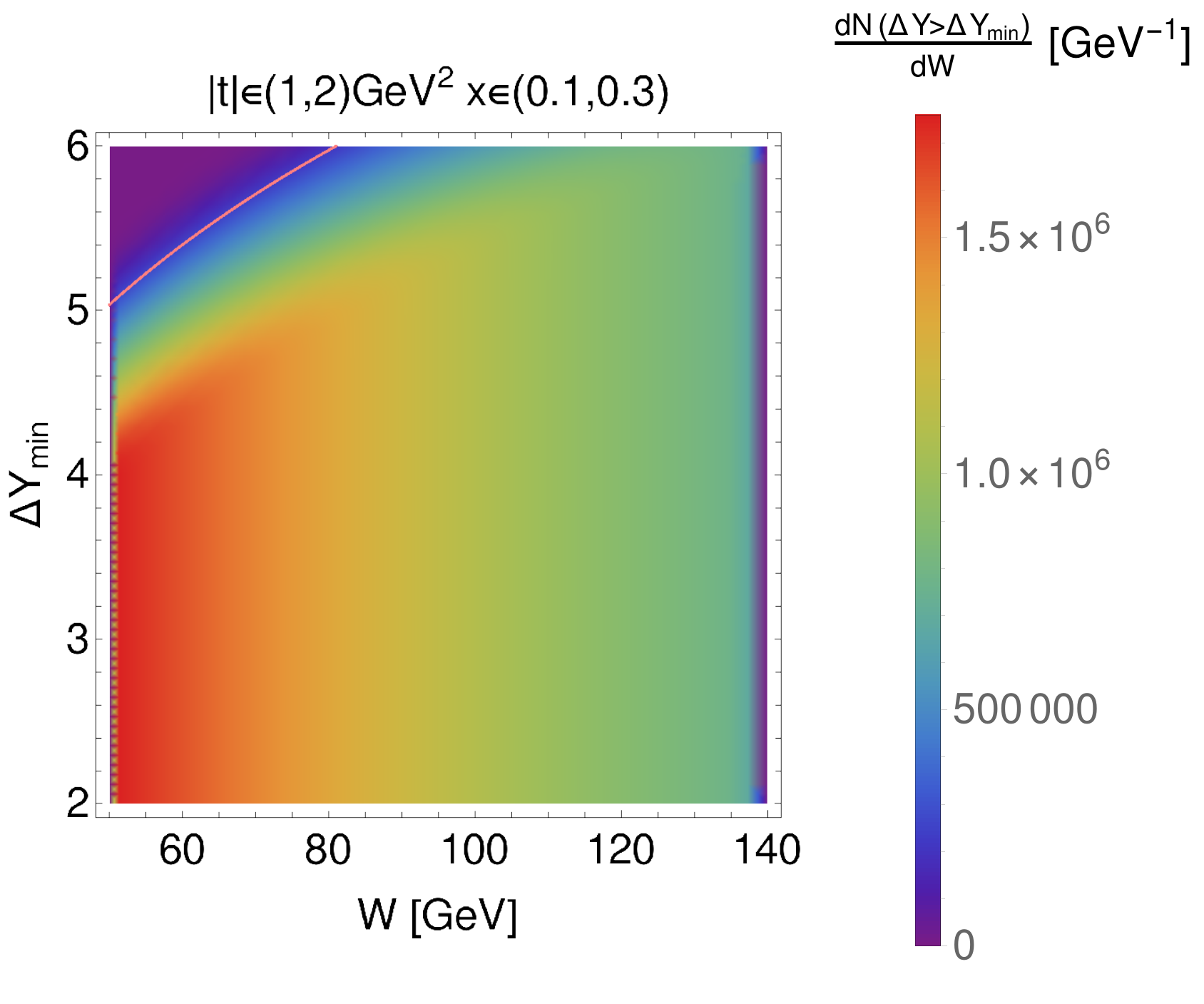}
\end{subfigure}
\hspace{2cm}
\begin{subfigure}{6cm}
\includegraphics[width=8cm, height=5.4cm]{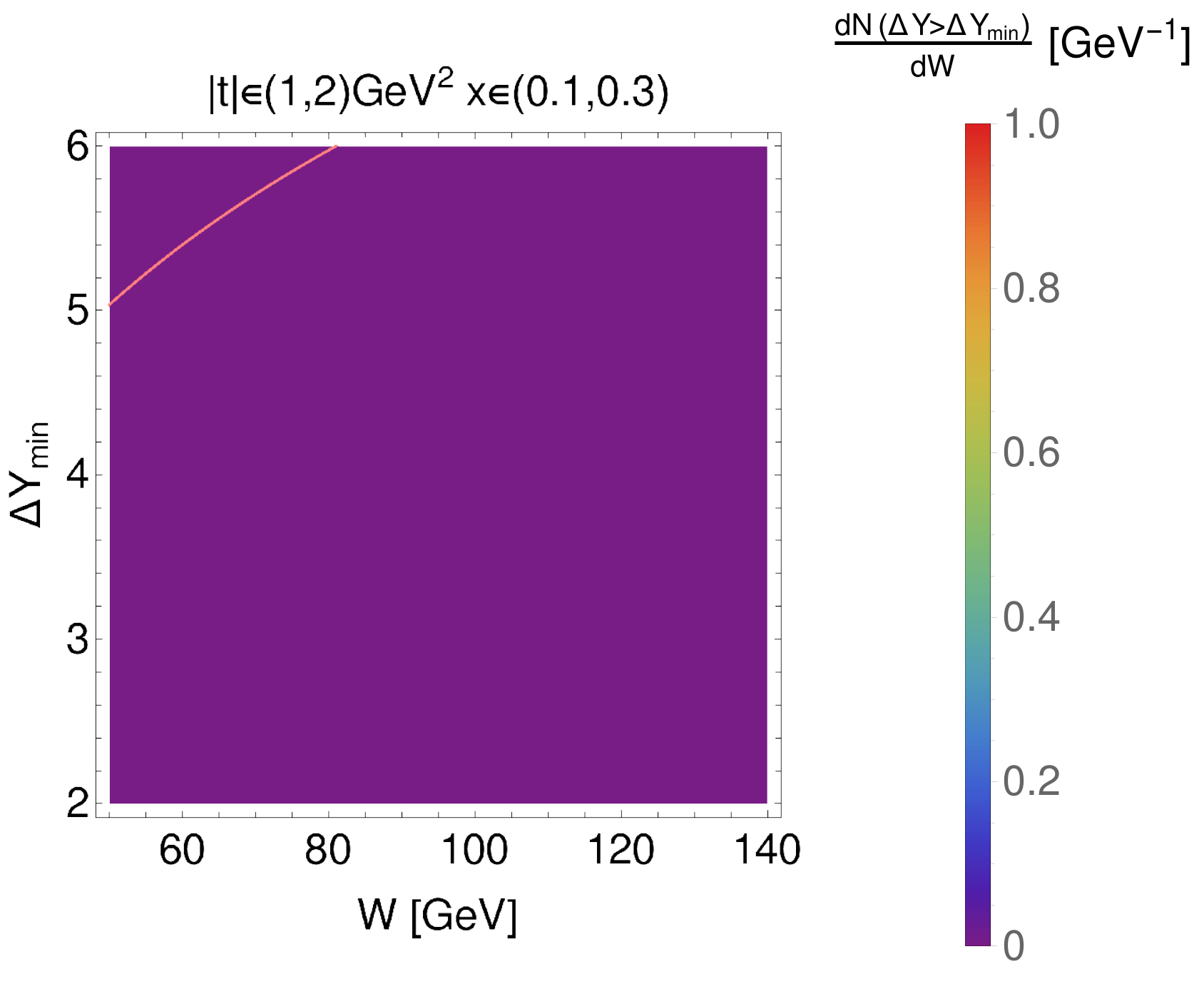}
\end{subfigure}}
\hspace*{-2cm}\centering{
\begin{subfigure}{6cm}
\includegraphics[width=8cm, height=5.4cm]{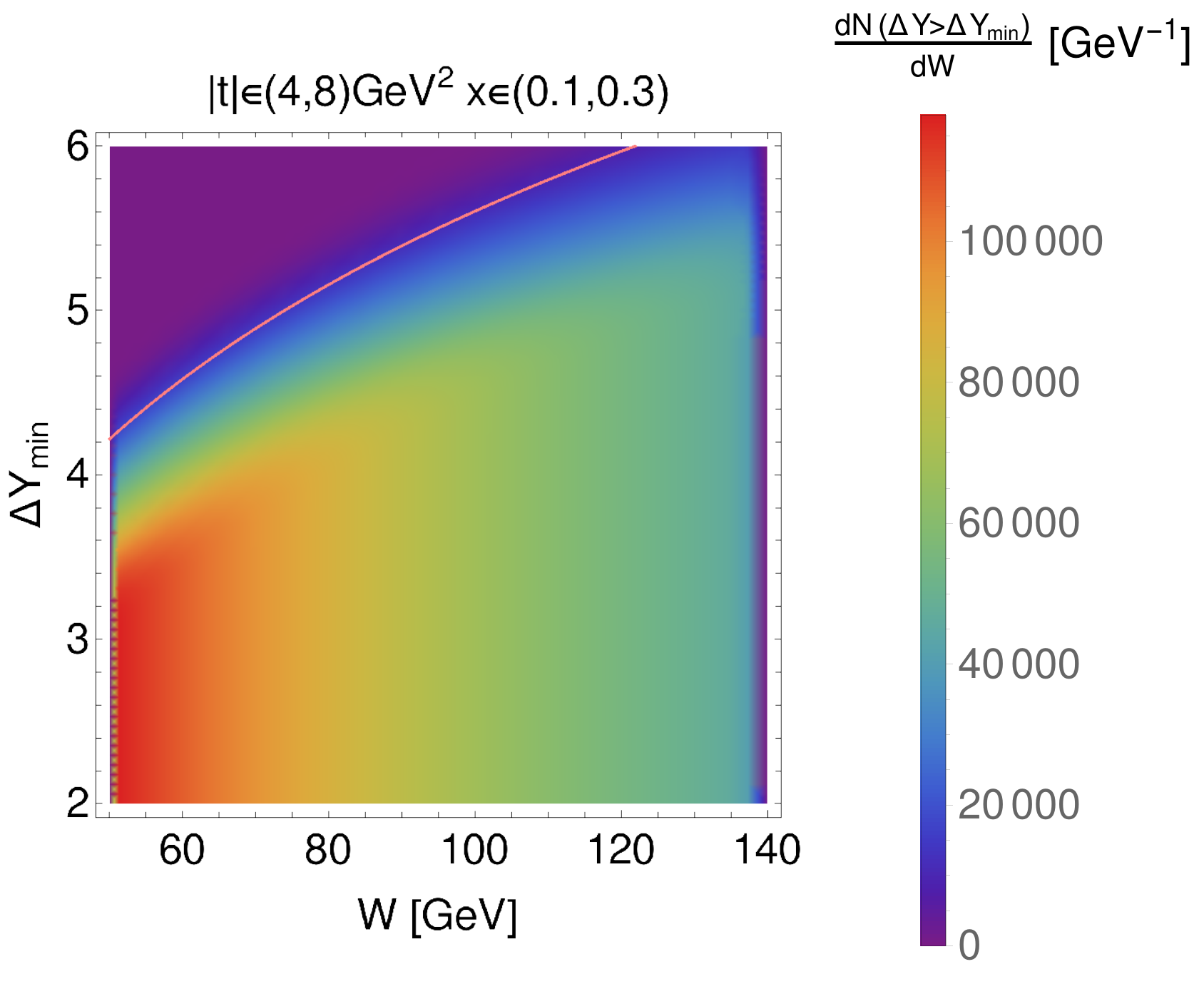}
\end{subfigure}
\hspace{2cm}
\begin{subfigure}{6cm}
\includegraphics[width=8cm, height=5.4cm]{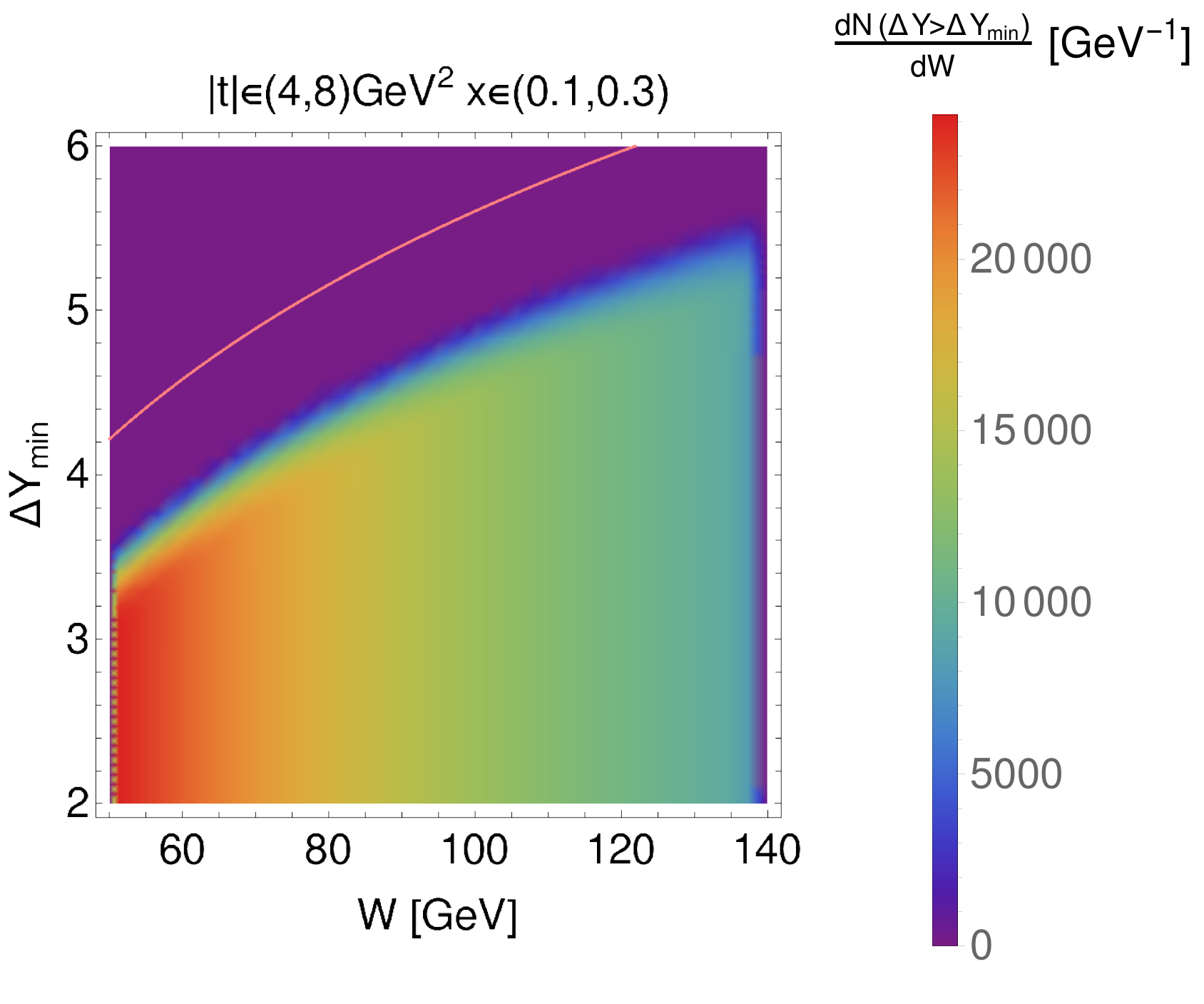}
\end{subfigure}}
\caption{Differential number of events in $W$ in bins in $t$ and $x$ as a two-dimensional function of $W$ and $\Delta Y_{\rm min}$. Left column: no cuts on angle, right column: restriction on angles $4^{\circ}$. Bin in  $x\in\left(0.1,0.3\right)$.  Upper row: bin in $|t|\in (1,2) { \rm GeV^2}$, lower row:   bin in $|t|\in (4,8) { \rm GeV^2}$.
 Integrated luminosity ${\cal L }=10 \;\rm fb^{-1}$. 
}
\label{fig:tbinnoacut3}
\end{figure}

This behavior is also illustrated in two-dimensional plots in Figs.~\ref{fig:tbinnoacut1}-\ref{fig:tbinnoacut3}), where we show the structure of the phase space of vector meson production in the diffractive photoproduction process. Here, the   number of events  differential in $W$ as a function of $W$ and $\Delta Y_{\rm min}$ is plotted. To be more precise, what we see in the plot are values of multiplicity given by a cross section evaluated for given values of $W$ and $\Delta Y_{\rm min}$ integrated over $x$ and $t$ in given bins and integrated over $\Delta Y>\Delta Y_{\rm min}$. The pink line shows the exact kinematical limit - the area above the pink line is kinematically forbidden.

The  kinematical limit Figs.~\ref{fig:tbinnoacut1}-\ref{fig:tbinnoacut3} drawn as a pink line in $(W, \Delta Y)$-space is given by an equation for maximal rapidity allowed for given $W$, maximal $x$ in a given $x$-bin $x_{max}^b$ and maximal $t$ (note $t$ is negative) in a given $t$-bin $t_{max}^{b}$:
\begin{equation}
Y_{max}^{(b1,b2)}=\ln{\frac{x_{max}^{b1}W^2}{\sqrt{-t_{max}^{b2}\left(-t_{max}^{b2}+M_{V}^2\right)}}}\;.
\label{eq:Wintegrated_rates}
\end{equation}
On the left hand side of Figs.~\ref{fig:tbinnoacut1}-\ref{fig:tbinnoacut3} shown are the  plots without any restriction by the angular coverage of the detector. On the right hand side of Figs.~\ref{fig:tbinnoacut1}-\ref{fig:tbinnoacut3} activity in the detector above $4^{\circ}$ is required. Events with no activity in the detector except $J/\psi$ and recoiled electron are discarded.
We see that the addition of a cut on the additional activity in the detector - shown in Figs. ~\ref{fig:tbinnoacut1}-\ref{fig:tbinnoacut3} on the right hand side - reveals, that for certain bins in $x$ and $t$ this cut acts as a veto on the vector meson production in this process. Similar behavior can be seen in plots in Figs.~\ref{fig:intW1}-\ref{fig:intW1b}. The cut is more effective for bins of larger $x$ and smaller $t$, since large $x$ and small $t$ mean lower angle at which the $J/\psi$ meson is produced. This is particularly striking when comparing top and bottom rows in the second column of Fig.\ref{fig:tbinnoacut3}, which correspond to two different bins in $t$ and large values of $x$.

\begin{figure}[h]
\centering{
\begin{subfigure}{6cm}
\includegraphics[width=6cm, height=4cm]{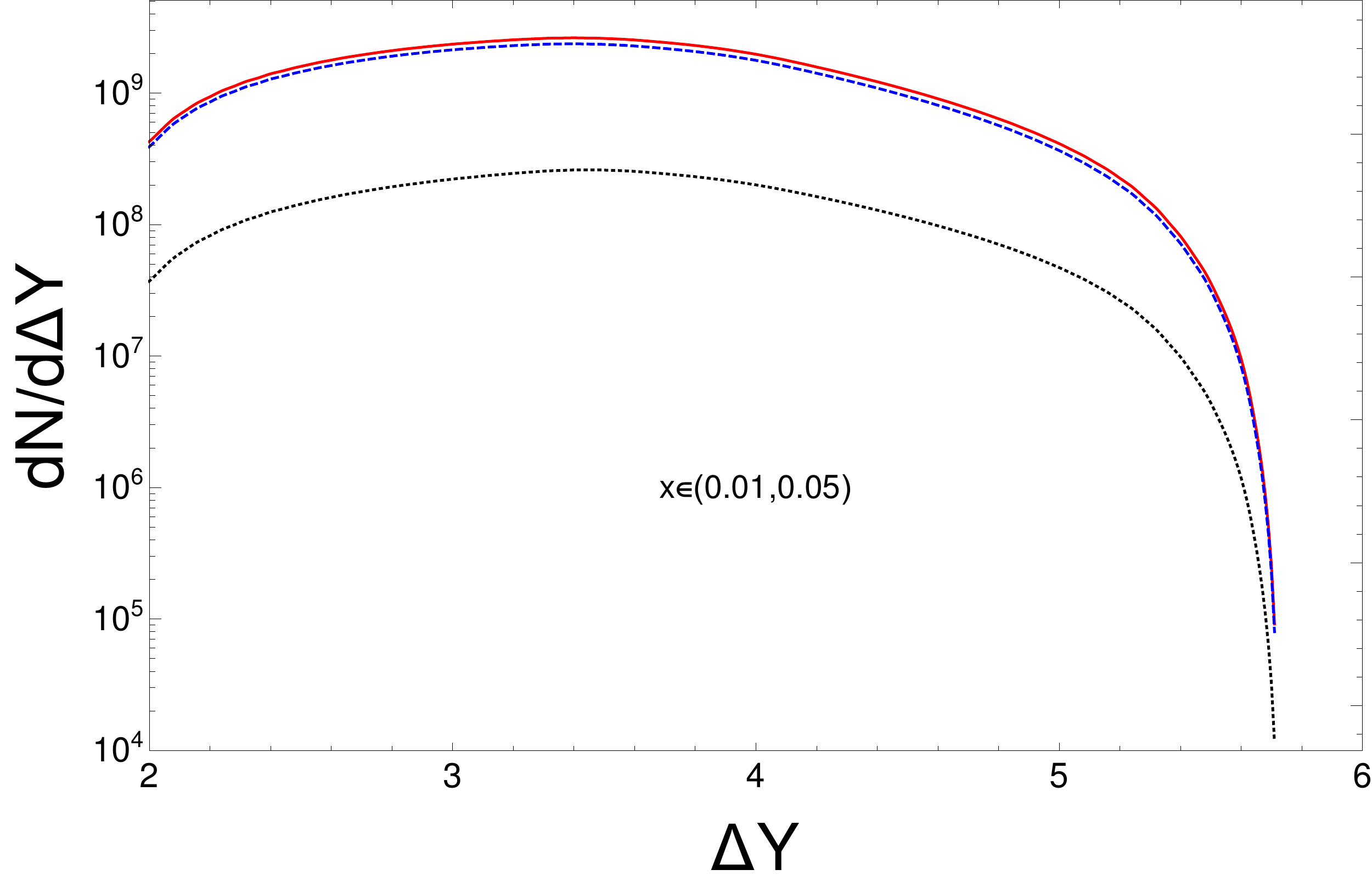}
\end{subfigure}
\hspace{1cm}
\begin{subfigure}{6cm}
	\includegraphics[width=6cm, height=4cm]{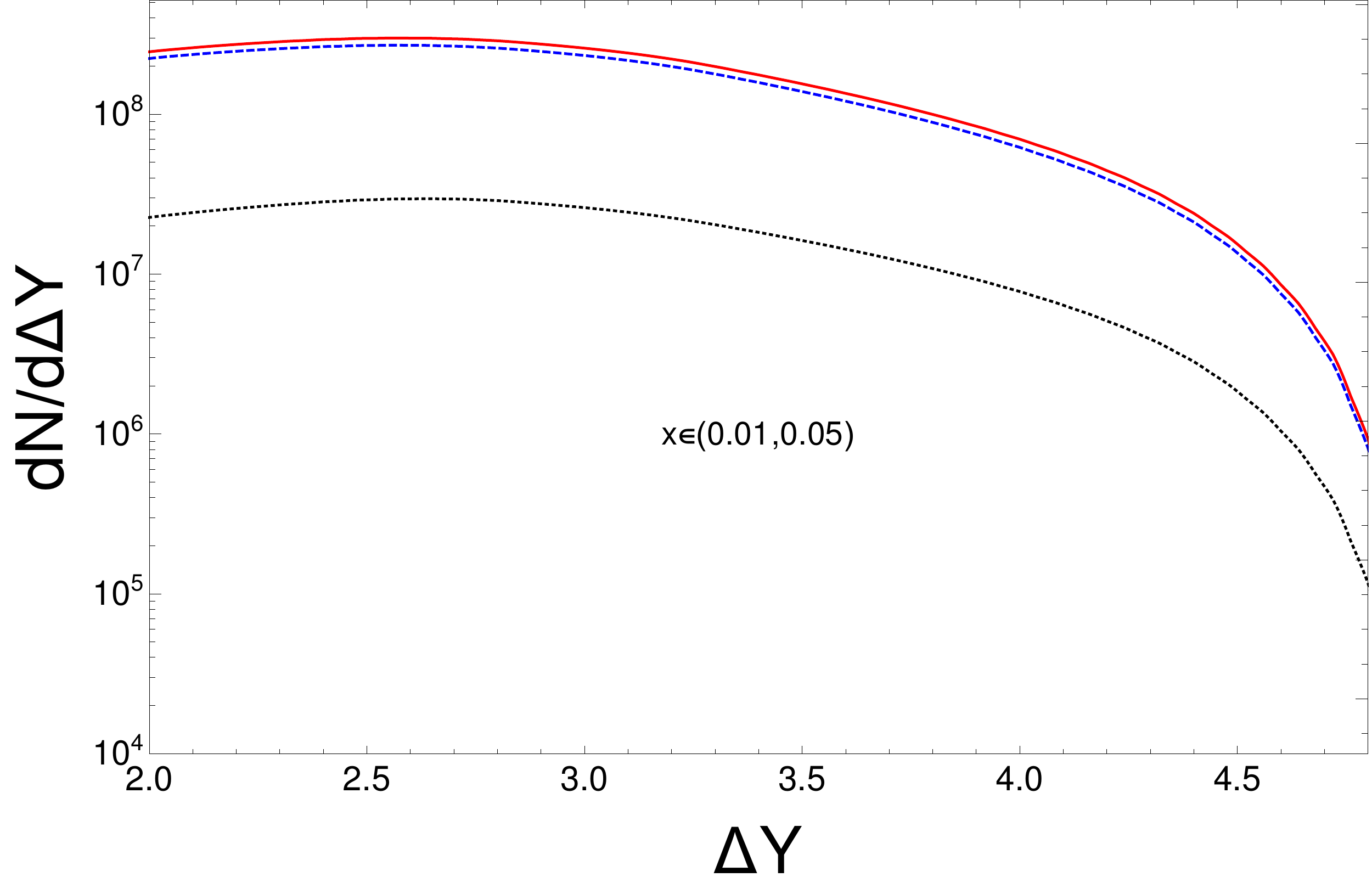}
\end{subfigure}}
\vspace*{0.5cm}
\centering{
\begin{subfigure}{6cm}
\includegraphics[width=6cm, height=4cm]{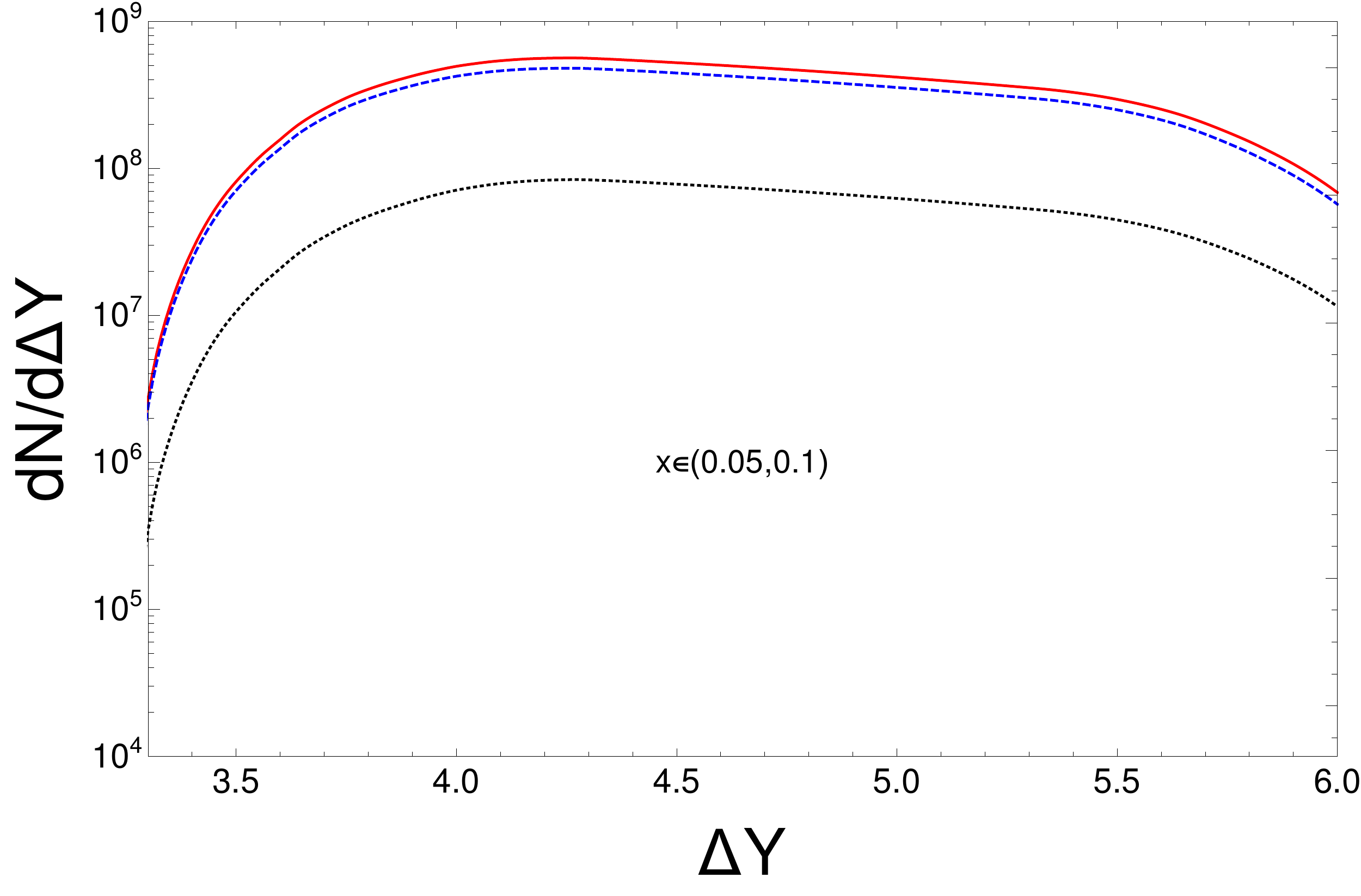}
\end{subfigure}
\hspace{1cm}
\begin{subfigure}{6cm}
	\includegraphics[width=6cm, height=4cm]{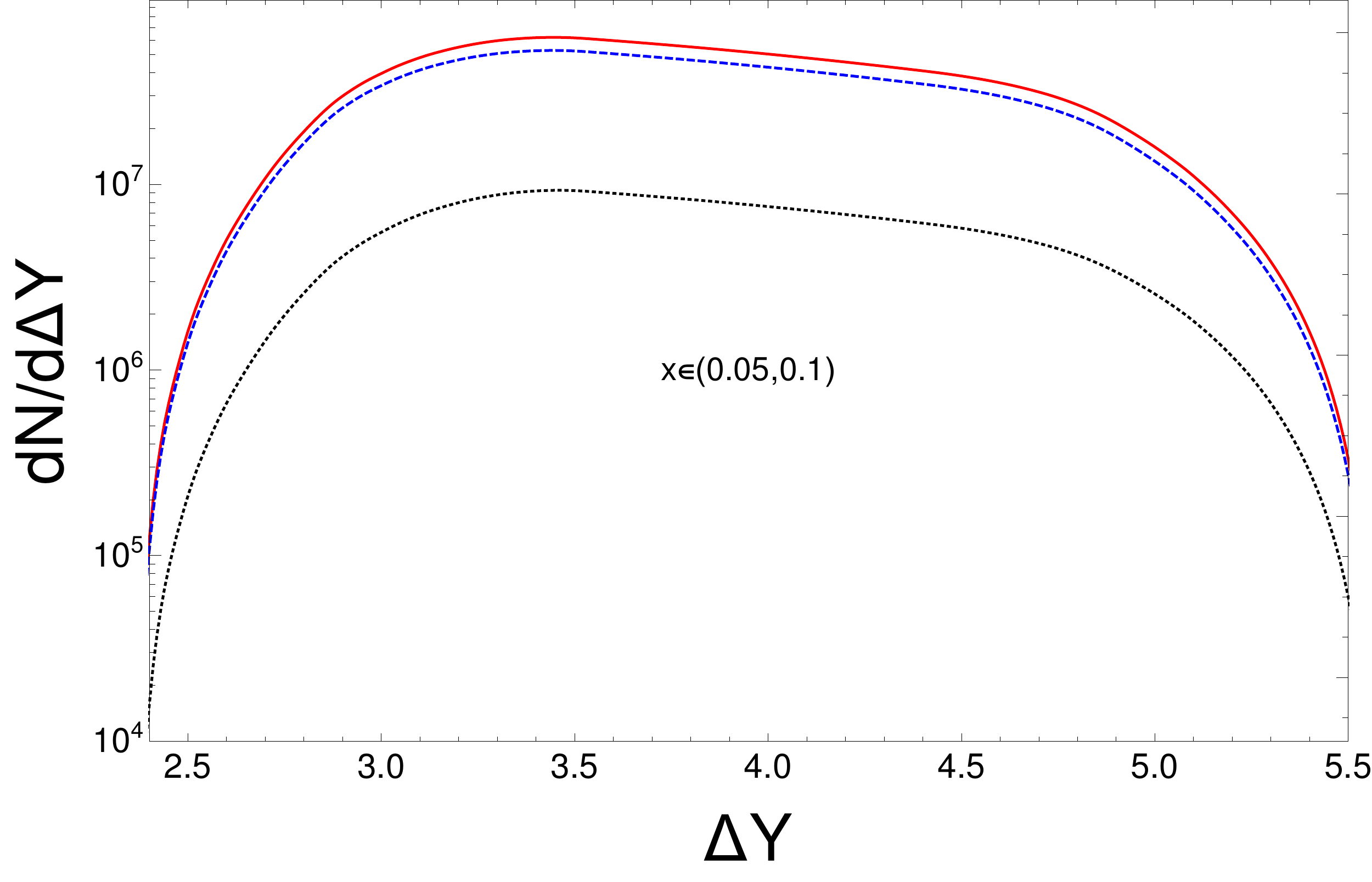}
\end{subfigure}}
\vspace*{0.5cm}
\centering{
\begin{subfigure}{6cm}
\includegraphics[width=6cm, height=4cm]{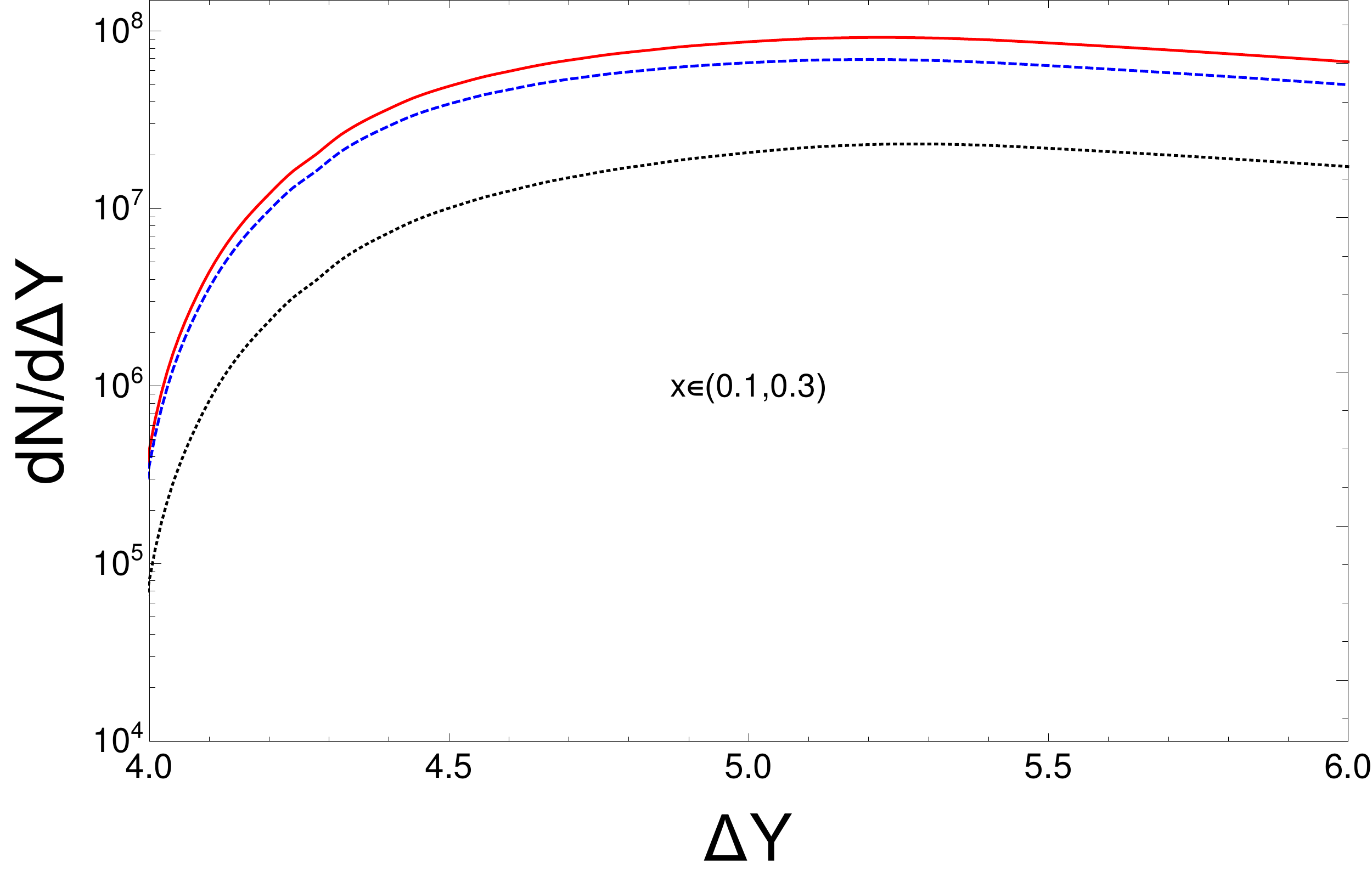}
\end{subfigure}
\hspace{1cm}
\begin{subfigure}{6cm}
	\includegraphics[width=6cm, height=4cm]{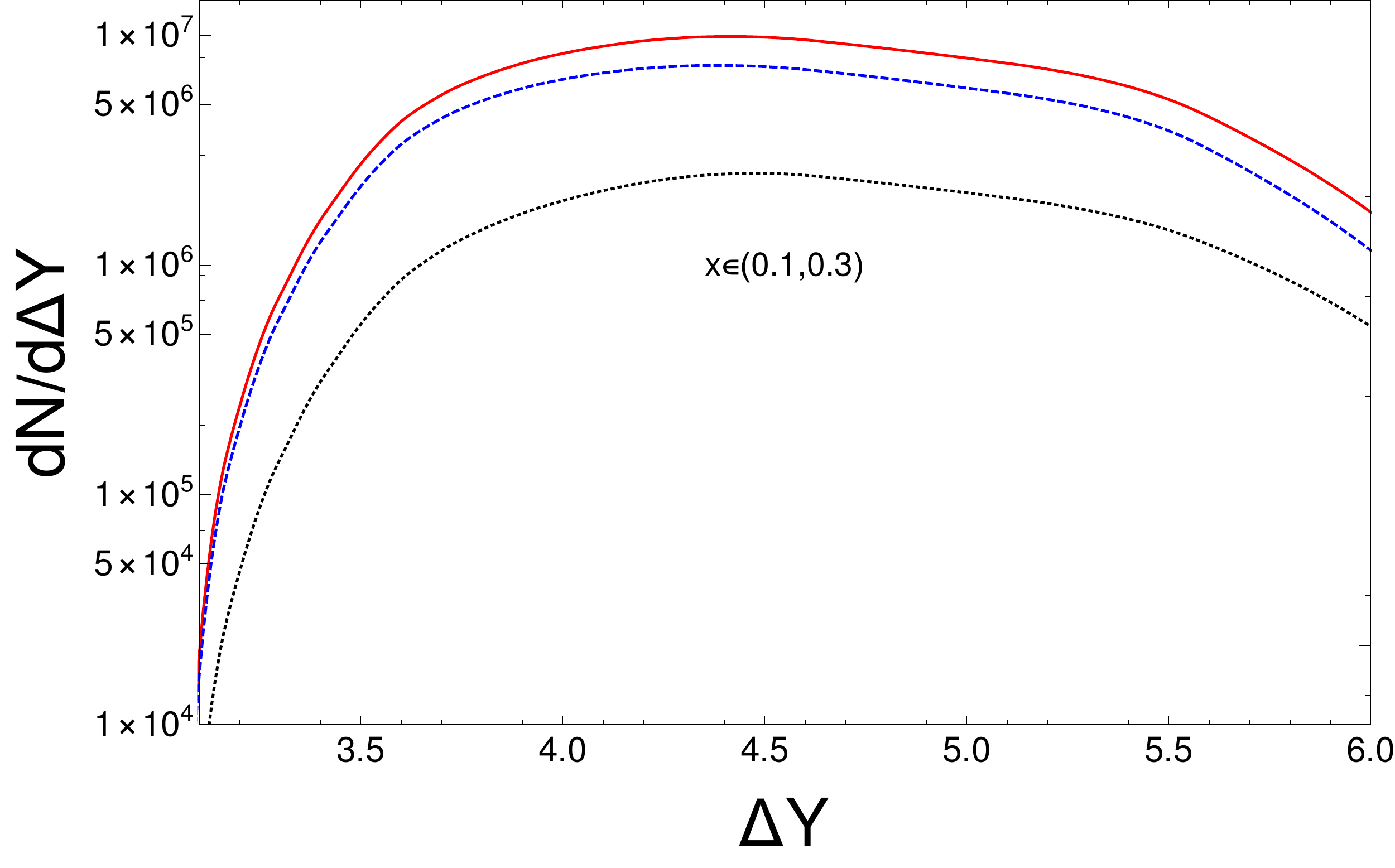}
\end{subfigure}}
\caption{Dependence of number of events on $\Delta Y$  over various bins in $x$ and $t$.
	Left column:  bin in $|t|\in\left(1,2\right)\;$GeV$^2$, right column:  bin in $|t|\in\left(4,8\right)\;$GeV$^2$. Rows from top to bottom bins in x: $x\in\left(0.01,0.05\right)$, $x\in\left(0.05,0.1\right)$, $x\in\left(0.1,0.3\right)$. Black line: quark contribution. Blue line: gluon contribution. Red line: sum of contributions. Integrated luminosity ${\cal L }=10 \;\rm fb^{-1}$. 
}
\label{fig:oDpdYY1fx}
\end{figure}

In Figs.~\ref{fig:oDpdYY1fx} the number of events as a function of the rapidity gap size $\Delta Y$ for various bins of $x$ and $t$. In these plots we can see not only the effect of the energy cuts ($W>50\;$GeV and $W<140\;$GeV) which manifest as sharp cut-offs (their position depending on $x$ and $t$) and the photon flux on the $\Delta Y$ dependence, but also the absolute contributions of the gluon channel and the quark channel on the multiplicity of the produced mesons.

\section{Conclusions}

In this work we have considered  $J/\psi$ production in the rapidity gap process at large momentum transfer $t$. In this kinematics diffusion in transverse momenta is suppressed and one can investigate QCD dynamics in the large rapidity gap region.
Two experimental strategies were considered - measuring  the gap size directly, by observing the activity in the forward detector , and putting a lower limit on the gap size. We have used a model based on the BFKL evolution  which describes the HERA data  to estimate the counting rates at the EIC.

We have found that  a much higher luminosity of the EIC than of HERA may partially compensate for a lower energy of the EIC. As a result one can test  at the EIC   dependence of the cross section 
on rapidity gap interval predicted by the    BFKL model for rapidity gaps up four units in rapidity.
The growth  by a  factor 1.6 per unit of rapidity is predicted which should be easy to measure if a detector has a good acceptance in the discussed kinematics.

This is possible with the present setup of the EIC detector which projects the coverage up to $3.5$ units of rapidity. However, the extension of the detector up to higher rapidity, for example to $4.5$ would provide even better lever arm. 
Importance of a  good detector acceptance in the nucleon fragmentation region for such studies is crucial. 

Though we considered only process of the $ J/\psi$ production, a  rapidity gap production of $\rho$-mesons maybe feasible in an even broader $t$ range since at large $|t|$  the rates of production of $J/\psi$ and $\rho$ become comparable, while the probabilities of the two body decay modes ($\pi ^+\pi^-$) and $(e^+e^-) + (\mu^+\mu^-)$ differ by a factor  of 9.
As mentioned earlier EIC will also perform $eA$ collisions in addition to $ep$, and thus it will offer possibility to investigate rapidity gaps in the presence of nuclei \cite{Frankfurt:2008et}. 
 Estimates for the 	LHeC and FCC-eh should also be performed, to test the range of rapidity gaps and inform the detector designs. 
	Finally, studies of rapidity gap  in the $J/\psi$ process are feasible also in ultra peripheral pA and AA collisions at the LHC \cite{Baltz:2007kq, Frankfurt:2008er}. Detailed analysis of this kinematics will be considered elsewhere.

\section*{Acknowledgments}

This work  was supported by the  Department of Energy Grants No. DE-SC-0002145 and DE-FG02-93ER40771, as well as the National Science Centre, Poland, Grant No.\ 2019/33/B/ST2/02588.

\newpage


%

\clearpage

\bibliography{mybib}
\bibliographystyle{unsrt}

\end{document}